[1994/12/01]

\newif\ifpdf\ifx\pdfoutput\undefined\pdffalse\else\pdfoutput=1\pdftrue\fi

\ifpdf

\documentclass[pdftex]{amsart}
\usepackage[dvips]{graphicx}
\usepackage[usenames,dvipsnames]{color}
\usepackage[pdftex,colorlinks,citecolor=blue]{hyperref} 
\else
\documentclass[colordvi]{amsart}
\usepackage{color}
\usepackage[hypertex,citecolor=blue]{hyperref} 
\fi

\usepackage{amscd}    
\usepackage{amsthm}   
\usepackage{amsfonts}
\usepackage{amssymb}
\usepackage{amsmath}
\usepackage{verbatim} 
\usepackage{alltt}    

\newcommand{\TODO}[1]{} 

\newcommand{\field}[1]{\ensuremath{\mathbb{F}_{#1}}}        
\newcommand{\ket}[1]{\ensuremath{|#1\rangle}}               
\newcommand{\bra}[1]{\ensuremath{\langle #1 |}}             
\newcommand{\braket}[2]{\ensuremath{\langle #1 |#2\rangle}} 
\newcommand{\brakett}[3]{\ensuremath{\langle #1 |#2| #3 \rangle}} 

\newtheorem{hwk}{Homework}[section]
\newtheorem{thm}{Theorem}[section]
\newtheorem{theorem}[thm]{Theorem}
\newtheorem{defn}[thm]{Definition}
\newtheorem{lemma}[thm]{Lemma}
\newtheorem{cor}[thm]{Corollary}
\newtheorem{note}[thm]{Note}
\newtheorem{conj}[thm]{Conjecture}

\newcommand{\minA}{\ensuremath{65}}
\newcommand{\minB}{\ensuremath{735}}
\newcommand{\tbound}{\frac{2}{\pi}\sqrt{\frac{22\ln^2
N}{L}+\frac{32N^2}{LM}}}
\newcommand{\mr}{\frac{\sqrt{N}}{\epsilon}}
\newcommand{\round}[1]{\ensuremath{\left\lfloor #1\right\rceil}}  
\newcommand{\floor}[1]{\ensuremath{\left\lfloor #1\right\rfloor}} 
\newcommand{\ceil}[1]{\ensuremath{\left\lceil #1\right\rceil}}    

\begin{document}

\title{The Hidden Subgroup Problem - Review and open problems}
\author{Chris Lomont, Cybernet}
\date{Oct 2004}
\thanks{clomont@cybernet.com, clomont@math.purdue.edu}

\begin{abstract}
An overview of quantum computing and in particular the Hidden
Subgroup Problem are presented from a mathematical viewpoint.
Detailed proofs are supplied for many important results from the
literature, and notation is unified, making it easier to absorb
the background necessary to begin research on the Hidden Subgroup
Problem. Proofs are provided which give very concrete algorithms
and bounds for the finite abelian case with little outside
references, and future directions are provided for the nonabelian
case. This summary is current as of October 2004.
\end{abstract}

\maketitle


\tableofcontents

\section{Introduction}

The main purpose of this paper is to give a self contained
explanation of the Hidden Subgroup Problem in quantum computing. A
second goal is to bring the interested reader to the forefront of
research in the area, so that a wider audience can attack the
problems. The final goal is to present this at a level accessible
to graduate students in math, physics, and computer science.
Prerequisites are some abstract algebra, linear algebra, and an
understanding of (classical) computation. However almost any
mathematically inclined reader should be able to learn something
from this presentation.

\subsection{Importance}
The importance of the Hidden Subgroup Problem (from now on
labelled the HSP) is that it encompasses most of the quantum
algorithms found so far that are exponentially faster than their
classical counterparts. Research in this area is centered on
extending the families of groups for which the HSP can be
efficiently solved, which may improve other classically
inefficient algorithms, such as determining graph isomorphism or
finding the shortest vector in a lattice. Finally, there are many
group theoretic algorithms that are more efficient on a quantum
computer, such as finding the order of a finite group given a set
of generators.

\subsection{History}
In 1994, Shor \cite{Shor94}, building on the work of Deutsch
\cite{Deu85} and Simon \cite{simon94power}, found a quantum
algorithm that could factor integers exponentially faster than any
known classical method, and opened the floodgates on quantum
computing research. Efficient integer factoring breaks the
ubiquitous RSA cryptosystem. Shor also gave an algorithm solving
the Discrete Log Problem (DLP), which is used in several other
cryptosystems. Kitaev \cite{Kitaev95} noted that these algorithms
as well as others fit in a framework of finding subgroup
generators from a group using a function that ``hides" the
subgroup, and thus the Hidden Subgroup Problem was born. For more
history, the book by Chuang and Neilsen \cite{ChuangNielsen}
contains a wealth of information, as well as the quantum physics
archives at \href{http://arxiv.org/archive/quant-ph}{{\tt{
http://arxiv.org/archive/quant-ph}}}.

\subsection{Notation}\label{s:notation1}
Here we fix some notation used throughout this paper. All $\log$s
are base 2 unless otherwise specified. $\mathbb{C}$ denotes the
field of complex numbers. $\mathbb{Z}$ is the ring of integers,
and for a positive integer $N$ we let $\mathbb{Z}_N$ denote the
ring of integers mod $N$. For each integer $N>0$ let
$\omega_N=\exp(2\pi i/N)$, a principal $N^{\text{th}}$ root of
unity. Quantum mechanics specific notation is in section
\ref{s:QuantumOverview} and appendix \ref{s:QuantumDetails}.

\subsection{Layout}
The layout of this paper is as follows. Section 2 covers the
necessary quantum mechanics and notation used therein. It also
introduces a quantum computing model well suited to present the
rest of the topics in this paper. Section 3 explains the algorithm
solving the abelian case of the HSP efficiently, describing in
detail the mathematics making it work. Section 4 generalizes the
examples from section 3 to give the a more general form of the
HSP, suitable for any finite group. Section 5 covers recent
results, and what is currently known about the HSP, as well as
quantum algorithms for other group related problems. Section 6
concludes. Much of the background and details are included in
numerous appendices, giving details on topics such as the
necessary background for the graph isomorphism reduction,
generating groups from random samples, number theory results, etc.


\section{Quantum Computing Model}\label{s:QuantumOverview}

\subsection{The Rules and Math of Quantum Mechanics}
Here we define the rules of quantum mechanics (from a mathematical
perspective). Details can be seen in Appendix
\ref{s:QuantumDetails}.

First some notation used in quantum mechanics. We define the
following symbols:

$\ket{\psi}$ represents a column vector in some complex Hilbert
space $V$, of finite dimension for this paper. For this section,
let this dimension be $N$. Quantum mechanics forces us to use an
orthonormal basis for $V$, so we fix the orthonormal
\emph{standard basis} $\mathcal{B}=\{\ket{0},\ket{1},
\dots,\ket{N-1}\}$. Then $\bra{\psi}$ denotes the conjugate
transpose row vector, often viewed as the dual to $\ket{\psi}$
with respect to $\mathcal{B}$. For example, we compute as follows:

If $\ket{\psi}=\sum_ia_i\ket{i}$, then $\bra{\psi}=\sum_i
a_i^*\bra{i}$, where $*$ denotes complex conjugation.
$\bra{i}\;\ket{j}$, written $\braket{i}{j}$, equals 1 if $i=j$,
otherwise \braket{i}{j} equals 0. A basis for linear operators on
$V$ can be written as a $\mathbb{C}$-linear combination of the
operators $\ket{i}\bra{j}$, which is the matrix with a 1 in the
$(i,j)$ entry, and 0's elsewhere. Thus any linear operator $A$ on
$V$ in the basis $\mathcal{B}$ can be written in the form
$A=\sum_{i,j}a_{i,j}\ket{i}\bra{j}$, which is the matrix with the
value $a_{i,j}$ in the $i,j$ entry, and acting on the left of a
column vector \ket{\psi}. $\brakett{\psi}{A}{\phi}$ is the inner
product of $\psi$ and $A\ket{\phi}$. Later the basis will often be
indexed with elements from a group $G$, viewed as fixing an
orthonormal basis and an injection mapping elements of $G$ to this
basis.

\subsubsection{The Postulates of Quantum Mechanics}
Now on to the physical content of quantum mechanics, abstracted to
a mathematical formalism. The content of quantum mechanics can be
summarized by 4 postulates, which we take as the definition of
quantum mechanics. They are:\footnote{Postulates are taken
verbatim from Neilsen and Chuang \cite{ChuangNielsen}.}

\textbf{Quantum Mechanics Postulate 1: State Space:} Associated to
an isolated physical system is a complex vector space with inner
product (a Hilbert space) known as the \emph{state space} of the
system. The system is completely described by its \emph{state
vector}, which is a unit vector in the system's state space.

\textbf{Quantum Mechanics Postulate 2: State Evolution:} The
evolution of a \emph{closed} quantum system is described by a
\emph{unitary transformation}\footnote{Recall a unitary operator
$U$ satisfies $UU^\dag=I=U^\dag U$ where $\dag$ is conjugate
transpose. In particular, unitary operators are invertible,
implying quantum computation is \emph{reversible}, which differs
significantly from classical computing.}. That is, the state of a
system \ket{\psi} at time $t_1$ is related to the state
\ket{\psi'} at time $t_2$ by a unitary operator $U$ which depends
only on the times $t_1$ and $t_2$,
\begin{equation}
\ket{\psi}=U\ket{\psi'}
\end{equation}

\textbf{Quantum Mechanics Postulate 3: State Measurement:} Quantum
measurements are described by a collection $\{M_m\}$ of
\emph{measurement operators}. These are operators acting on the
state space of a system being measured. The index $m$ refers to
the measurement outcomes that may occur in the experiment. If the
state of the system is \ket{\psi} immediately before the
measurement, then the probability that result $m$ occurs is given
by
\begin{equation}
p(m) = \brakett{\psi}{M_m^\dag M_m}{\psi}
\end{equation}
and the state of the system after the measurement is
\begin{equation}
\frac{M_m\ket{\psi}}{\sqrt{p(m)}}
\end{equation}
The measurement operators satisfy the \emph{completeness equation}
\begin{equation}
\sum_m M_m^\dag M_m = I
\end{equation}

\textbf{Quantum Mechanics Postulate 4: State Combining:} The state
space of a composite physical system is the tensor product of the
state spaces of the component systems. Moreover, if we have
systems numbered 1 through $n$, and system number $j$ is prepared
in the state \ket{\psi_j}, then the joint state of the total
system is
$\ket{\psi_1}\otimes\ket{\psi_2}\otimes\dots\ket{\psi_n}$.

We will explain briefly how these postulates are used in practice
for quantum computing.

\subsubsection{Qubits and Operators}
Analogous to the bit being the basic block in classical computing,
the qubit is the basic building block in quantum computing.
Formally we define
\begin{defn}[Qubit]A \emph{\textbf{qubit}} (or quantum-bit) is a unit vector in
$\mathbb{C}^2$. We fix an orthonormal basis of column vectors
denoted as $\ket{0}=\left(\begin{array}{c}1\\0\end{array}\right)$
and $\ket{1}=\left(\begin{array}{c}0\\1\end{array}\right)$,
corresponding to classical bits 0 and 1.
\end{defn}

\begin{defn}[State vector] The \emph{\textbf{state}} of a quantum
system is a (column) vector in some vector space, written
\ket{\psi}.
\end{defn}

By postulate 4, we can combine single qubits as follows.
\subsubsection{Qubits Galore} Similar to concatenating $n$ classical
bits to ``bitstrings", we concatenate qubits to get larger
systems. Two qubits form a space spanned by the four vectors
\begin{equation}
\ket{0}\otimes\ket{0}, \;\;\ket{0}\otimes\ket{1},
\;\;\ket{1}\otimes\ket{0}, \;\;\text{and }\ket{1}\otimes\ket{1}
\end{equation}
where the tensor product is the usual vector space tensor. See
Appendix \ref{s:QuantumDetails} for details. Shorthand for the
above expressions is
\begin{equation}
\ket{00}, \;\;\ket{01}, \;\;\ket{10}, \;\;\text{and }\ket{11}
\end{equation}

Now we can check the second basis element (dictionary ordering)
\begin{eqnarray}
\ket{01} &=
\ket{0}\otimes\ket{1}\\
&=\binom{1}{0}\otimes\binom{0}{1}\\
&=\left(\begin{matrix}1\binom{0}{1}\\0\binom{0}{1}\end{matrix}\right)
&=\left(\begin{matrix}0\\1\\0\\0\end{matrix}\right)
\end{eqnarray}
and we get the second usual basis element of $\mathbb{C}^4$. This
works in general; that is, the vector corresponding to the state
$\ket{n}$ where $n$ is a binary number, is the $(n+1)^\text{th}$
standard basis element. We frequently use decimal shorthand:
$\ket{32}$ is the 33rd standard basis vector in some space which
would be clear from context.

Thus the the state of an $n$-qubit system is a unit vector in
$\mathbb{C}^{2^n}$. Note that the state of $n$ classical bits is
described by $n$ elements each either 0 or 1, while the state of
$n$ qubits requires $2^n$ complex numbers to describe. Thus it
seems qubits contain much more ``information" than classical bits.
Unfortunately we cannot retrieve all this ``information" from the
state; we are limited by quantum mechanics due to the fact that
measuring the state destroys information.

\subsubsection{Measurement}

The final operation we need to understand about qubits is
measurement, the process of getting information out of a quantum
state. There are several equivalent ways to think about it. We
will cover the easiest to understand, intuitively and
mathematically. However, to gain precise control over
measurements, often one has to resort to an equivalent, yet more
complicated, measurement framework\footnote{This is the ``Positive
Operator-Valued Measure" (POVM) formalism.}, which we do not
discuss here. See Nielsen and Chuang \cite[Ch. 2]{ChuangNielsen}.

We will do our measurements in the \emph{computational} basis
$\{\ket{0},\ket{1},\dots,\ket{2^n-1}\}$ over an $n$-qubit system.
Suppose we have the state
$\ket{\psi}=\sum_{j=0}^{2^n-1}a_j\ket{j}$, which is a unit vector
in $\mathbb{C}^{2^n}$. Measuring in the computational basis has
the following effect: it returns the state $\ket{j}$ with
probability $p_j=|a_j|^2$, and after the measurement, the state
becomes $\ket{\psi'}=\ket{j}$. Thus measuring ``collapses" the
waveform, returning a state with probability the square of its
coefficient (\emph{amplitude}), and the resulting state is the one
returned by the measurement. Thus from a given state, we return
one answer depending on the basis we measure, and destroy all
other information about the state.

Finally we note that cascaded measurements (one after the other)
can always be replaced by a single measurement.

\subsubsection{The No Cloning Theorem}
As an example of the using above postulates, we prove an important
and surprising theorem:
\begin{thm}{\textbf{The No Cloning Theorem.}} It is impossible to
build a machine that can clone any given quantum state.
\end{thm}
This is in stark contrast to the classical case, where we copy
information all the time. It is the tip of the iceberg for the
differences between quantum and classical computing.
\begin{proof}
Suppose we have a machine with two slots: $A$ for the quantum
state \ket{\psi} to be cloned, and $B$ in some fixed initial state
\ket{s}, and the machine makes a copy of the quantum state $A$. By
the rules of quantum mechanics, the evolution $U$ is unitary, so
we have
\begin{equation}
\ket{\psi}\otimes\ket{s}\xrightarrow{U}\ket{\psi}\otimes\ket{\psi}
\end{equation}
Now suppose we have two states we wish to clone, \ket{\psi} and
\ket{\varphi}, giving
\begin{eqnarray*}
U\left(\ket{\psi}\otimes\ket{s}\right) =& \ket{\psi}\otimes\ket{\psi}\\
U\left(\ket{\varphi}\otimes\ket{s}\right) =&
\ket{\varphi}\otimes\ket{\varphi}
\end{eqnarray*}
Taking the inner product of these two equations, and using $U^\dag
U=I$:
\begin{eqnarray*}
\left(\bra{\varphi}\otimes\bra{s}\right)U^\dag U\left(\ket{\psi}\otimes\ket{s}\right) &=& \left(\bra{\varphi}\otimes\bra{\varphi}\right)\left(\ket{\psi}\otimes\ket{\psi}\right)\\
\braket{\varphi}{\psi}\braket{s}{s} &=& \braket{\varphi}{\psi}\braket{\varphi}{\psi}\\
\braket{\varphi}{\psi} &=& \left(\braket{\varphi}{\psi}\right)^2
\end{eqnarray*}
This has solutions if and only if \braket{\varphi}{\psi} is 0 or
1, so cloning cannot be done for general states.\footnote{There is
a lot of research on precisely what can be cloned, how to
approximate cloning, and what other limitations there are to
duplicating quantum states.}
\end{proof}

\subsection{Efficient Quantum Computation}\label{s:EfficientQC}
\subsubsection{Quantum Computing} Quantum sates are transformed by
applying unitary operators to the state. So where classical
computing can be viewed as applying transforms to $n$-bit systems,
quantum computation proceeds by constructing an $n$-qubit machine,
applying unitary operators to the state until some desired state
is found, and then measuring the result. This paper will avoid the
physical construction of such machines, and focus on the unitary
transformations that seem likely to be physically realizable, and
the computational outcomes of such systems. Again, for an
introduction to the physical issues, see \cite[Ch.
7]{ChuangNielsen} and the references therein.

\subsubsection{Circuit Model} Similar to being able to construct
any classical circuit with NAND gates, there are finite
\footnote{There are many ways to choose them. See for example
\cite{BBCD95}.} sets of quantum gates that allow the construction
of any unitary operator to a desired precision. Kitaev
\cite{Kitaev97} shows that these approximations can be done with
minimal overhead, allowing quantum computation to be modelled with
simple ``quantum circuits". A final note on quantum circuits is
that Deutsch's Quantum Turing Machine \cite{Deu85} and the circuit
model used more recently were shown equivalent by Yao
\cite{Yao93}. We will use a few quantum gates that operate on 1,2,
or 3 qubits at a time, defined later. The intuitive description is
that quantum computations are built of quantum circuits, which are
composed of quantum gates, and each quantum gate operates on only
a few qubits at a time. This statement mirrors the classical one
with ``quantum" removed and qubits replaced with bits.

\subsubsection{Quantum Circuit Size}\label{s:sizedepth} In loose terms, efficient
classical computations are done on small circuits, in the sense
that as the problem size grows, the size of the circuit required
to solve the problem grows at a certain rate, usually bounded
polynomially in the size of the problem. We want to make this
precise in the quantum context.

The following is just a mathematically precise way to say our
``elementary operations" only operate on a few qubits at a time,
which is desirable since it makes quantum computation physically
plausible. Some definitions:

\begin{defn}
Given a $2^n$-dimensional vector space $V$ with basis
$\mathcal{B}$, and a $2^m\times 2^m$ matrix $U$ with $m\leq n$, an
\emph{\textbf{expansion of $U$ relative to $\mathcal{B}$}} is any
matrix of the form $G(U\otimes I_{2^{n-m}})G^{-1}$ where $G$
permutes the basis, and $I_k$ is the $k\times k$ identity matrix.
\end{defn}
This just says each expansion of $U$ operates on $m$ of the $n$
qubits in a $n$-qubit machine. In general $m$ will be small, $n$
will vary, and we will build computations by composing these
operators.
\begin{defn}
Given a $2^n$-dimensional vector space $V$, an orthonormal basis
$\mathcal{B}$ of $V$, and a finite set
$\mathcal{U}=\left\{U_1,U_2,\dots\,U_k\right\}$ of unitary
matrices of dimensions dividing $2^n$, then the set of
\emph{\textbf{elementary operations relative to
$(\mathcal{B},\mathcal{U})$}} consists of all expansions of the
$U_i$ relative to $\mathcal{B}$.
\end{defn}
This just allows us to consider all operations on any subset of
$n$ qubits generated from our initial set of ``elementary
operations". Note $U$ unitary and $\mathcal{B}$ orthonormal
implies expansions of $U$ relative to $\mathcal{B}$ are unitary.

For our use $V$ will be the state space of a quantum system, clear
from context, and $\mathcal{B}$ will be the standard orthonormal
basis of $V$. We fix a specific generating set
$\mathcal{U}_\tau=\left\{H,CNOT,CCNOT,P\right\}$ relative to such
a fixed basis to be the matrices
\begin{eqnarray}
H&=&\frac{1}{2}\left(\begin{matrix}1 & 1 \\ 1 &
-1\end{matrix}\right) \text{  the Hadamard matrix}\\
CNOT&=&\left(\begin{matrix}1&0&0&0\\0&1&0&0\\ 0&0&0&1\\0&0&1&0\end{matrix}\right) \text{  the controlled NOT}\\
CCNOT&=&\left(a_{ij}\right)\text{ with }a_{ii}=1,i=1,...,6,
a_{87}=a_{78}=1, \\&&\text{ the rest }=0, \text{  the controlled controlled NOT}\notag\\
P&=&\left(\begin{matrix}e^{i\frac{\theta}{2}}&0\\0&e^{-i\frac{\theta}{2}}\end{matrix}\right)
\text{  the phase matrix, where }\cos\theta=\frac{3}{5}.
\end{eqnarray}
For any $n>2$ and using the standard basis $\mathcal{B}$ defined
earlier, the elementary operations from this set of 4 matrices
generates a group dense in $U(2^n)$, the space of legal quantum
operations on an $n$-qubit machine\footnote{From chapter 4
exercises in \cite{ChuangNielsen}.}. So from now on one can assume
these 4 matrices and associated elementary operations are the
legal set of elementary operations on any $n$-qubit machine. The
definitive paper on elementary gates for quantum computing is
\cite{BBCD95}.

\begin{defn}
A \emph{\textbf{quantum circuit}} is a unitary matrix built from
composing elementary operations from $\mathcal{U}_\tau$
\end{defn}

Now mathematically, quantum computing becomes the following. We
have an initial state \ket{0} in the $n$-qubit space
$\mathbb{C}^{2^n}$. Applying unitary transformations that are
products of the elementary transformations, we want to obtain a
quantum state (unit vector \ket{\psi}) that, when measured, has a
high probability of returning some useful answer. We want to know
how ``efficient" such transformation are. We restrict legal
quantum operations to those obtained from the elementary
operations from some finite set, such as $\mathcal{U}_\tau$.

\begin{defn}
The \emph{\textbf{size}} of a quantum circuit will be the minimal
number of elementary operations composed to obtain it.
\end{defn}
This gives us a way to measure the complexity of a quantum
operation. From here on we can assume all quantum operation
complexities are measured against our set of elementary operations
coming from $\mathcal{U}_\tau$ and a corresponding $V$ and
$\mathcal{B}$ taken from context.

Often it is possible to rearrange the elementary operations and
obtain the same quantum circuit. For example if adjacent
operations affect disjoint sets of qubits, these two operations
can be swapped obtaining the same circuit (the matrices commute).
Similar to parallelizing classical circuits, this reordering
allows us to partition the sequence of elementary operations into
ordered lists of operations, where within each list a qubit is
affected by at most one operation. This leads to the notion of
depth:
\begin{defn}
The \emph{\textbf{depth}} of a quantum circuit is the minimal
length of a partition of the ordered elementary operations
composing the circuit into ordered lists where each qubit is
affected at most once per list.
\end{defn}
As a result, we always have \textbf{depth}$\leq$\textbf{size}.
{~}\\

To parallel the quantum to classical terminology, we sometimes
call a state (or part of a state) a quantum register. Physically a
quantum state is basically constructed using $n$ particles which
can be either of two states 0 or 1 when measured. If we take a
subset of these particles, and operate on them, it is convenient
to call this subset a register.
\begin{defn}
A \emph{\textbf{register}} in a quantum computer is a subset of
the total set of qubits. We often write $\ket{a}\ket{b}$ to denote
that the first register is in state \ket{a} and the second in
state \ket{b}, the number of qubits in each set being understood
from context.
\end{defn}

\subsubsection{Efficient Quantum Computation}
Most of this paper is concerned with \emph{efficient} quantum
computation. Sometimes this has two components: needing an
efficient quantum process, and an efficient classical computing
method to post-process the data output from the quantum process to
obtain the desired answer. We will see these two are (often)
separate issues.

Given a problem to solve on a quantum computer, we need a way to
represent the problem as a quantum state, a unitary operation $U$
built from elementary operations to convert this quantum state to
a final state, and a way to process the final state to obtain the
desired answer. Although a precise definition of ``efficient"
takes us too far afield, we will make it precise in special cases
throughout this paper. The general idea is that as the ``size" of
the input grows (the number of qubits required to represent the
problem, say), the size of the necessary quantum operator $U$
should grow polynomially in the size of the input.

An example: suppose we want to determine the order of finite
abelian groups given a generating set for each one. Given a group
$|G|$, we can represent each element using roughly $\log |G|$
qubits. To call a quantum algorithm efficient for this problem
would mean the size of the quantum circuit computing the order of
$G$ should be of size polynomial in $\log |G|$, as $G$ varies
throughout the \emph{family} of finite abelian groups.

As a final technical point, we require what is called a ``uniform
class of algorithms," meaning that, for a problem of size $n$,
there is a Turing machine that given $n$, can produce the circuit
description in number of steps equal to a polynomial in $n$. This
ensures that we can (in theory) construct an explicit machine to
solve each problem in time polynomial in the size of the problem.

For more information on quantum complexity, see
\cite{BV97,Cleve99}.

\subsubsection{A Note on Probabilistic Algorithms}
Quantum computers are probabilistic, meaning that algorithms tend
to be of the form ``Problem A is solved with probability 80\%."
For those used to thinking that algorithms solve problems with
certainty (such as algorithms encountered in a first algorithms
class), note that probabilistic algorithms suffice in practice. We
just run the experiment a few times, and take the majority result.
This returns the correct answer with probability exponentially
close to 1 in the number of trials. Precisely we use the following
theorem:
\begin{thm}[The Chernoff Bound]
Suppose $X_1,X_2,\dots,X_n$ are independent and identically
distributed random variables, each taking the value 1 with
probability $1/2+\epsilon$ and 0 with probability $1/2-\epsilon$.
Then
\begin{equation}
p\left(\sum_{i=1}^n X_i\leq\frac{n}{2}\right) \leq
e^{-2\epsilon^2n}.
\end{equation}
\end{thm}
Thus the majority is wrong very rarely. For example, we will make
most algorithms succeed with probability 3/4, so our
$\epsilon=1/4$. Although it sounds like a lot, taking 400
repetitions of the algorithm causes our error to drop below
$10^{-20}$, at which point it is more likely our computer fails
than the algorithm fails. And since the algorithms we are
considering are usually exponentially faster than classical ones,
there is still a net gain in performance. If we do 1000 runs, our
error drops below $10^{-55}$, at which point it is probably more
likely you'll get hit by lightning while reading this sentence
than the algorithm itself will fail. For completeness, here is a
proof of the Chernoff Bound.
\begin{proof}
Consider a sequence $(x_1,x_2,\dots,x_n)$ containing at most $n/2$
ones. The probability of such a sequence is maximized when it
contains $\lfloor n/2 \rfloor$ ones, so
\begin{eqnarray}
p\left(X_1=x_1,X_2=x_2,\dots,X_n=x_n\right)&\leq&
\left(\frac{1}{2}-\epsilon\right)^{\frac{n}{2}}
\left(\frac{1}{2}+\epsilon\right)^{\frac{n}{2}}\\
=\frac{(1-4\epsilon^2)^{\frac{n}{2}}}{2^n}.
\end{eqnarray}
There can be at most $2^n$ such sequences, so
\begin{equation}
p\left(\sum_{i=1}^n X_i\leq\frac{n}{2}\right) \leq
2^n\times\frac{(1-4\epsilon^2)^{\frac{n}{2}}}{2^n} =
(1-4\epsilon^2)^{\frac{n}{2}}.
\end{equation}
From calculus, $1-x\leq \exp(-x)$, so
\begin{equation}
p\left(\sum_{i=1}^n X_i\leq\frac{n}{2}\right) \leq e^{-4\epsilon^2
n/2} = e^{-2\epsilon^2 n}
\end{equation}

\end{proof}


\section{The Abelian Hidden Subgroup Problem}
We will detail the Hidden Subgroup Problem (HSP), starting with
some illustrative and historically earlier examples, before
covering the most general cases and research problems. The
simplest groups considered are the finite cyclic groups, followed
by finite abelian groups. Kitaev \cite{Kitaev95} examines a
similar problem over finitely generated abelian groups, but we
will not cover that here. The finite abelian case was first used
to spectacular effect by Shor \cite{Shor94} and Simon
\cite{simon94power}. The HSP for finite nonabelian groups is
currently researched for the reasons given in sections
\ref{s:GeneralProblem} and \ref{s:NonabelianResults}.

Related to the HSP over finite groups is the Abelian Stabilizer
Problem, in Kitaev \cite{Kitaev95}.

\subsection{Definition of the Hidden Subgroup
Problem}\label{s:HSPDefinition}

In order to set the stage for the rest of the paper, we make a
general definition of the Hidden Subgroup Problem, which we will
abbreviate HSP for the rest of this paper, and then attempt to
determine for which groups $G$ and subgroups $H$ we can solve the
HSP efficiently. We will also discuss partial results on groups
for which efficient HSP algorithms are not known.

\begin{defn}[Separates cosets]
Given a group $G$, a subgroup $H\leq G$, and a set $X$, we say a
function $f:G\rightarrow X$ \textbf{separates cosets} of $H$ if
for all $g_1,g_2\in G$, $f(g_1)=f(g_2)$ if and only if
$g_1H=g_2H$.
\end{defn}

\begin{defn}[The Hidden Subgroup Problem]
Let $G$ be a group, $X$ a finite set, and $f:G\rightarrow X$ a
function such that there exists a subgroup $H<G$ for which $f$
separates cosets of $H$. Using information gained from evaluations
of $f$, determine a generating set for $H$.
\end{defn}

For any finite group $G$, a classical algorithm can call a routine
evaluating $f(g)$ once for each $g\in G$, and thus determine $H$
with $|G|$ function calls. A central challenge of quantum
computing is to reduce this naive $O(|G|)$ time algorithm to
$O(\text{poly}(\log|G|))$ time (including oracle calls and any
needed classical post-processing time). This can be done for many
groups, which gives the exponential speedup found in most quantum
algorithms.

We assume an efficient encoding of $G$ and $X$ to basis states of
our quantum computer. We also assume a quantum ``black-box" that
operates in unit time for performing the unitary transform
$U_f\ket{g}\ket{x}=\ket{g}\ket{x\oplus f(g)}$, for $g\in G$, $x\in
X$, and $\oplus$ bitwise addition on the state indices.


\subsection{The Fast Fourier Transform}
The Fast Fourier Transform (FFT) of Cooley and Tukey
\cite{Cooley65} reduced the cost of doing Fourier transforms from
the naive $O(n^2)$ down to $O(n\log n)$, allowing a large class of
problems to be attacked by computers. Mikhail
Atallah\footnote{Private comment.} remarked the FFT is the most
important algorithm in computer science. The success of the FFT is
that so many other problems can be reduced to a Fourier transform,
from multiplication of numbers and polynomials to image processing
to sound analysis to correlation and convolution\footnote{Lomont
\cite{Lomont03a} has shown that there can be no quantum
correlation or convolution algorithms that parallel the quantum
Fourier transform.}. More references are Beth\cite{Beth87},
Karpovsky\cite{Kar77}, and Maslen and Rockmore \cite{MasRock01}.

Most, if not all, quantum algorithms that are exponentially faster
than their classical counterparts rely on a quantum Fourier
transform (QFT), and much of the rest of this document deals with
the QFT. For more information beyond this paper on the QFT see
Ekert and Jozsa\cite{EkJ96}, Hales and Hallgren\cite{HalHal99},
and Jozsa\cite{Joz98}.

Just as the FFT turned out to be a big breakthrough in classical
computing, exploiting the QFT so far is the central theme in
quantum algorithms. The main reason quantum algorithms are
exponentially faster is the QFT can be done exponentially faster
than the classical FFT. However there are limitations due to the
probabilistic nature of quantum states. 

\subsection{The Basic Example}\label{s:BasicExample}
Fix an integer $N>1$. Let $X$ be a finite set, and let
$G=\left<\mathbb{Z}_N,+\right>\;$ be the additive group of
integers mod $N$. Suppose we have a function (set map)
$f:G\rightarrow X$ such that there is a subgroup
$H=\left<d\right>$ of $G$, such that $f$ is constant on $H$ and
distinct on cosets of $H$, that is, $f$ separates cosets of $H$.
Let $M=|H|$. We assume we have a quantum machine\footnote{Recall
\ket{x}\ket{y} merely means $\ket{x}\otimes\ket{y}$ and is used as
shorthand.} capable of computing the unitary transform on two
registers $f:\ket{x}\ket{y}\rightarrow\ket{x}\ket{f(x)\oplus y}$,
where $\oplus$ is (qu)bitwise addition\footnote{Check this is
unitary, thus an allowable quantum operation.}. We do not assume
we know $M$ or $d$ or $H$; we only know $G$ and have a machine
computing $f$. We want to determine a generating set for $H$,
calling the ``black-box" function $f$ as few times as possible.
For now we ignore the size of the quantum circuit and focus on the
math making the whole process work. Later we will deal with
efficiency.

\begin{defn}[Quantum Fourier Transform (QFT)]
The quantum Fourier transform $F_N$ is the operator on a register
with $n\geq \log N$ qubits given by
\begin{eqnarray}\label{e:FNdefn}
F_N=\frac{1}{\sqrt{N}}\sum_{j,k=0}^{N-1}e^{\frac{2\pi i j
k}{N}}\ket{k}\bra{j}\end{eqnarray}
\end{defn}

Note later we will define the QFT over other groups, so this one
is actually the \emph{cyclic} QFT.

The $\frac{1}{\sqrt{N}}$ factor is required to make this a unitary
transformation\footnote{Homework!}, so it is a valid quantum
transformation. Map the group, which we view as integers added
$\bmod N$, into the basis of the quantum state, that is,
$G=\{\ket{0},\ket{1},\dots,\ket{N-1}\}$ and
$H=\{\ket{0},\ket{d},\ket{2d},\dots,\ket{(M-1)d}\}$. Compute on
two registers:
\begin{eqnarray}
\ket{0}\ket{0}&\xrightarrow{F_N\text{ on 1st}}& \frac{1}{\sqrt{N}}\sum_{j=0}^{N-1}\ket{j}\ket{0}\\
&\xrightarrow{\text{apply
}f}&\frac{1}{\sqrt{N}}\sum_{j=0}^{N-1}\ket{j}\ket{f(j)}
\end{eqnarray}
Measuring the second register to obtain some value $f(j_0)$
collapses the state, leaving only those values in the first
register that have $f(j_0)$ in the second register, namely the
coset $H+j_0$. This is where we needed that $f$ separates cosets
of $H$. This ``entanglement" is not present in classical
computation, and seems to be one source of the increased
computational power of quantum computing, another source being the
ability to do computations on $2^n$ state coefficients in parallel
by manipulating only $n$ qubits. We now drop the second register
which remains $\ket{f(j_0)}$.
\begin{eqnarray}
&\xrightarrow{\text{measure}}&\frac{1}{\sqrt{M}}\sum_{h\in H}\ket{j_0+h}\\
&=&\frac{1}{\sqrt{M}}\sum_{s=0}^{M-1}\ket{j_0+sd}\\
&\xrightarrow{\text{apply }F_N}&\frac{1}{\sqrt{M}}\sum_{s}\frac{1}{\sqrt{N}}\sum_{k=0}^{N-1}e^\frac{2\pi i (j_0+sd)k}{N}\ket{k}\\
&=&\frac{1}{\sqrt{MN}}\sum_{k=0}^{N-1}e^\frac{2\pi i
j_0k}{N}\ket{k}\sum_{s=0}^{M-1}e^\frac{2\pi i
sdk}{N}\label{e:abelian1}
\end{eqnarray}
Using $\frac{d}{N}=M$, evaluate the geometric series
\begin{eqnarray}
\sum_{s=0}^{M-1}e^\frac{2\pi i
sdk}{N}&=&\sum_{s=0}^{M-1}\left(e^\frac{2\pi i k}{M}\right)^s\\
&=&\left\{\begin{array}{ll}0&\text{ if }M\nmid k
\\M&\text{ if }M\mid k\end{array}\right.
\end{eqnarray}
So in expression \ref{e:abelian1}, only those values of $k$ that
are multiples of $M$ remain, simplifying to the superposition
\begin{eqnarray}
\ket{\psi_f}&=&\frac{1}{\sqrt{d}}\sum_{t=0}^{d-1}e^\frac{2\pi i
j_0 tM}{N}\ket{tM}\label{e:abelian2}
\end{eqnarray}
Now measuring at this point gives a multiple of $M$ in
$\{0,M,\dots,(d-1)M\}$ with uniform probability. All that remains
is to repeat this to get several multiples of $M$, and to take the
GCD to obtain $M$ with high probability. Computing the GCD with
the Euclidean algorithm\footnote{This is the oldest known
algorithm \cite{Knuth2}.} has complexity $O(\log^2|N|)$, where
$\log|N|$ is the number of digits in $N$.

To estimate how many trials we need, suppose we have obtained $k$
multiples of $M$, say the (possibly repeated) multiples
$t_1,\dots,t_k\in T=\{0,1,\dots,d-1\}$. We want to estimate the
probability that $\gcd(t_1,t_2,\dots,t_k)=1$, which would
guarantee we would obtain the true value of $M$, and hence
determine $H$ properly. By lemma \ref{l:trials} in appendix
\ref{s:GCDProbabilities},
$$\text{prob}\left(\gcd(t_1,t_2,\dots,t_k)=1\right)\geq
1-\left(\frac{1}{2}\right)^{k/2}$$

Thus a few runs of the algorithm determines $H$ with high
probability, for any size $N$ and $d$. To understand the complete
cost of the algorithm, we need the computational cost of the QFT,
which is shown next in section \ref{s:CyclicFT}. Then we show how
these pieces can be used to find hidden subgroups in any finite
abelian group in section \ref{s:AbelianGen}, and finally in
section \ref{s:StandardProbs} we show some applications.

Above we assume infinitely precise values in the operations making
the QFT. Since this is not physically reasonable, work has been
done to cover the case of slight errors in the precision of the
computations. Kitaev \cite{Kitaev95} and the error correction
methods of Calderbank and Shor \cite{CalSho96} are good places to
start, and show that it is still possible to sample multiples of
$M$ with high probability even with errors in the QFT, so the
process works.


\subsection{Computing the Fourier Transform on $\mathbb{Z}_N$ Efficiently}\label{s:CyclicFT}
In this section we want to show how to compute the quantum Fourier
transform $F_N$ on the cyclic group $\mathbb{Z}_N$ efficiently, or
at least approximate it to as high a precision as necessary. We
will do this in two steps: first we do it for the case $N=2^n$,
and then use this in the second step to do it for general $N$.
$F_N$ will be used to construct HSP algorithms for general finite
abelian groups. We make the next definition for general groups,
but reserve the more general QFT definition until section
\ref{s:GeneralQFT}.

\begin{defn}
A family of quantum circuits $\{U_i\}$ computing the quantum
Fourier transform over a family of finite groups $\{G_i\}$ is
called \emph{\textbf{efficient}} if $U_i$ has size polynomial in
$\log |G_i|$ for all $i$.
\end{defn}

Efficient quantum circuits for the Fourier transform over
$\mathbb{Z}_N$ are well studied. Kitaev \cite{Kitaev95} gives an
approximate method. Mosca and Zalka \cite{MZ03} use ``amplitude
amplification" \cite{BHT98} to give an exact method, but claim it
is unlikely to be of practical use. Mosca's \cite{Mos99} thesis
and Hales' thesis \cite{Hal02} both contain circuit descriptions.
Hales and Hallgren \cite{HalHal00} give the algorithm used in
appendix \ref{s:GenCyclicFT} for the general case. For practical
implementations of Shor's algorithm the ``semiclassical" version
given by Griffiths and Niu \cite{GriNiu95} would probably be the
best known choice. Cleve and Watrous \cite{CleWat00} have given
parallel algorithms, showing even more speed increases. Shor
\cite{Shor94} did the cyclic case for ``smooth" values of $N$, and
Coppersmith \cite{Copper94} gave an efficient algorithm for the
case $N=2^n$ as well as an approximate version. Brassard and
H{\o}yer \cite{BH97} show how to solve Simon's problem, and have a
useful framework for analyzing the general finite abelian HSP.

It has been said \cite{Hallgren00b} that ``The efficient algorithm
for the abelian HSP is folklore." This section attempts to clear
that up with precision.

\subsubsection{Reduction to Odd Order and $2^n$ Order.}
As mentioned in Mosca's thesis \cite[Appendix A.4]{Mos99}, it is a
fact that the Fourier transform $F_N$ over a composite $N=AB$,
with $(A,B)=1$, can be computed efficiently from the efficient
Fourier transforms over $A$ and $B$. We show this briefly.

We assume $(A,B)=1$, and we have efficient QFT algorithms for
$F_A$ and $F_B$. Let $U_B$ be the unitary transform $\ket{x\bmod
A}\xrightarrow{U_B}\ket{xB\bmod A}$, and similarly $\ket{y\bmod
B}\xrightarrow{U_A}\ket{yA\bmod B}$. Both $U_A$ and $U_B$ are
efficiently computable, since they are just multiplication,
followed by a remainder operation.

The main idea comes from the ring isomorphism $\mathbb{Z}_N\cong
\mathbb{Z}_A\times \mathbb{Z}_B$, given in one direction by
$j\rightarrow (j \bmod A,j\bmod B)$, and in the other direction by
$(j_1,j_2)\rightarrow j_1 B B^{-1}+j_2 A A^{-1}$, where
$AA^{-1}\equiv 1 \mod B$ and $B B^{-1}\equiv 1 \mod A$. These
statements required $(A,B)=1$. With this notation it is
instructive to check
\begin{align}
F_N&=\left(U_B\otimes U_A\right)\left(F_A\otimes F_B\right).
\end{align}

This reduces the general QFT over $\mathbb{Z}_N$ for general $N$
to the cases $N=2^n$ and $N$ odd. Finding QFT algorithms with time
complexity of $O(\text{poly}\log N)$ for each case thus results in
such an algorithm for any $N$, since $U_A$ and $U_B$ are
efficient.

Thus for our purposes it is enough to show how to compute $F_N$
efficiently for $N$ a power of two and for $N$ odd.

\subsubsection{The Case $N=2^n$}\label{s:powerNCase}
We start with the easiest case: $N=2^n$. We show an explicit
construction of the Fourier transform $F_N$, where $N=2^n$. This
presentation follows \cite[Ch. 5]{ChuangNielsen}, which in turn is
adapted from sources mentioned in their book.

We use the notation from section \ref{s:BasicExample}, specialized
to the case $N=2^n$. We write the integer $j$ in binary as
$j=j_12^{n-1}+j_22^{n-2}+\dots+j_n2^0$, or in shorthand, as
$j=j_1j_2\dots j_n$. We also adopt the notation $0.j_lj_{l-1}\dots
j_m=j_l/2+j_{l+1}/4+\dots+j_m/2^{m-l+1}$. Note the
Fourier\footnote{Note that the Fourier coefficients can be viewed
as group homomorphisms $\omega_N^{k}:\mathbb{Z}_N\rightarrow
\mathbb{C}^{*}$, taking $a\rightarrow \omega_N^{ka}$. This
viewpoint generalizes well.} operator $F_N$ sends a basis element
$\ket{j}$ to
$\frac{1}{\sqrt{N}}\sum_{k=0}^{N-1}\omega_N^{jk}\ket{k}$. The
inverse transform has $\omega_N^{-1}$ instead of $\omega_N$. Then
we can derive a formula giving an efficient way to compute the
Fourier transform:
\begin{eqnarray}
\quad\quad F_N\ket{j}&=&\frac{1}{\sqrt{N}}\sum_{k=0}^{2^n-1}e^{\frac{2\pi i j k}{2^n}}\ket{k}\\
&=&\frac{1}{\sqrt{N}}\sum_{k_1=0}^1\sum_{k_2=0}^1\dots\sum_{k_n=0}^1e^{2\pi
i j \sum_{l=1}^{n}k_l2^{-l}}\ket{k_1k_2\dots k_n}\\
&=&\frac{1}{\sqrt{N}}\sum_{k_1=0}^1\sum_{k_2=0}^1\dots\sum_{k_n=0}^1\bigotimes_{l=1}^{n}e^{2\pi
i j k_l2^{-l}}\ket{k_l}\\
&=&\frac{1}{\sqrt{N}}\bigotimes_{l=1}^{n}\left[\sum_{k_l=0}^1e^{2\pi
i j k_l2^{-l}}\ket{k_l}\right]\\
&=&\frac{1}{\sqrt{N}}\bigotimes_{l=1}^{n}\left[\ket{0}+e^{2\pi
i j 2^{-l}}\ket{1}\right]\\
&=&\frac{\left(\ket{0}+e^{2\pi i
0.j_n}\ket{1}\right)\left(\ket{0}+e^{2\pi i
0.j_{n-1}j_n}\ket{1}\right)\dots\left(\ket{0}+e^{2\pi i
0.j_1j_2\dots j_n}\ket{1}\right)}{\sqrt{N}}\label{e:FNout}
\end{eqnarray}
where in the last step we used $\exp\left(2\pi i j
2^{-l}\right)=\exp\left(2 \pi i j_0j_1\dots j_{n-l}.j_{n-l+1}\dots
j_n\right)=\exp\left(2 \pi i 0.j_{n-l+1}\dots j_n\right)$. Using
this expression, we exhibit a quantum circuit (unitary operator)
using $O((\log N)^2)$ elementary operations that transforms the
state $\ket{j}$ into the one shown in equation \ref{e:FNout}.

We need two types of unitary\footnote{A careful reader should
check these are unitary.} operations, $H^{(a)}$ and $R_k^{(a,b)}$,
where $a$ and $b$ index the qubits in the quantum machine, as
follows\footnote{Note Chuang and Nielsen in \cite{ChuangNielsen}
denote $R_k$ as a single qubit operator, ours is what they would
call a controlled $R_k$.}: Let
$H^{(a)}=\frac{1}{\sqrt{2}}\left(\begin{matrix}1 & 1
\\ 1 & -1\end{matrix}\right)$ be the standard Hadamard operator, applied to qubit $a$, and let
$R_k^{(a,b)}$ be the operator on qubits $a$ and $b$ given by
$$R_k^{(a,b)}=\left(\begin{matrix}1 & 0 & 0 & 0\\0 & 1 & 0 & 0\\0 & 0 & 1 & 0\\0 & 0 & 0 &
\omega_{2^k}\end{matrix}\right)$$ where $\omega_N=e^{\frac{2 \pi
i}{N}}$ is the standard primitive $\text{N}^{\text{th}}$ root of
unity. $R_k^{(a,b)}$ has the effect of multiplying the phase of
the \ket{1} component of qubit $b$ by $\omega_{2^k}$ if and only
if qubit $a$ is \ket{1}, and is called a controlled phase change.
For example, looking at the two-qubit state,
\begin{align}
\left(\alpha\ket{0}+\beta\ket{1}\right)\;\ket{1}&\xrightarrow{R_5^{(2,1)}}&
\left(\alpha\ket{0}+\beta e^{2\pi i/2^5}\ket{1}\right)\ket{1}
\end{align}
Note each $H^{(a)}$ and $R_k^{(a,b)}$ is a local interaction on
the quantum state, so we will count the number of them needed to
implement a Fourier transform.

Apply to the state $\ket{j_1j_2\dots j_n}$ the operator
$R_n^{(n,1)}R_{n-1}^{(n-1,1)}\dots R_2^{(2,1)}H^{(1)}$. We have
\begin{eqnarray}
\ket{j_1j_2\dots
j_n}&\xrightarrow{H^{(1)}}&\frac{1}{\sqrt{2}}\left(\ket{0}+e^{2\pi
i 0.j_1}\ket{1}\right)\ket{j_2j_3\dots
j_n}\\
&\xrightarrow{R_2^{(2,1)}}&\frac{1}{\sqrt{2}}\left(\ket{0}+e^{2\pi
i 0.j_1j_2}\ket{1}\right)\ket{j_2j_3\dots
j_n}\\
&\dots&\\
&\xrightarrow{R_2^{(n,1)}}&\frac{1}{\sqrt{2}}\left(\ket{0}+e^{2\pi
i 0.j_1j_2\dots j_n}\ket{1}\right)\ket{j_2j_3\dots j_n}
\end{eqnarray}
This required $n$ local operations.

Apply to the state $\ket{j_2j_3\dots j_n}$ the operator
$R_n^{(n,2)}R_{n-1}^{(n-1,2)}\dots R_2^{(3,2)}H^{(2)}$, which
changes only the second qubit, resulting similarly in
\begin{equation}
\frac{1}{\sqrt{2}}\left(\ket{0}+e^{2\pi i 0.j_1j_2\dots
j_n}\ket{1}\right)\frac{1}{\sqrt{2}}\left(\ket{0}+e^{2\pi i
0.j_2\dots j_n}\ket{1}\right)\ket{j_3j_4\dots j_n}
\end{equation}
which required $n-1$ operations. Repeating this process uses
$1+2+\dots+n=\frac{n(n+1)}{2}$ local operations and results in the
state
\begin{equation}
\frac{\left(\ket{0}+e^{2\pi i 0.j_1j_2\dots
j_n}\ket{1}\right)\frac{1}{\sqrt{2}}\left(\ket{0}+e^{2\pi i
0.j_2\dots j_n}\ket{1}\right)\dots\left(\ket{0}+e^{2\pi i
0.j_n}\ket{1}\right)}{\sqrt{N}}
\end{equation}
Noting this is similar to equation \ref{e:FNout}, we finish the
Fourier transform by reversing the order of the qubits with
approximately $\lfloor\frac{n}{2}\rfloor$ unitary qubit swaps.
Thus the total number of operations, each affecting at most 2
qubits, is $O(n^2)=O(\log^2 N)$. We get an exact $F_N$ transform
with this method.

Most discussions avoid the following point. Notice as $N$ grows,
so the number of basic operations $R_k$ grows as $\log N$, and
this seems like cheating. If for each $N=2^n$ we use only $H$ and
$R_n$, we may construct $R_m,\; 0\leq m\leq n$ as $(R_n)^{(n-m)}$,
thus upping the complexity to $O(\log^3 N)$, which seems more fair
from a complexity standpoint. Also, the $H^{(a)}$ were in list of
elementary operations from section \ref{s:EfficientQC}, but the
$R_k^{(a,b)}$ were not. We remark they can be approximated in a
manner leaving the overall QFT circuit efficient.

So this shows how to get an exact transform in $O(\log^2N)$ or
$O(\log^3N)$ operations, depending on one's viewpoint. Since
physical realizations will have error, we would be fine just
approximating the QFT, a viewpoint detailed in Coppersmith
\cite{Copper94}, where he shows how to approximate the transform
within any $\epsilon>0$ in time $O(\log N (\log \log N + \log
1/\epsilon))$. See appendix \ref{s:GenCyclicFT} for more
information on this result.

\subsubsection{The Case $N$ Odd}\label{s:oddNCase}
We use the algorithm over powers of 2 to get one for an odd $N$.
The details of the proof are lengthy, and are left to Appendix
\ref{s:GenCyclicFT}. The main result however gives

\newtheorem*{thmA*}{Theorem \ref{t:MainBound}}

\begin{thmA*}
Given an odd integer $N\geq 13$, and any $\sqrt{2}\geq\epsilon>0$.
Then $F_N$ can be computed with error bounded by $\epsilon$ using
at most $\ceil{12.53+3\log\frac{\sqrt{N}}{\epsilon}}$ qubits. The
algorithm has operation complexity
\begin{equation}
O\left(\log \mr\left(\log \log \mr + \log1/\epsilon\right)\right)
\end{equation}

The induced probability distributions $\mathcal{D}_v$ from the
output and $\mathcal{D}$ from $F_N\ket{u}\otimes\ket{\psi}$
satisfy
\begin{equation}
\left|\mathcal{D}_v-\mathcal{D}\right|\leq 2\epsilon+\epsilon^2
\end{equation}
\end{thmA*}

This says we can approximate the QFT very well. For odd $N<13$ we
can also design circuits using the methods in the proof. It is not
currently known how to construct an exact QFT for odd cyclic
groups, so this is as good as it (currently) gets.

\subsubsection{Final result: the Cyclic HSP
Algorithm}\label{s:CyclicHSPAlgorithm} Combining sections
\ref{s:powerNCase} and \ref{s:oddNCase} with the reasoning in
section \ref{s:BasicExample}, we end up with the cyclic HSP
algorithm:

\textbf{The Hidden Subgroup Algorithm, Cyclic Abelian Case}
\begin{description}
\item[Given] The group $G=\mathbb{Z}_N$ for a positive integer
$N$, and a quantum black-box that evaluates a function
$f:\ket{x}\ket{y}\rightarrow\ket{x}\ket{f(x)\oplus y}$, which we
assume requires constant time\footnote{Even if the time to compute
$f$ is not constant, if $f$ can be computed efficiently, the
overall algorithm is still efficient since $f$ is called only a
few times.}.

\item[Promise] $f$ is constant on a subgroup $H=\left<d\right>$ of
$G$, and is distinct on cosets of $H$.

\item[Output] The integer $d$, in time $O(\log^2 N)$ with
probability at least $\frac{3}{4}$, and using at most
$O(\text{poly}(\log N))$ qubits.
\end{description}~\\
We proceed as follows
\begin{enumerate}
\item Do the following steps for $8$ trials, obtaining samples
$t_1,t_2,...,t_8$.
\begin{enumerate}

\item On the initial state \ket{0}\ket{0} apply the quantum
Fourier transform $F_N$ (as earlier), with an approximation error
of at most $\epsilon=0.01$.

\item Apply $f$ in constant time.

\item Sample the registers in constant time, obtaining $t_j$, a
multiple of $M=|H|$.
\end{enumerate}

\item Compute $M=\gcd(t_1,t_2,\dots,t_8)$ using the Euclidean
algorithm\footnote{The GCD complexity follows from $O(\log N)$
time algorithms for division in \cite{BCH84} and that the most
steps used in the Euclidean algorithm happens when the input is
two consecutive Fibonacci numbers multiplied by an integer.} in
time $O(\log^2 N)$.

\item Output the answer $d=N/M$.
\end{enumerate}

The probability of any one run returning a valid sample is at
least $1-(2\epsilon+\epsilon^2)$. We fix $\epsilon=0.01$. We
require 8 good samples, at which point the probability of them
returning the correct GCD is at least $1-(1/2)^4$, so the
probability of success is then $(1-(.0201))^8(15/16)>3/4$. Oddly
enough, the Euclidean Algorithm to compute the GCD requires more
time than the QFT, and the result follows.


\subsection{The General Finite Abelian Group}\label{s:AbelianGen}

We want to generalize the cyclic case algorithm to all finite
abelian groups. This discussion is a mixture of \cite{BH97} and
\cite{Dam01}, with unified notation, and minor changes and
corrections.

A basic result about finite abelian groups is the following
structure theorem (Lang \cite{Lang93}):
\begin{thm}\label{t:abelianStructure}
Every finite abelian group $G$ is a direct sum of cyclic groups.
\end{thm}
That is, $G\cong
\mathbb{Z}_{N_1}\oplus\mathbb{Z}_{N_2}\oplus\dots\mathbb{Z}_{N_k}$.
Given generators for $G$, finding the $N_i$ is hard classically,
but Cheung and Mosca \cite{CM02} (Theorem \ref{t:AbelDecomp}
below) give an efficient quantum algorithm to find the $N_i$. For
example, given the cyclic group $\mathbb{Z}_N$, there is no known
efficient classical algorithm to find the decomposition of the
multiplicative group $\mathbb{Z}_N^*$ of integers relatively prime
to $N$. Yet classically we can compute within this group
efficiently.

So from now on, we assume we know the decomposition of our finite
abelian group $G$, and can compute in $G$ efficiently both
classically (and hence) quantum mechanically.

Let $G=\mathbb{Z}_{N_1}\oplus\dots\oplus\mathbb{Z}_{N_k}$ be a
finite additive abelian group, and assume we have a function $f$
from $G$ to a finite set $X$, such that there is a subgroup $H<G$
such that $f$ separates cosets of $H$ as in section
\ref{s:HSPDefinition}. Denote elements of $G$ as $k$-tuples:
$g=(g_1,\dots,g_k)$, where we view $g_j$ either as an integer mod
$N_j$ or an integer $\in\{0,1,\dots,N_j-1\}$. Write $-g$ for the
(additive) inverse of $g\in G$.

To generalize the cyclic group Fourier transform $F_N$ to an
arbitrary finite abelian group, we need some representation
theory, specifically character theory, and to this area we now
turn. See also section \ref{s:Representation} for representation
theory basics.

\subsubsection{Character Theory of Finite Abelian Groups}
To define a Fourier transform over $G$, we need to generalize the
$\omega_N^{jk}$ terms from the cyclic case, basically by putting
one such term for each entry in the $k$-tuple description of $G$.
\begin{defn}
A \emph{\textbf{character}} of a group $G$ is a group homomorphism
from $G$ to the multiplicative group of nonzero complex numbers
$\mathbb{C}^*$.
\end{defn}
Recall this is then just a map of sets $\chi:G\rightarrow
\mathbb{C}^*$ such that
\begin{equation}\label{e:gpHom}
\chi(g_1+g_2)=\chi(g_1)\chi(g_2)
\end{equation}
We think of $G$ as having an additive structure, and
$\mathbb{C}^*$ a multiplicative structure. From this simple
definition we derive some tools and facts which will allow us to
finish the HSP discussion for finite abelian groups.

Our first task is to describe all characters
$\chi:G\rightarrow\mathbb{C}^*$. Denote the identity of $G$ by
$e=(0,0,\dots,0)$; the identity of $\mathbb{C}^*$ is 1.

Let $\chi:G\rightarrow\mathbb{C}^*$ be a character (so
$\chi(ng)=\chi(g)^n$ for any integer $n$ and group element $g$).
Let $\beta_1=(1,0,0,\dots,0)\in G$, $\beta_2=(0,1,0,0\dots,0)\in
G,\dots, \beta_k=(0,0,\dots,0,1)\in G$. Then for any element
$g=(g_1,g_2,\dots,g_k)$ we have
\begin{eqnarray}
\chi(g)&=&\chi\left(\sum_{j=1}^k g_j\beta_j\right)\\
&=&\prod_{j=1}^k\chi(\beta_j)^{g_j}
\end{eqnarray}
so $\chi$ is completely determined by its values on the $\beta_j$.
Since $\beta_j$ has order $N_j$, $\chi(\beta_j)$ must have order
dividing $N_j$, for each $j$. Then we must have\footnote{Recall
$\omega_N$ is a primitive $\text{N}^{\text{th}}$ root of unity,
from section \ref{s:notation1}.} that
$\chi(\beta_j)=\omega_{N_j}^{h_j}$ for some integer $h_j$. It is
sufficient to consider $h_j\in\{0,1,\dots,N_j-1\}$ since the
values of $\omega_{N_j}^{h_j}$ are periodic, so any given
character $\chi:G\rightarrow\mathbb{C}^*$ is determined by a
$k$-tuple $(h_1,h_2,\dots,h_k)$, which may be viewed as an element
$h\in G$. This allows labelling each distinct character $\chi$ by
an element of $G$: for each $g\in G$ define the character
$\chi_g:G\rightarrow\mathbb{C}^*$ via
$\chi_g(h)=\prod_{j=1}^k\omega_{N_j}^{g_jh_j}$, for $h\in G$. From
this definition we notice that for all $g,h\in G$
\begin{eqnarray}
\chi_g(h)&=&\chi_h(g)\\
\chi_g(-h)&=&\frac{1}{\chi_g(h)}
\end{eqnarray}

Let $\chi(G)$ denote the set of all such homomorphisms, which is a
group under the operation $\chi_{g_1}\chi_{g_2}=\chi_{g_1+g_2}$
with identity $\chi_e$. Then we prove
\begin{theorem}\label{t:charisom}
For a finite abelian group $G$, $\chi(G)\cong G$.
\end{theorem}
\begin{proof}
From the discussion above, there is a set bijection between the
two sets given (in one direction) by $\alpha:g\rightarrow \chi_g$,
which is also a group isomorphism. The identity
$e=(0,0,\dots,0)\in G$ is sent to the identity $\alpha(e)=\chi_e$
in $\chi(G)$, and
$\alpha(g_1+g_2)=\chi_{g_1+g_2}=\chi_{g_1}\chi_{g_2}=\alpha(g_1)\alpha(g_2)$,
making $\alpha$ a group homomorphism and a set bijection, thus an
isomorphism.
\end{proof}

In the cyclic QFT algorithm, we sampled elements that were
multiples of the generator of the subgroup $H$, and to generalize
this to the finite abelian case where there may not be a single
generator, we introduce \emph{orthogonal elements}. For any subset
$X\subseteq G$, we say an element $h\in G$ is \emph{orthogonal} to
$X$ if $\chi_h(x)=1$ for all $x\in X$. Then for any subgroup $H<G$
we define the \emph{orthogonal subgroup}
\begin{eqnarray}
H^\bot=\left\{g\in G | \chi_g(h)=1\text{ for all }h\in H\right\}
\end{eqnarray}
as the set of all elements in $G$ orthogonal to $H$. $H^\bot$ is a
subgroup of $G$ as follows: the identity $e\in G$ is in $H\bot$
since $\chi_e(g)=1$ for all $g\in G$, and if $a,b\in H^\bot$ then
for any $h\in H$ we have $\chi_h(a-b)=\chi_h(a)/\chi_h(b)=1$ so
$a-b\in H^\bot$, and $H^\bot$ is a subgroup of $G$.

\begin{note} \emph{These orthogonal subgroups are not quite like
orthogonal subspaces. For example, we could have nontrivial $H\cap
H^\bot$, unlike the vector space example. Here is an example
following \cite{Dam01} where $H=H^\bot\neq G$. Let
$G=\mathbb{Z}_{4}$, $H=\{0,2\}$. Then $H^\bot=\{(a)\in G|
(i)^{ah}=1 \text{ for all }(h)\in H\}=\{(a) | (-1)^a=1\}=H$. This
can be extended to give examples of varying weirdness.}
\end{note}

Another useful fact is
\begin{theorem}\label{t:charortho}
Let $G$ be a finite abelian group, and $\chi\in\chi(G)$ a fixed
character, and $\chi_e$ the identity character sending
$G\rightarrow 1$. Then
\begin{equation}
\sum_{g\in G}\chi(g)=\left\{\begin{array}{cl}|G|&\text{ if
}\chi=\chi_e\\0&\text{ if }\chi\neq \chi_e\end{array}\right.
\end{equation}
\end{theorem}
\begin{proof}
Fix $G\cong\mathbb{Z}_{N_1}\oplus\dots\oplus\mathbb{Z}_{N_k}$, and
by theorem \ref{t:charisom} fix $h\in G$ with $\chi=\chi_h$. Using
the notation above,
\begin{eqnarray}
\sum_{g\in
G}\chi_h(g)&=&\sum_{g_1\in\mathbb{Z}_{N_1}}\sum_{g_2\in\mathbb{Z}_{N_2}}\dots
\sum_{g_k\in\mathbb{Z}_{N_k}}\prod_{j=1}^k\omega_{N_j}^{h_jg_j}\\
&=&\left(\sum_{g_1\in\mathbb{Z}_{N_1}}\omega_{N_1}^{h_1g_1}\right)
\left(\sum_{g_2\in\mathbb{Z}_{N_2}}\omega_{N_2}^{h_2g_2}\right)
\dots\left(\sum_{g_k\in\mathbb{Z}_{N_k}}\omega_{N_k}^{h_kg_k}\right)
\end{eqnarray}
If some $\omega_{N_j}^{h_j}\neq 1$, then the geometric series
$\sum_{g_j\in
\mathbb{Z}_{N_j}}\left(\omega_{N_j}^{h_j}\right)^{g_j}=0$, making
the entire product 0. This happens if and only if
$\chi_h\neq\chi_e$. If $\chi_h=\chi_e$ then the sum is $|G|$.
\end{proof}

We now prove some relations between $H$ and $H^\bot$.
\begin{theorem}\label{t:GHH}
With the notation above,
\begin{eqnarray}
G/H&\cong&H^\bot\\
H^{\bot\bot}&=&H
\end{eqnarray}
\end{theorem}
\begin{proof}
Using theorem \ref{t:charisom}, we already have
$H^\bot\cong\chi(H^\bot)$ and $\chi(G/H)\cong G/H$, so it is
enough to prove $\chi(H^\bot)\cong\chi(G/H)$. For any element
$g\in G$ let $\overline{g}$ denote the image in $G/H$ under the
projection map $\pi:G\rightarrow G/H$. Note that any character
$\chi_{h'}\in\chi(H^\bot)$ coming from an element $h'\in H^\bot$
can also be viewed as a character on $G$, since $h'$ is also in
$G$. Then define a map $\alpha:\chi(H^\bot)\rightarrow\chi(G/H)$
via
$$\left(\alpha\chi\right)(\overline{g})=\chi_{h'}(g)$$
where $\overline{g}\in G/H$ and $g$ is any coset representative,
i.e., $\overline{g}=g+H$. We will show $\alpha$ is a group
isomorphism.

$\alpha$ is well defined since if $g_1$ and $g_2$ are different
representations of the same coset $\overline{g_1}=\overline{g_2}$,
then there is an $h\in H$ with $g_1-g_2=h$, giving
$(\alpha\chi_{h'})(\overline{g_1})=\chi_{h'}(g_1)*1=\chi_{h'}(g_1+h)=\chi_{h'}(g_2)=(\alpha\chi_{h'})(\overline{g_2})$.
For the identity $\chi_e\in\chi(H^\bot)$ and any $\overline{g}\in
G/H$ we have $(\alpha\chi_e)(\overline{g})=\chi_e(g)=1$, so
$\alpha\chi_e$ is the identity in $\chi(G/H)$. Also for
$\overline(g)\in G$ $(\alpha(\chi_{h_1}\chi_{h_2}))(\overline{g})=
(\alpha(\chi_{h_1+h_2}))(\overline{g})=\chi_{h_1+h_2}(g)=
\chi_{h_1}(g)\chi_{h_2}(g)=((\alpha\chi_{h_1})(\alpha\chi_{h_2}))(\overline{g})$,
so $\alpha$ is a group homomorphism.

To show $\alpha$ is injective, suppose for some $h'\in H^\bot$
that $\alpha\chi_{h'}$ is the identity in $\chi(G/H)$. Take any
$g\in G$. $\alpha\chi_{h'}(\overline{g})=1$ implies
$\chi_{h'}(g)=1$, and since this is for any $g\in G$, we have
$\chi_{h'}=\chi_e$. $G\cong\chi(G)$ then gives $h'=e$, and thus
$\alpha$ is injective.

Now all we need is to show alpha is surjective. Let
$\overline{\chi}\in\chi{G/H}$. The composite map with the
projection $\pi:G\rightarrow G/H$ gives a homomorphism
$\chi=\overline{\chi}\circ\pi:G\xrightarrow{\pi}G/H
\xrightarrow{\overline{\chi}}\mathbb{C}^*$, thus is a character,
say $\chi_t$, for some fixed $t\in G$. For $h\in H$ this evaluates
to $\chi_t(h)=\overline{\chi}(\overline{e})=1$, so $t\in H^\bot$,
and $\chi_t\in\chi(H^\bot)$. To show
$\alpha\chi_t=\overline{\chi}$, let $\overline{g}\in G/H$, and
compute:
$(\alpha\chi_t)(\overline{g})=\chi_t(g)=\overline{\chi}\pi(g)
=\overline{\chi}(\overline{g})$. Thus $\alpha$ is surjective and
thus a group isomorphism.

To show $H^{\bot\bot}=H$ start with the isomorphism already
proven: $|G/H|=|H^\bot|$ gives $|G/H^\bot|=|H|$ and also implies
$|G/H^\bot|=|H^{\bot\bot}|$, giving $|H|=|H^{\bot\bot}|$. Fix
$h\in H$. By definition $H^{\bot\bot}=\{g\in G | \chi_g(h')=1
\text{ for all }h'\in H^\bot\}$. In particular
$\chi_h(h')=\chi_{h'}(h)=1$ for all $h'\in H^\bot$ by the
definition of $H^\bot$, so we have $h\in H^{\bot\bot}$, giving
$H\subseteq H^{\bot\bot}$. Thus $H=H^{\bot\bot}$.
\end{proof}

\subsubsection{The General Finite Abelian Group Quantum Fourier Transform}
We continue the notation from the previous section. Similar to the
cyclic QFT algorithm returning multiples of the generator of $H$
(which is really the orthogonal subgroup), this general finite
abelian QFT algorithm will return elements of the orthogonal
subgroup $H^\bot$. We start with the Fourier transform.

We define three quantum operators over the group G: the
\emph{Fourier transform} $F_G$ over $G$, the \emph{translation
operator} $\tau_t$ for a $t\in G$, and the \emph{phase-change
operator} $\phi_h$ for $h\in G$ as
\begin{eqnarray}
F_G&=&\frac{1}{\sqrt{|G|}}\sum_{g,h\in
G}\chi_g(h)\ket{g}\bra{h}\\
\tau_t&=&\sum_{g\in G}\ket{t+g}\bra{g}\\
\phi_h&=&\sum_{g\in G}\chi_g(h)\ket{g}\bra{g}
\end{eqnarray}

Note that for cyclic $G=\mathbb{Z}_N$ the Fourier transform is the
same as earlier in section \ref{s:BasicExample}, since then
$\chi_h(g)=e^{\frac{2\pi i h g}{N}}$, and we recover the earlier
algorithm.

First we check that the Fourier transform maps a subgroup $H$ to
its orthogonal subgroup $H^\bot$.
\begin{theorem}\label{t:FHHPerp}
\begin{equation}
F_G\ket{H}=\ket{H^\bot}
\end{equation}
\end{theorem}
\begin{proof}
Recall from the definition of a subset
$\ket{H}=\frac{1}{\sqrt{|H|}}\sum_{h\in H}\ket{h}$. Then
\begin{eqnarray}
F_G\ket{H}&=&\frac{1}{\sqrt{|G|}}\sum_{g,h'\in
G}\chi_{g}(h')\ket{g}\bra{h'}
\frac{1}{\sqrt{|H|}}\sum_{h\in H}\ket{h}\\
&=&\frac{1}{\sqrt{|G||H|}}\sum_{\substack{g,h'\in G\\h\in H}}\chi_g(h')\ket{g}\braket{h'}{h}\\
&=&\frac{1}{\sqrt{|G||H|}}\sum_{\substack{g\in G\\h\in
H}}\chi_g(h)\ket{g}\\
&=&\frac{1}{\sqrt{|G||H|}}\sum_{g\in G}\left(\sum_{h\in
H}\chi_g(h)\right)\ket{g}\label{e:FGH}
\end{eqnarray}
Now consider the coefficient $\sum_{h\in H}\chi_g(h)$ of the ket
\ket{g}. The $G$ character $\chi_g$ is also a character of $H$, so
by theorem \ref{t:charortho} the sum is 0 unless the character is
the identity on $H$, in which case the sum is $|H|$. $\chi_g$ is
the identity on $H$ precisely when $\chi_g(h)=1$ for all $h\in H$,
i.e., $g\in H^\bot$. So equation \ref{e:FGH} becomes
\begin{eqnarray}
\frac{1}{\sqrt{|G||H|}}\sum_{g\in
H^\bot}|H|\ket{g}&=&\sqrt{\frac{|H|}{|G|}}\sum_{g\in
H^\bot}\ket{g}\\
&=&\ket{H^\bot}
\end{eqnarray}
where we used theorem \ref{t:GHH} to get
$\frac{|H|}{|G|}=\frac{1}{|H^\bot|}$.
\end{proof}

We also have
\begin{thm}[\textbf{Commutative laws of the
$G$-operators}]\label{t:Gcommute} For every $h,t\in G$ we have
\begin{eqnarray}
\chi_h(t)\tau_t\phi_h&=&\phi_h\tau_t\\
F_G\phi_h&=&\tau_{-h}F_G\\
F_G\tau_t&=&\phi_tF_G
\end{eqnarray}
\begin{proof}
We prove the last one, which is the only one we explicitly use.
The rest are similar. We use the identity $I=\sum_{g\in
G}\ket{g}\bra{g}$.
\begin{eqnarray*}
F_G\tau_t&=&\left(\frac{1}{\sqrt{|G|}}\sum_{g,h\in
G}\chi_g(h)\ket{g}\bra{h}\right)\left(\sum_{g'\in G}\ket{t+g'}\bra{g'}\right)\\
&=&\frac{1}{\sqrt{|G|}}\sum_{g,g',h\in G}\chi_g(h)\ket{g}\braket{h}{t+g'}\bra{g'}\\
&=&\frac{1}{\sqrt{|G|}}\sum_{g,g'\in
G}\chi_g(t+g')\ket{g}\bra{g'}\\
&=&\frac{1}{\sqrt{|G|}}\sum_{g,g'\in
G}\chi_g(t)\chi_g(g')\ket{g}\bra{g'}\\
&=&\frac{1}{\sqrt{|G|}}\sum_{a,g,g'\in
G}\chi_g(t)\chi_g(g')\ket{a}\braket{a}{g}\bra{g'}\\
&=&\frac{1}{\sqrt{|G|}}\sum_{a,g,g'\in
G}\chi_a(t)\chi_g(g')\ket{a}\braket{a}{g}\bra{g'}\\
&=&\left(\sum_{a\in
G}\chi_{a}(t)\ket{a}\bra{a}\right)\left(\frac{1}{\sqrt{|G|}}\sum_{g,g'\in
G}\chi_g(g')\ket{g}\bra{g'}\right)\\
&=&\phi_tF_G
\end{eqnarray*}
\end{proof}

\end{thm}

Then the algorithm becomes:
\begin{enumerate}
\item Apply the quantum Fourier transform\footnote{Usually the
\emph{inverse} transform is applied here, but this has the same
effect for the \ket{0} state. \cite[Lemma 8]{Dam01} allows quicker
setting of these superposed states with high probability.} to the
first register of the zero state on two registers:
$$
\ket{0}\ket{0}\xrightarrow{F_G}\frac{1}{\sqrt{|G|}}\sum_{g\in
G}\ket{g}\ket{0}
$$
obtaining a superposition over all elements of $G$.

\item Apply the coset separating function $f$:
$$
\xrightarrow{f}\frac{1}{\sqrt{|G|}}\sum_{g\in G}\ket{g}\ket{f(g)}
$$
and as before, $f$ constant and distinct on cosets allows the
simplification
\begin{eqnarray*}
&=&\frac{1}{\sqrt{|T|}}\sum_{t\in T}\ket{t+H}\ket{f(t)}\\
&=&\frac{1}{\sqrt{|T|}}\sum_{t\in T}\tau_t\ket{H}\ket{f(t)}\\
\end{eqnarray*}
where $T=\{t_1,\dots,t_m\}$ is a transversal (\emph{set of coset
representatives}) for $H$ in $G$.

\item Apply the Fourier transform $F_G$ to the first register, and
apply theorems \ref{t:FHHPerp} and \ref{t:Gcommute}
\begin{eqnarray*}
&\xrightarrow{F_G}&\frac{1}{\sqrt{|T|}}\sum_{t\in T}F_G\tau_t\ket{H}\ket{f(t)}\\
&=&\frac{1}{\sqrt{|T|}}\sum_{t\in T}\phi_tF_G\ket{H}\ket{f(t)}\\
&=&\frac{1}{\sqrt{|H^\bot|}}\sum_{t\in
T}\phi_t\ket{H^\bot}\ket{f(t)}
\end{eqnarray*}
We used that $|T|=|G|/|H|=|H^\bot|$ by theorem \ref{t:GHH}. Note
we could have measured the second register as in the cyclic case,
but a fact called ``The Principle of Deferred Measurement" allows
us to measure at the end\footnote{As you will see we still get the
desired outcome whether or not we measure twice, or only once at
the end.}.

\item Measure the first register, obtaining a random element
(uniformly distributed) of $H^\bot$. Note that the phase $\phi_t$
does not affect amplitudes, so we could measure the second
register first if we desired, fixing a $t_0$, as mentioned in the
previous step.
\end{enumerate}

This algorithm returns uniformly distributed random elements of
$H^\bot$. Since $(H^\bot)^\bot=H$, determining a generating set
for $H^\bot$ determines $H$ uniquely. The following discussion
comes from \cite{Dam01}, with details not mentioned there to make
the results precise.

Theorem \ref{t:randomGroupGen} in appendix \ref{s:RandomGroup}
proves that choosing $t+\lceil\log|G|\rceil$ uniformly random
elements of a finite group $G$ will generate $G$ with probability
greater than $1-\frac{1}{2^t}$.

For the moment assume we have chosen a generating\footnote{Here
the exponent does not denote power, but is used since later we
will use subscripts on these elements.} set $g^1,g^2,\dots,g^t$
for $H^\bot$. We want to find efficiently a generating set for
$H$, finishing the algorithm. Since $H^{\bot\bot}=H$, an element
$h\in H$ if and only if $\chi_h(h_j')=1$ for all $j=1,2,\dots,t$.
Next we make these relations linear.

Let $d = \text{LCM}\{N_1,N_2,\dots,N_k\}$. Set $\alpha_l=d/N_l$,
giving $\omega_{N_l}=\omega_{d}^{\alpha_l}$. Then
$\chi_h(g^j)=\prod_{l=1}^k\omega_{d}^{\alpha_l h_lg^j_l}=1$ if and
only if $\sum_{l=1}^k \alpha_l h_lg^j_l \equiv 0 \pmod d$. So to
find elements of $H$, we find random solutions to the system of
$t$ linear equations
\begin{equation}\label{e:linSystem}
\begin{matrix}
\alpha_1 g^1_1 X_1 + \alpha_2 g^1_2 X_2 +\dots+\alpha_k g^1_k X_k
&\equiv& 0 \pmod d\\
\alpha_1 g^2_1 X_1 + \alpha_2 g^2_2 X_2 +\dots+\alpha_k g^2_k X_k
&\equiv& 0 \pmod d\\
\vdots&& \vdots \\
 \alpha_1 g^t_1 X_1 + \alpha_2 g^t_2 X_2
+\dots+\alpha_k g^t_k X_k &\equiv& 0 \pmod d
\end{matrix}
\end{equation}

We do the following. Run the algorithm $T=t_1+\lceil\log|G|\rceil$
times, giving elements $g^1,g^2,\dots,g^T\in H^\bot$. Since
$H^\bot\subseteq G$, these elements generate $H^\bot$ with
probability $p_1\geq 1-1/2^{t_1}$. We want to sample solutions to
the system of equations \ref{e:linSystem} randomly and uniformly,
to get $S=t_2+\lceil\log|G|\rceil$ samples of $H$, which would
generate $H$ with probability $p_2\geq 1-1/2^{t_2}$. To sample the
solutions, view the equations in matrix form $AX\equiv 0\pmod d$,
and then compute the Smith normal form\footnote{\cite{Stor96}
shows how to compute the Smith Normal $D=UAV$ form of an $m\times
n$ integer matrix $A$ $\mod d$ in time $O(n^2m)$, and recover the
$U$ and $V$ in time $O(n^2m \log^c(n^2m))$, for some constant
$c>0$.} of $A$, that is, a diagonal matrix $D$ such that $D=UAV$
with $U$ and $V$ being integer valued invertible matrices. Then we
can uniformly randomly find solutions to $DY\equiv 0\pmod d$ by
solving simple linear congruences, and then compute $X=VY$, which
is a uniformly randomly selected solution to the system of
equations \ref{e:linSystem}. This determines generators of $H$
with probability at least
$(1-\frac{1}{2^{t_1}})(1-\frac{1}{2^{t_2}})$.

Note that $F_G=\otimes_{j=1}^k F_{N_j}$, so we compute it by using
the cyclic case algorithm from section \ref{s:CyclicHSPAlgorithm},
with the time complexity listed there. Choosing
$t_1=t_2=\ceil{\log|G|}+1$ gives a probability of success at least
$1-\frac{1}{|G|}$. After obtaining the system of equations
\ref{e:linSystem}, we compute $D$ and $V$ in time
$O(\log|G|\log\log|G|)$ as in \cite{Stor96}. Then we sample the
resulting system $O(\log|G|)$ times, and convert the answers to
solutions to \ref{e:linSystem}, totaling a time $O(\text{poly}
(\log|G|))$.

Thus we have proven the following (partially stated in Ettinger
and H{\o}yer \cite{EH00}, theorem 2.2.)

\begin{theorem}[Finite abelian HSP
algorithm]\label{t:FiniteAbelianHSPAlgorithm} Given a finite
abelian group $G$, a finite set $X$, and a function
$f:G\rightarrow X$ that separates cosets of $H$ for some subgroup
$H<G$, then there exists a quantum algorithm that outputs a subset
$S\subseteq H$ such that $S$ is a generating set for $H$ with
probability at least $1-1/|G|$. The algorithm uses $O(\log|G|)$
evaluations of $f$, and runs in time polynomial in $\log|G|$ and
in the time required to compute $f$, using a quantum circuit of
size $O(\log |G|\log\log|G|)$.
\end{theorem}


\subsection{The Standard Problems}\label{s:StandardProbs}
Now that we can efficiently find hidden subgroups of finite
abelian groups, we show a few examples of how to use the
algorithms. For a longer list of examples, see \cite[Figure
5.5]{ChuangNielsen}. We merely mention some algorithms that fall
into this framework: Deutsch's algorithm \cite{Deu85} (modified by
Cleve), Deutsch and Jozsa's algorithm \cite{DeuJoz92}, Simon's
algorithm \cite{simon94power}, Shor's factoring and discrete log
algorithms \cite{Shor97}, hidden linear function algorithms, and
the abelian stabilizer algorithm \cite{Kitaev95}. Now on to two
examples.

\subsubsection{Simon's Algorithm}\label{s:SimonAlgorithm}

Simon's \cite{simon94power} algorithm distinguishes a trivial
subgroup from an order 2 subgroup over the additive group
$\mathbb{Z}_2^n$. He showed that a classical probabilistic oracle
requires \emph{exponentially} many (in $n$) more oracle queries
than a quantum algorithm to distinguish the two subgroup types
with probability greater than 1/2, giving a major boost to the
argument that quantum computers may be more powerful than
classical ones. He posed the following problem in 1994 (modified
somewhat to fit our discussion):

\textbf{GIVEN } a function $f:\mathbb{Z}^n_2\rightarrow
\mathbb{Z}^m_2$ with $m\geq n$, and such that there is a constant
$s\in\mathbb{Z}^n_2$ for which $f(x)=f(x')$ if and only if
$x=x'\oplus s$, where $\oplus$ is componentwise (binary) addition.

\textbf{FIND} $s$.

Here the subgroup is $H=\{0,s\}<G=\mathbb{Z}^n_2$, and so we can
find it quickly with high probability using the algorithm from
theorem \ref{t:FiniteAbelianHSPAlgorithm}. However, to solve this
classically, one may have to call $f$ $O(|G|)$ times, evaluating
$f$ on many points, to find the value $s$.

\subsubsection{Shor's Factoring
Algorithm}\label{s:ShorFactoringAlgorithm} Shor \cite{Shor94}
generalizes Simon's algorithm to obtain an integer factorization
(and discrete log) algorithm. A good explanation is also in
\cite{Joz00}. Integer factorization is classically very hard (see
Lenstra and Pomerance \cite{LenPom92}), and is the basis of the
widely used public key cryptography algorithm RSA. Shor's Integer
Factorization Algorithm reduces to finding the order $r$ of an
integer $x\bmod N$, that is, the smallest $r$ such that $x^r\equiv
1\bmod N$. We wish to factor a composite integer $N>0$, and it
suffices to find a non-trivial solution to $x^2\equiv 1 \bmod N$,
then $x+1$ or $x-1$ is a factor of $N$. A randomly chosen $y$
relatively prime to $N$ is likely to have even order, giving the
solution $x=y^{(r/2)}$. All of this, except the order finding
part, is efficient classically. Thus the hard part of the problem
is to find the order of a given $x$ modulo $N$. In other words,
$f(a)=x^a\bmod N$, so $f(a+r)=f(a)$ for all $a$, and the HSP finds
the generator $r$ of the subgroup
$\left<r\right>=H<G=\mathbb{Z}_N$.


\subsection{Conclusion}\label{s:AbelianConclusion}
In conclusion, we have shown that for any finite abelian group
$G$, and any efficiently computable function $f$ that separates
cosets of some subgroup $H<G$, we can efficiently find a
generating set for $H$ with high probability. This was summarized
in theorem \ref{t:FiniteAbelianHSPAlgorithm}.

In the process of doing this we isolated a few items needed to
construct an efficient HSP algorithm for a group $G$:
\begin{enumerate}
\item An efficient way is needed to compute the quantum Fourier
transform over the group $G$. This evolved from the simple Fourier
transform, through a more abstract one involving character theory,
and in the general setting will involve representation
theory\footnote{See section \ref{s:Representation} for
representation theory basics.} to define the Fourier transform
over nonabelian groups.

\item An efficient way is needed to compute the coset separating
function $f$. For Shor's algorithm this is raising an integer to a
power mod $N$, which is efficient classically. Simon's algorithm
had bitwise addition as the function, which also is efficient
classically.

\item Finally, these HSP algorithms needed some post processing to
extract the desired information from the randomly sampled elements
of the orthogonal subgroup. This will turn out to be hard for
nonabelian groups. Groups are known with efficient quantum Fourier
transforms, but no known polynomial time algorithm is available to
reconstruct hidden subgroups.
\end{enumerate}

With that said, let's begin analyzing the general (nonabelian
case).



\section{The General Hidden Subgroup
Problem}\label{s:GeneralProblem}
Why do we want to find hidden subgroups of nonabelian groups? An
efficient abelian HSP algorithm yielded an integer factoring
algorithm which is exponentially faster than any known classical
algorithm. Similarly, finding efficient HSP algorithms over
certain nonabelian groups would yield algorithms faster than any
known classical ones for several important problems, two of which
we now explain.

\subsection{Importance}
One of the main reasons much research has been done into the HSP
problem for nonabelian groups is the desire to find an efficient
algorithm for the Graph Isomorphism problem: when are two graphs
isomorphic? This algorithm has eluded researchers for over thirty
years \cite{KST93,Math79}. Appendix \ref{s:Graph} shows
equivalences between several graph related algorithms, and
describes several reductions. One reduction shown in appendix
\ref{s:Graph} gives that if the HSP could be solved efficiently
for the symmetric group $S_n$, then we would have a polynomial
time algorithm for the Graph Isomorphism Problem.

Another reason is that an efficient algorithm for solving the HSP
for the dihedral group $D_n$ would yield a fast algorithm for
finding the shortest vector in a lattice, first shown by Regev
\cite{Regev02}. This would yield another algorithm whose classical
counterpart is much less efficient than the quantum version.
Finding the shortest lattice vector has many uses, including
applications to cryptography.

Before we cover the nonabelian HSP, we need to generalize the QFT
algorithm, which is what the rest of this section will do. Then
section \ref{s:NonabelianResults} will list the main results known
so far for the nonabelian HSP.

\subsection{Representation Theory
Overview}\label{s:Representation}

To generalize the abelian QFT algorithm, we need the nonabelian
analogue of the Fourier transform. The method explained in section
\ref{s:AbelianGen} shows the general machinery: we need
representations of the group $G$. What follows is a brief overview
of representation theory, which can be seen in detail in either of
the excellent texts Fulton-Harris \cite{HarFul91} or Serre
\cite{Serre77}. We only cover enough of the definitions and facts
to define precisely the quantum Fourier transform for finite
groups. Some definitions and facts:

\textbf{Representation.} A representation $\rho$ of a group $G$ is
a group homomorphism $\rho:G\rightarrow GL(V)$ where $V$ is a
vector space over a field $\field{}$. For our purposes $G$ will be
finite, $V$ will be finite dimensional of (varying) dimension $d$,
and the field $\field{}$ will be the complex numbers $\mathbb{C}$.
Fixing a basis of $V$, each $g\in G$ gives rise to a $d\times d$
invertible matrix $\rho(g)$, which we can take to be unitary. The
\emph{dimension} $d_\rho$ of the representation is the dimension
$d$ of $V$. We will often use the term \textbf{\emph{irrep}} as
shorthand for an irreducible representation.

We say two representations $\rho_1:G\rightarrow GL(V)$ and
$\rho_2:G\rightarrow GL(W)$ are \emph{isomorphic} when there is a
linear vector space isomorphism $\phi:V\cong W$ such that for all
$g\in G$ and $v\in V$, $\rho_1(g)(v)=\rho_2(g)(\phi(v))$. In this
case we write $\rho_1\cong\rho_2$.

\textbf{Irreducibility.} We say a subspace $W\subseteq V$ is an
\emph{invariant} subspace of a representation $\rho$ if
$\rho(g)W\subseteq W$ for all $g\in G$. Thus the zero subspace and
the total space $V$ are invariant subspaces. If there are no
nonzero proper subspaces, the representation is said to be
\emph{irreducible}.

\textbf{Decomposition.} When a representation does have a nonzero
proper subspace $V_1\varsubsetneq V$, it is always possible to
find a complementary invariant subspace $V_2$ so that $V=V_1\oplus
V_2$. The restriction of  $\rho$ to $V_i$ is written $\rho_i$, and
these give representations $\rho_i:G\rightarrow GL(V_i)$. Then
$\rho=\rho_1\oplus\rho_2$, and there is a basis of $V$ so that
each matrix $\rho(g)$ is in block diagonal form with a block for
each $\rho_i$.

\textbf{Complete reducibility.} Repeating the decomposition
process, we obtain for any representation a decomposition
$\rho=\rho_1\oplus\dots\oplus\rho_k$, where each representation
$\rho_i$ is irreducible. This is unique up to permutation of
isomorphic factors.

\textbf{Complete set of irreducibles.} Given a group $G$, there
are a finite number of irreducible representations up to
isomorphism. We label this set $\hat{G}$. Then we have the fact
\begin{equation}\label{t:repDim}
|G|=\sum_{\rho\in\hat{G}}d_\rho^2.
\end{equation}

\textbf{Characters}. To a representation $\rho$ is associated a
\emph{character} $\chi_\rho$ defined by
$\chi_\rho(g)=\text{\textbf{tr}}(\rho(g))$, where \textbf{tr} is
the trace of the matrix. It is basis independent. An alternative,
equivalent description is that a character is a group homomorphism
$\chi:G\rightarrow \mathbb{C}^*$ where $\mathbb{C}^*$ denotes
complex numbers of unit length, and the operation in $C$ is
multiplication, as we saw in section \ref{s:AbelianGen}.
Characters are fixed on conjugacy classes, which follows easily
from the second definition:
$\chi(hgh^{-1})=\chi(h)\chi(g)\chi(h^{-1})=\chi(g)$.

\textbf{Orthogonality of characters}. For two functions
$f_1,f_2:G\rightarrow \mathbb{C}$, there is a natural \emph{inner
product} $\left<f_1,f_2\right>_G=\frac{1}{|G|}\sum_{g\in
G}f_1(g)f_2(g)^*$ where $*$ denotes complex conjugation. The main
fact is: given the character $\chi_\rho$ of a representation
$\rho$ and the character $\chi_i$ of an irreducible representation
$\rho_i$, the inner product $\left<\chi_\rho,\chi_i\right>_G$ is
exactly the number of times the representation $\rho_i$ appears in
the decomposition of $\rho$ into irreducibles. Taking each $\rho$
as unitary simplifies the inner product to
$$\left<\chi_\rho,\chi_i\right>_G=\frac{1}{|G|}\sum_{g\in
G}\chi_\rho(g)\chi_i(g^{-1})$$

\textbf{Orthogonality of the second kind}. Let $C$ be a conjugacy
class of $G$. Since a character $\chi_\rho$ is fixed on a
conjugacy class, let this value be $\chi_\rho(C)$. Then
$$\sum_{\rho\in\hat{G}}\left|\chi_\rho(C)\right|^2=\frac{|G|}{|C|}$$

\textbf{The Regular Representation}. Take $\dim V=|G|$, and fix a
basis of $V$ indexed by elements of $G$, labelling the basis as
$e_g$. Then the regular representation $\rho_G:G\rightarrow GL(V)$
is defined by $G$ permuting the basis elements, i.e.,
$\rho_G(g)e_x=e_{gx}$, extended $\mathbb{C}$-linearly. Thus the
dimension of the regular representation is $|G|$. Another way to
view this representation is as the group algebra $\mathbb{C}[G]$.

The regular representation contains as subrepresentations every
irreducible representation of $G$. If $\rho_1,\dots,\rho_k$ are
all the possible irreducible representations of $G$, then
$$\rho_G=\rho_1^{\oplus d_{\rho_1}}\oplus\dots\oplus
\rho_k^{\oplus d_{\rho_k}}$$ that is, each irreducible $\rho_i$ is
contained exactly $d_{\rho_i}$ times. This yields the important
relation in equation \ref{t:repDim}. Taking the character
associated to this gives, for $g\in G$, the ``regular character"
\begin{equation}\label{e:regChar}
\chi_G(g)=\sum_{\rho\in\hat{G}}d_{\rho}\chi_\rho(g)=\left\{\begin{matrix}0
\text{ if }g\neq e\\N\text{ if }g=e\end{matrix}\right.,
\end{equation}
where the last equality is obtained by noting that $\rho(g)$ acts
on $\mathbb{C}[G]$ by permuting basis elements, so the trace is 0
if $g\neq e$ (all basis elements are permuted by any non-identity
element $g$, so the diagonal is all 0's) and is otherwise $N$.

\textbf{The Induced Representation.} Given a representation
$\rho:H\rightarrow\text{GL}(W)$ of a subgroup $H$ in a group $G$,
we can define a way to extend this to a representation on $G$
written $\textbf{Ind}_H^G\rho:G\rightarrow\text{GL}(V)$, unique up
to isomorphism. The idea is to make copies of $W$ for each coset
of $H$ in $G$, and let cosets permute the copies. So let $\Lambda
= \{e,\tau_1,\dots,\tau_k\}$ be a complete set of coset
representatives, and let $V=\oplus_{\tau\in\Lambda}W_\tau$. Then
any $g\in G$ can be written $g=\tau_g h_g$ for some representative
$\tau_g\in\Lambda$ and $h_g\in H$, which acts on $V$ via
$\tau_gh_g\left(\oplus W_{\tau}\right)=\oplus
h_gW_{\tau_g\tau}$.\\

For representation theory on various groups, most notably the
symmetric group $S_n$, see James and Kerber \cite{JamKer82},
Kerber \cite{Ker71,Ker75}, and Simon \cite{Simon96}. A package for
constructive representation theory is \cite{EgPu98}.


\subsection{The General Fourier
Transform}\label{s:GeneralQFT}
With the machinery above, we can define the general Fourier
transform which works for any finite group, abelian or nonabelian.

\begin{defn}[Fourier Transform over a finite group]
Let $G$ be a finite group of order $N$, $f:G\rightarrow
\mathbb{C}$ any map of sets. For an irreducible representation
$\rho$ of $G$ of dimension $d_\rho$, define \textbf{the Fourier
transform of $f$ at $\rho$} to be
\begin{equation}
\hat{f}(\rho)=\sqrt{\frac{d_\rho}{N}}\sum_{g\in G}f(g)\rho(g)
\end{equation}
Let $\hat{G}$ be a complete set of irreducible representations of
$G$. We define \textbf{the inverse Fourier transform of $\hat{f}$}
to be
\begin{equation}
f(g)=\sqrt{\frac{1}{N}}\sum_{\rho\in\hat{G}}\sqrt{d_\rho}\text{tr}\left(\hat{f}(\rho)\rho(g^{-1})\right)
\end{equation}
\end{defn}

To ensure this definition makes sense, we check that the $f(g)$ in
the definition of the inverse is actually the $f$ we started with,
by substituting the definition of $\hat{f}$ in the definition for
the inverse, and swapping the order of summation, obtaining
\begin{equation}
\frac{1}{N}\sum_{g'\in G}f(g')\sum_{\rho\in\hat{G}}d_\rho
\text{tr}\left(\rho(g'g^{-1})\right) = f(g),
\end{equation}
where we note the rightmost sum is 0 by equation \ref{e:regChar}
unless $g'=g$, in which case that sum is $N$, so the equality
follows. Thus the definition agrees with the initial $f$.

To understand this as a Fourier transform, we associate $f$ and
$\hat{f}$ with vectors in $\mathbb{C}^N$, and examine the map
$\Gamma:f\rightarrow \hat{f}$. To do this, fix an ordering of
$G=\{g_1,g_2,\dots,g_N\}$, and then $f$ is equivalent to a vector
we also label $f$,
$$f=\left(f(g_1),f(g_2),\dots,f(g_N)\right)\in\mathbb{C}^N.$$
To view $\hat{f}$ as a vector in $\mathbb{C}^N$, we need more
choices. Fix an ordering
$\hat{G}=\left\{\rho_1,\rho_2,\dots,\rho_m\right\}$, let
$d_k=d_{\rho_k}$, and for each $\rho_k:G\rightarrow
\text{GL}(\mathbb{C}^{d_k})$ fix a basis of $\mathbb{C}^{d_k}$, so
each $\hat{f}(\rho_k)$ is a $d_k\times d_k$ matrix. We choose each
basis as explained in the following paragraph so each
$\hat{f}(\rho_k)$ is a unitary matrix. This is required to make
the final transform unitary, thus an allowable quantum transform.
Since $\sum_\rho d_\rho^2=N$, there are $N$ matrix entries, which
we order. For brevity label the matrix entry
$\hat{f}(\rho_k)_{ij}=\hat{f}_{ijk}$. Then we can associate
$\hat{f}$ with a vector
$$\hat{f}=\left(\hat{f}_{111},\hat{f}_{121},\dots,\hat{f}_{d_Nd_Nm}\right)\in\mathbb{C}^N.$$
Viewing $\Gamma:f\rightarrow\hat{f}$ as a map from $\mathbb{C}^N$
to itself, it is not hard to show $\Gamma$ is linear. It is a good
exercise to show $\Gamma$ is a unitary transformation when viewed
this way.

In order to make the final operation unitary, which is required by
quantum mechanics, we need to choose each of the bases needed
above so that each $\hat{f}(\rho_k)$ is a unitary matrix. This is
possible, and can be worked out from exercises in Harris and
Fulton \cite{HarFul91}. The rough idea is as follows: On each
$\mathbb{C}^{d_k}$ take the standard basis, and the standard
Hermitian product $\left<v,w\right>_H=\sum_{i=1}^{d_k}v_iw_i^*$.
Average over $G$ to make a $G$-invariant Hermitian norm,
$\left<v,w\right>=\sum_{g\in G}\left<gv,gw\right>_H$. Finally use
Gram-Schmidt with this $G$-invariant norm to get a new orthonormal
basis (relative to the new norm). Use this basis change to get a
matrix for $\rho_k$, which will be unitary. Then the final matrix
for the entire Fourier transform will be unitary, as desired.

Note in the finite abelian case each irreducible representation is
one dimensional, so each $d_\rho=1$, and then the only
representations are given by the characters in section
\ref{s:AbelianGen}. Then $\Gamma$ becomes the finite abelian
Fourier transform, and this definition generalizes the definition
given earlier.


\subsection{The Standard HSP Algorithm - Quantum Fourier Sampling}\label{s:StandardAlgorithm}
We now cover what is called the standard algorithm for finding
hidden subgroups of a given group. The complexity and qubit
requirements depend on the group in question; we will cover what
is known in section \ref{s:NonabelianResults}. This section
follows Hallgren \cite{Hallgren00b} and Grigni, Schulman,
Vazirani, and Vazirani \cite{GSVV01}.

The process about to be described is called \textbf{Quantum
Fourier Sampling}, or \textbf{QFS} for short. It is the process of
preparing a quantum state in a uniform superposition of states
indexed by a group, then performing an oracle function, then a
quantum Fourier transform, and finally sampling the resulting
state to gather information about subgroups hidden by the oracle.

{~} 

We first note the standard finite abelian group case can be
summarized as:

\label{a:Algorithm1}[\textbf{Algorithm 1}]
\begin{enumerate}
\item Compute $\frac{1}{\sqrt{|G|}}\sum_{g\in G}\ket{g}\ket{f(g)}$
and measure the second register $f(g)$. The resulting
superposition is then $\frac{1}{\sqrt{|H|}}\sum_{h\in
H}\ket{ch}\ket{f(ch)}$ for some uniformly chosen coset $cH$ of
$H$.

\item Compute the Fourier transform of the coset state, obtaining
in the first register
$$
\sum_{\rho\in \hat{G}}\frac{1}{\sqrt{|G||H|}}\sum_{h\in
H}\rho(ch)\ket{\rho}
$$
where $\hat{G}$ is the set of (irreducible)
representations\footnote{In the abelian case these are the same as
the characters.} $\{\rho:G\rightarrow\mathbb{C}\}$.

\item Measure the register, and observe a representation $\rho$.
This gives information about $H$.

\item Classically process the information from the previous step
to determine the hidden subgroup $H$.
\end{enumerate}

We can generalize this to handle the nonabelian and abelian cases
in one framework via

\label{a:Algorithm2}[\textbf{Algorithm 2}]
\begin{enumerate}
\item Compute $\frac{1}{\sqrt{|G|}}\sum_{g\in G}\ket{g}\ket{f(g)}$
and measure the second register $f(g)$. The resulting
superposition is then $\frac{1}{\sqrt{|H|}}\sum_{h\in
H}\ket{ch}\ket{f(ch)}$ for some uniformly chosen coset $cH$ of
$H$.

\item Compute the Fourier transform of the coset state, obtaining
in the first register
$$
\sum_{\rho\in
\hat{G}}\sum_i^{d_\rho}\sum_j^{d_\rho}\frac{\sqrt{d_\rho}}{\sqrt{|G||H|}}\left(\sum_{h\in
H}\rho(ch)\right)_{i,j}\ket{\rho,i,j}
$$
where $\hat{G}$ is the set of (irreducible) representations
$\{\rho:G\rightarrow\mathbb{C}\}$.

\item \emph{\textbf{Weak form}}: Measure the register, and observe
a
representation $\rho$. This gives information about $H$.\\
\emph{\textbf{Strong form}}: Measure the register, and observe a
representation $\rho$ as well as matrix indices $i$ and $j$. This
gives information about $H$.

\item Classically process the information from the previous step
to determine the hidden subgroup $H$.
\end{enumerate}

This algorithm gives information useful for finding generators of
the hidden subgroup $H$. Ignoring the problem of engineering the
physical quantum computer, there are three theoretical obstacles
to making this algorithm efficient for a given family of
nonabelian groups. They are:
\begin{enumerate}
\item We need an efficient way to compute the QFT over the groups
in question, similar to the way that equation \ref{e:FNout} led to
an efficient quantum circuit computing the QFT over
$\mathbb{Z}_{2^n}$. Beals \cite{Beal97} constructs an efficient
QFT for the symmetric groups, and Diaconis and Rockmore
\cite{DiaRoc90} construct efficient classical Fourier transforms
over many other groups. For more information on the QFT see
\cite{MasRock95,MasRock97,MasRock97b,MasRockPrep} and section
\ref{s:NonabelianResults}. Efficient QFT quantum circuits are not
known for all finite groups.

\item We need to choose a basis for the irreducible
representations $\rho\in\hat{G}$. For the abelian case, the
irreducible representations are one dimensional characters, so the
basis choices are canonical, so this step is trivial. However, in
the nonabelian case some bases may give better results. For
example, it is known the standard method cannot solve the HSP over
$S_n$ if the basis choice is random - it will take a clever basis
choice for the irreducibles to obtain an efficient algorithm.

\item We need an efficient way to reconstruct the subgroup
generators for $H$ from the irreducible representations returned.
For the abelian case this is efficient since they are canonical
and computing the GCD and solving linear systems $\bmod\; d$ are
efficient classically as explained in section \ref{s:AbelianGen}.
However, this reconstruction is harder in the nonabelian case. For
example, Ettinger, H{\o}yer, and Knill \cite{EHK99} have shown
only polynomially many calls in $\log|G|$ to the oracle
distinguishes subgroups for any group $G$ information
theoretically, but it is currently unknown how to extract
generators for $H$ without \emph{exponential} classical
postprocessing time.
\end{enumerate}

One immediate question is if the weak and strong forms are
equivalent. Section \ref{s:NonabelianResults} shows the strong
form can distinguish between certain subgroups which the weak form
cannot. The reason is roughly that conjugate subgroups determine
the same statistics on representations, but not on rows and
columns, which gives more information. However, there are still
cases where the weak form is good enough.

The next question is to ask which groups have efficient HSP
algorithms, and are there any groups for which the HSP cannot be
solved efficiently?

These questions are ongoing research problems, and there are
partial results showing which groups are likely to be efficiently
solvable, and some negative results showing limitations of this
approach. The next section covers many known results and current
research directions.

For more reading on the (classical) computation of FFT's over
finite groups, see Babai and Ronyai \cite{BabRon90}, Baum
\cite{Bau91}, Baum and Clausen \cite{BauCl91,BaumClau93,BauCl93},
Baum, Clausen, and Tietz \cite{BauClT91}, Rockmore
\cite{Rockmore90,Rockmore94,Rockmore95}, and Terras \cite{Terras}.



\section{Nonabelian Results}\label{s:NonabelianResults}
\subsection{Overview}
\newcommand{\pl}[1]{\ensuremath{\text{poly}\log #1}}

In this section we present results about the HSP over finite
nonabelian groups. Throughout this section we fix notation: $G$ is
a member of a family of finite groups $\textbf{G}=\{G_i\}$ that
should be clear from context, and $H$ is a subgroup of $G$. The
size $n$ of the problem is $n=\lceil\log|G|\rceil$ or sometimes
$n=O(\log|G|)$, also clear from context. We say a quantum
algorithm is \emph{efficient} in either case if the circuit size
is polynomial in $n$ as $G$ varies through the family.

We also divide families of groups into three classes (following
Moore, Rockmore, Russell, and Schulman\cite{MRRS02}):

\begin{enumerate}
\item[\textbf{I.}] \textbf{Fully Reconstructible}. Subgroups of a
family of groups $\textbf{G}=\{G_i\}$ are \emph{fully
reconstructible} if the HSP on $G_i$ can be solved with
probability $>\frac{3}{4}$ by a quantum circuit of size polynomial
in $\log|G_i|$.

\item[\textbf{II.}] \textbf{Measurement Reconstructible}.
Subgroups of a family of groups $\textbf{G}=\{G_i\}$ are
\emph{measurement reconstructible} if the solution to the HSP on
$G_i$ is determined information-theoretically using the fully
measured result of a quantum circuit of size polynomial in
$\log|G_i|$.

\item[\textbf{III.}] \textbf{Query Reconstructible}. Subgroups of
a family of groups $\textbf{G}=\{G_i\}$ are \emph{query
reconstructible} if the solution to the HSP for $G_i$ is
determined by the quantum state resulting from a quantum circuit
of size polynomial in $\log|G_i|$, in the sense that there is a
POVM that yields the subgroup $H$ with constant probability. There
is no guarantee that this POVM can be implemented by a small
quantum circuit.
\end{enumerate}

 A primary goal of quantum algorithm research is to
move groups into lower numbered classes, and the driving force is
to place all finite groups in class I. Currently very few group
families are class I, but we will see all finite groups are in
class III, with some moving up to class II and I. This is
contrasted with what we saw above: all finite abelian groups are
in class I. We will see examples of each of the three classes
below.

\subsection{A Necessary Result}
In order to find an efficient quantum algorithm for a given
family, it is necessary that $O(\text{poly}(n))$ oracle queries
suffices. Fortunately this has been shown possible for any finite
group by Ettinger, H{\o}yer, and Knill\cite{EHK99,EHK04}. They
prove that polynomially many oracle queries in $n$ distinguishes
subgroups information theoretically. They do this by creating the
state
\begin{equation}
\ket{\psi}=\frac{1}{\sqrt{\left|G\right|^m}}\sum_{\left(g_1,g_2,\dots,g_m\right)\in
G^m}\ket{g_1,g_2,\dots,g_m} \ket{f(g_1),f(g_2),\dots,f(g_m)}
\end{equation}
which requires $m$ oracle queries. Taking $m=\lceil 4n+2\rceil$
results in a state from which $H$ can be extracted with high
probability, unfortunately requiring $O(|G|)$ operations to do so.
Precisely, they prove
\begin{thm}
Let $G$ be a finite group, and $f$ an oracle function on $G$ which
separates a subgroup $H$. Then there exists a quantum algorithm
that calls the oracle function $\lceil 4\log|G|+2\rceil$ times and
outputs a subset $X\subseteq G$ such that $X=H$ with probability
at least $1-1/|G|$.
\end{thm}
So for any finite group $G$ and subgroup $H$ it is possible to
gather enough information to determine $H$ using only
$O(\text{poly}(\log|G|))$ queries of $f$, thus placing all finite
groups in class III. Their proof is reproduced in section
\ref{s:hidden} since it is foundational.


\subsection{The Dihedral Group $D_N$}

Many attempts have been made to find an efficient HSP algorithm
for the dihedral groups. One reason is that it one of the
``simplest'' nonabelian groups and is easily studied. Another
reason is that they have exponentially many (in $n$) subgroups of
small order, making classical algorithms infeasible\footnote{For
example, it takes exponentially many evaluations of $f$ just to
determine if $H$ is nontrivial with probability bounded above 1/2.
This holds for the reasons in Simon\cite{Simon97}}. A better
reason is that an efficient HSP algorithm for the dihedral groups
gives efficient algorithms for solving some classically hard
lattice problems \cite{Regev02}, which is covered below. Recall
$\mathbb{Z}_N$ is a cyclic group on $N$ elements\footnote{We could
abstractly call $C_N$ the cyclic group on $N$ elements, but then
$C_N\cong\mathbb{Z}_N$, not always canonically. We choose the
concrete $\mathbb{Z}_N$.}. Then we define the dihedral group
$D_N=\mathbb{Z}_2\ltimes \mathbb{Z}_N$ with $2N$ elements and with
relations
\begin{equation}
x^N=y^2=yxyx=1.
\end{equation}

\subsubsection{Equivalent Problems}\label{s:EquivalentDHSP} Before we start on dihedral group algorithms, we
remark Kuperberg \cite{Kup03} lists equivalences between the
Dihedral HSP (DHSP) and other problems. Precisely we define the
DHSP as finding a hidden subgroup $H$ that is either trivial or
generated by a reflection $H=\left<x^sy\right>$. This is
equivalent to the general problem of determining subgroups of
$D_N$ as we outline below in section \ref{s:dihedralRed}.

Next we define the \emph{abelian hidden shift problem} to be:
given an abelian group $A$, a set $S$, and two injective functions
$f,g:A\rightarrow S$ that differ by a hidden shift $s$
\begin{equation}
f(v)=g(v+s)
\end{equation}
and are otherwise distinct, then determine $s$ (using quantum
oracles $f$ and $g$).

The DHSP is equivalent to the abelian hidden shift problem with
$A=\mathbb{Z}_N$. If we define $h:D_N\rightarrow S$ by
\begin{equation}\label{e:DHSPshift}
h(x^n)=f(n)\;\;\;\;\;\;h(x^ny)=g(n),
\end{equation}
then $h$ hides the reflection $x^sy$. Solving the DHSP for $h$
gives $s$, solving the shift problem. Conversely, given the DHSP
$h:D_N\rightarrow S$, define $f,g$ as in equation
\ref{e:DHSPshift}. Then a solution to the abelian hidden shift
problem for $f$ and $g$ determines $s$, which determines the
subgroup $H$ hidden by $h$.

Generally, the a solution to the HSP on $G=\mathbb{Z}_2\ltimes A$
where $\mathbb{Z}_2$ acts by inversion on $A$ is equivalent to the
abelian hidden shift problem on $A$.

The \emph{cyclic hidden reflection problem} is:
$h:\mathbb{Z}_N\rightarrow S$ satisfies
\begin{equation}
h(n)=h(s-n)
\end{equation} and otherwise takes distinct values. We want to find $s$. This problem is equivalent to
the DHSP; we show it equivalent to the abelian hidden shift
problem as follows.

It reduces to the shift problem by defining the ordered pairs
\begin{equation}
f(n)=(h(-n),h(-n-1)) \;\;\;\;\;\; g(n)=(h(n),h(n+1)).
\end{equation}
We need pairs to ensure $f$ and $g$ are injective. Then $f(n)=
g(s+n)$ and are distinct otherwise, giving the reduction.

Conversely, if $f,g:\mathbb{Z}_N\rightarrow S$ are injective and
\begin{equation}
f(n)=g(s+n)
\end{equation}
then we can define the unordered pairs
\begin{equation}
h(n)=\left\{f(-n),g(n)\right\}.
\end{equation}
which reduces the hidden reflection problem to the shift problem.
Note
$h(n)=\left\{f(-n),g(n)\right\}=\left\{g(s-n),f(n-s)\right\}=\left\{f(-(s-n)),g(s-n)\right\}=h(s-n)$.

\subsubsection{Dihedral Results}\label{s:dihedralRed}
Now we cover what is known about the DHSP.

Ettinger and H{\o}yer \cite{EH00} show an algorithm that produces
data sufficient to determine any hidden subgroup $H$ in a dihedral
group $D_N$, but it is unknown if this data can be post processed
in $O(\text{poly}(n))$ time to reconstruct the subgroup $H$. This
is stronger than the result in \cite{EH99} since it returns the
classical data from the quantum state. \cite{EH99} only
constructed a state determining $H$, but required exponential time
to extract that information to classical information. Their
algorithm exploits the normality of the (abelian) cyclic group
$\mathbb{Z}_N<D_N$, and uses the abelian QFT to gather information
which is then extended to determine the subgroup $H$. They reduce
to the case of finding a subgroup $H$ generated by a reflection.
They prove
\begin{thm}
Let $f$ be a function that separates $H$ in the dihedral group
$D_N$. There exists a quantum algorithm that uses $\Theta(\log N)$
evaluations of $f$ and outputs a subset $X\subseteq H$ such that
$X$ is a generating set for $H$ with probability at least
$1-\frac{2}{N}$.
\end{thm}

Following Ettinger and H{\o}yer \cite{EH00}, we outline the proof
that it is sufficient to solve the DHSP for the simpler case where
$H$ is either trivial or generated by a reflection. We want to
find the hidden subgroup $H<D_N$, where we view $D_N$ as the
semidirect product $\mathbb{Z}_N\rtimes \mathbb{Z}_2$. Using the
abelian QFT algorithm, we find $H_1=H\cap \{\mathbb{Z}_N\times
\{0\}\}$, which is normal in $D_N$. Then we work on the quotient
group $D_N/H_1\cong D_M$ with $M=[\mathbb{Z}_N\times\{0\}:H_1]$,
and find $H/H_1$ which is either generated by a reflection $r+H_1$
or is trivial. Precisely,
\begin{thm}
Let $f$ be a function that separates $H$ in the dihedral group
$D_N$, and suppose we are promised that $H=\{0\}$ is trivial or
$H=\{0,r\}$ is generated by a reflection $r$. Then there exists a
quantum algorithm that given $f$, outputs either ``trivial" or the
reflection $r$. If $H$ is trivial , the output is always trivial,
otherwise the algorithm outputs $r$ with probability at least
$1-\frac{1}{2N}$. The algorithm uses at most $89\log_2 (N) + 7$
evaluations of $f$ and it runs in time $O(\sqrt{N})$.
\end{thm}

Finally, Kuperberg \cite{Kup03} gives a subexponential time
quantum algorithm for solving the dihedral HSP, using time and
query complexity $O(\exp(C\sqrt{\log N}))$ for $D_N$. This is much
better than the classical query complexity of $O(\sqrt{N})$.
Unfortunately this algorithm requires $\Theta(\exp(C\sqrt{\log
N}))$ quantum space. Variants of this algorithm also work for the
abelian hidden shift problem described above and for the hidden
substring problem\footnote{See the paper for a precise definition.
It is basically a string matching algorithm.}. The main results
are
\begin{thm}
There is an algorithm that finds a hidden reflection in the
dihedral group $G=D_N$ (of order $2N$) with time and query
complexity $O(\exp(C\sqrt{\log N}))$.
\end{thm}
\begin{thm}
The abelian hidden shift problem has an algorithm with time and
query complexity $O(\exp(C\sqrt{n}))$ where $n$ is the length of
the output, uniformly for all finitely generated abelian groups.
\end{thm}
(Note this is even true for infinite groups; we only need finitely
generated!)
\begin{cor}
The $N\hookrightarrow 2N$ hidden substring problem has an
algorithm with time and query complexity $O(\emph{exp}(C\sqrt{\log
N})$.
\end{cor}


\subsection{Groups with an Efficient QFT}
Next we turn to some other groups with an efficient QFT. To use
the standard weak or strong form of the algorithm, we need to be
able to compute efficiently the Fourier transform of a function
over a given group. So in this section we list some of the groups
for which efficient quantum Fourier transform algorithms are
known.

Zalka \cite{Zalka} gives an algorithm for the HSP on wreath
product groups $G=\mathbb{Z}_2^n\wr\mathbb{Z}_2$. The idea is
similar to Ettinger and H{\o}yer \cite{EH00}, in that it finds
generators for an abelian subgroup in the desired subgroup, and
then extends it.

H{\o}yer \cite{Hoy97} shows how to construct QFT for many groups:
quaternions, a class of metacyclic\footnote{A group $G$ is
\emph{metacyclic} if it contains a cyclic normal subgroup $H$ such
that the quotient group $G/H$ is cyclic.} groups (up to phase),
and a certain subgroup $E_n$ of the orthogonal group $O(2^n)$
useful for quantum error correction \cite{CRSS97}.

Beth, P{\"{u}}schel, R{\"{o}}tteler, \cite{BPR99} show how to do
the QFT efficiently on a class of groups - solvable 2 groups
containing a cyclic normal subgroup of index 2 ($|G|$ is a power
of 2 and solvable): They give reference to the fact for $n\geq 3$
there are exactly 4 isomorphism classes of such nonabelian groups
of order $2^{n+1}$ with a cyclic subgroup of order $2^n$:
\begin{itemize}
\item the dihedral group
$D_{2^{n+1}}=\left<x,y\;|\;x^{2^n}=y^2=1,\;\;yxyx=1\right>$,

\item the quaternion group
$Q_{2^{n+1}}=\left<x,y\;|\;x^{2^n}=y^4=1,\;\;y^3xyx=1\right>$,

\item the quasi-dihedral group
$QD_{2^{n+1}}=\left<x,y\;|\;x^{2^n}=y^2=1,\;\;yxy=x^{2^{n-1}-1}\right>$,

\item the group
$QP_{2^{n+1}}=\left<x,y\;|\;x^{2^n}=y^2=1,\;\;yxy=x^{2^{n-1}+1}\right>$.
\end{itemize}

Beals \cite{Beal97} shows how to compute the QFT over $S_n$ in
time $O(\text{poly}(n))$, by adapting the methods of Clausen
\cite{Clau89} and Diaconis-Rockmore \cite{DiaRoc90} to the quantum
setting.

Moore, Rockmore, and Russell \cite{MRR04} show how to construct
efficient quantum Fourier transform circuits of size
$O(\text{poly}\log|G|)$ for many groups, including
\begin{itemize}
\item the Clifford groups $\mathbb{CL}_n$,

\item the symmetric group, recovering Beals algorithm
\cite{Beal97},

\item wreath products $G\wr S_n$, where $|G|=O(\text{poly}(n))$,

\item metabelain groups (semidirect products of two abelian
groups), including metacyclic groups such as the dihedral and
affine groups, recovering the algorithm of H{\o}yer \cite{Hoy97},

\item bounded extensions of abelian groups such as the generalized
quaternions, recovering the algorithm of P{\"{u}}schel et al.
\cite{BPR99}.
\end{itemize}
Their results also give subexponential size quantum circuits for
the linear groups $\text{GL}_k(q)$, $\text{SL}_k(q)$,
$\text{PGL}_k(q)$, $\text{PSL}_k(q)$, for a fixed prime power $q$,
finite groups of Lie type, and the Chevalley and Weyl groups.
Unfortunately, defining \emph{polynomially uniform}, \emph{adapted
diameter}, \emph{homothetic}, and \emph{multiplicity} would take
us too far afield; see their paper for details. These have to do
with certain group items being efficiently computable. But we
state their two main theorem anyway:
\begin{thm}
If $G$ is a polynomially uniform group with a subgroup tower
$G=G_m>G_{m-1}>\dots>{1}$ with adapted diameter $D$, maximum
multiplicity $M$, and maximum index $I=\max_i[G_i:G_{i-1}]$, then
there is a quantum circuit of size $\emph{poly}(I\times D\times
M\times \log|G|)$ which computes the quantum Fourier transform
over $G$.
\end{thm}
\begin{thm}
If $G$ is a homothetic extension of $H$ by an abelian group, then
the quantum Fourier transform of $G$ can be obtained using
$O(\text{\emph{poly}}\log |G|)$ elementary quantum operations.
\end{thm}


\subsection{HSP Algorithms and Groups}
\subsubsection{Group Definitions I}

$H$ is a subgroup of $G$; let $N(H)$ or $N_G(H)$ be the
\emph{normalizer} of $H$ in $G$. Let $M_G$ be the intersection of
all normalizers in $G$, i.e., $M_G=\bigcap_{H\leq G}N(H)$. $M_G$
is a subgroup of $G$ and can be taken to be the size of how
nonabelian $G$ is ($[G:M_G]=1$ for abelian groups). $H^G$ is the
largest subgroup of $H$ that is normal in $G$, and is called the
\emph{normal core} of $H$.
\begin{defn}[Wreath Product]
The \emph{\textbf{wreath product}} of two finite groups $G$ and
$H$ is defined as follows. For $|H|=n$, view $H$ a subgroup of the
symmetric group $S_n$ on $n$ items. Let $P=G\times\dots\times G$
be the direct product of $n$ copies of $G$. The wreath product
$G\wr H$ of $G$ with $H$ is a semidirect product $P\rtimes H$ with
multiplication
\begin{equation}
\left(g_1,\dots,g_n;\tau\right)\left(g_1',\dots,g_n';\tau'\right)
= \left(g_{\tau'(1)}g_1',\dots,g_{\tau'(n)}g_n;\tau\tau'\right)
\end{equation}
\end{defn}
That is, the permutations in $H$ are composed as usual, but the
right permutation permutes the left factors of $P$ and then the
$n$-tuple is multiplied pointwise. It is instructive to verify
this operation forms a group.

\subsubsection{Normal Subgroups Can be Found in Any Group}
Hallgren, Russell, and Ta-Shma (2002) \cite{Hallgren00b} prove
that the natural extension of the abelian case algorithm finds
$H^G$ efficiently, the normal core of $H$. This also gives that
normal subgroups can be found efficiently by the standard (weak or
strong version of the) algorithm. In particular, this allows
finding hidden subgroups in Hamiltonian groups (groups whose
subgroups are all normal); the nonabelian Hamiltonian groups are
of the form $\mathbb{Z}^k_2\times B\times Q$, where $Q$ is the 8
element quaternion group and $B$ is an abelian group with
exponent\footnote{Recall the exponent $a$ of a group $G$ is the
smallest integer $a$ such that $g^a=e$, the identity, for every
element $g\in G$, if such an integer exists.} $b$ coprime with 2.
See Rotman \cite[Exercise 4.28]{Rot95}. They show the probability
of measuring a representation $\rho$ is independent of the coset
of $H$.
\begin{thm}
The probability of measuring the representation $\rho$ in
 Algorithm 2 of section \ref{s:StandardAlgorithm} is $d_\rho\frac{|H|}{|G|}$
times the number of times $\rho$ appears in
$\emph{\textbf{Ind}}_H^G\text{\textbf{1}}_H$.
\end{thm}

They also obtain:
\begin{thm}
Let $H$ be an arbitrary subgroup of $G$, and let $H^G$ be the
largest subgroup of $H$ that is normal in $G$. With probability at
least $1-2\exp(-\log_2|G|/8)$, $H^G$ is determined by observing
$O(\log|G|)$ independent trials of QFS.
\end{thm}
In fact, if $\rho_1,\dots,\rho_m$ are the representations sampled
by $m$ repetitions of the algorithm, then
$H^G=\bigcap_{i}\ker\rho_i$ with high probability.

They also show that weak QFS does not distinguish between order 1
and 2 subgroups in $S_n$:
\begin{thm}
For $S_n$, there is a subgroup $H_n$ so that the weak QFS does not
distinguish (even information theoretically) the case that the
hidden subgroup is trivial from the case the hidden subgroup is
$H_n$. Specifically, the distributions induced on representations
in these two cases have exponentially small total variation
distance.
\end{thm}
\begin{thm}
Let $H$ be an arbitrary subgroup of $G$, and let $H^G$ be the
largest subgroup of $H$ that is normal in $G$. With probability at
least $3/4$, $H^G$ is uniquely determined by observing
$m=O(\log|G|)$ independent trials of Algorithm 2 of section
\ref{s:StandardAlgorithm} when $H$ is the hidden subgroup. When
$H$ is normal, $H^G=H$, and this determines $H$.
\end{thm}

\subsubsection{``Almost Abelian" Subgroups Can be Found and Measuring Rows is Strong Enough}
Grigni, Schulman, Vazirani, and Vazirani \cite{GSVV01} show
another class of groups for which the HSP has an efficient quantum
solution - what they call ``almost abelian" groups. These are
groups for which the intersection $M(G)$ of all the normalizers of
all subgroups of $G$ is large. For $n=\log|G|$, they require
$\left[G:M(G)\right]$ (called the Baer norm \cite{MRRS02}) to be
of order $\exp O(\log^{1/2}n)$, and then the HSP can be solved if
the QFT can be performed efficiently. In particular they show that
the subgroups of the semidirect product $\mathbb{Z}_m\ltimes
\mathbb{Z}_3$ for $m$ a power of 2 can be found efficiently.

Another useful result in their paper shows that measuring both the
row and column in the strong form of the QFT gives no more
information than measuring just one of them (depending on how one
lets the irreps act - left or right). This follows from the
quantum mechanical requirement that the irreps are unitary
matrices, and thus each matrix row (or column) has the same norm,
which gets ``absorbed."

Most importantly, they show that even using the strong form
\emph{with a random basis} for the irreps, the strong QFS
algorithm cannot distinguish between the case of a trivial
subgroup and an order two subgroup without exponentially many
oracle queries.

The restriction on the size of $M(G)$ was extended by Gavinsky
\cite{Gav03} to allow $\left[G:M(G)\right]$ to be of size
$O(\text{poly}(n))$, allowing the corresponding HSP to be solved
efficiently if the QFT over $G$ can be. These groups are labelled
``poly-near-hamiltonian groups." A final algorithm in this paper
shows how to solve the HSP efficiently on poly-near hamiltonian
groups \emph{even when the QFT over the group G is not known to be
efficient}, by using QFS over a hamiltonian group, which was shown
to be efficient by a result from above.

\subsubsection{Strong is Indeed Stronger}
Moore, Rockmore, Russell, and Schulman \cite{MRRS02} show that the
strong form is indeed stronger, by exhibiting semidirect products
$\mathbb{Z}_q\ltimes\mathbb{Z}_p$ (the $q$-\emph{hedral} groups,
which include the affine groups $A_p\cong
\mathbb{Z}_p^*\ltimes\mathbb{Z}_p$) , where $q|(p-1)$ and
$q=p/\text{polylog}(p)$, such that the strong form can determine
hidden subgroups efficiently, but the weak form and ``forgetful"
abelian form cannot. They also prove a closure property for the
class of groups over which the HSP can be solved efficiently:
\begin{thm}
Let $H$ be a group for which hidden subgroups are fully
reconstructible, and $K$ a group of size polynomial in $\log|H|$.
Then hidden subgroups in any extension of $K$ by $H$, i.e. any
group $G$ with $K\triangleleft G$ and $G/K\cong H$, are fully
reconstructible.
\end{thm}
They also place some groups in class I.
\begin{thm}
Let $p$ be a prime, $q$ a positive integer, and
$G=\mathbb{Z}_q\ltimes\mathbb{Z}_p$. Then
\begin{enumerate}
\item if $q$ is prime and $q=(p-1)/\text{\emph{polylog}}(p)$, then
subgroups of $G$ are fully reconstructible (class I),

\item if $q$ divides $p-1$, then hidden conjugates of $H$ in $G$
are fully reconstructible (class I) if $H$ has index
$\text{\emph{polylog}}(p)$,

\item if $q$ divides $p-1$, then hidden conjugates of $H$ in $G$
are measurement reconstructible (class II),

\item if $q$ divides $p-1$, then subgroups the $q$-hedral groups
$G$ are measurement reconstructible (class II). In particular, the
subgroups of the affine groups $A_p=\mathbb{Z}^*_{p-1}\ltimes
\mathbb{Z}_p$ are measurement reconstructible (class II).
\end{enumerate}
\end{thm}

For another direction studying the HSP over infinite groups, see
Lomonaco and Kauffman \cite{KaufLom03}. They consider a version of
the HSP for finding periods of functions over the real numbers
$\mathbb{R}$, although it is not clear if these could be
physically implemented due to $\mathbb{R}$ being an infinite set.
They have a good overview of the HSP in \cite{KaufLom02}.

R{\"{o}}tteler and Beth \cite{RottBeth98} give an efficient
algorithm solving the HSP on wreath products
$W_n=\mathbb{Z}_2^n\wr\mathbb{Z}_2$ (like Zalka) by giving quantum
circuits for the QFT and showing how to reconstruct the subgroup
efficiently from samples. It uses $O(n)$ queries of $f$ and
$O(\text{poly}(n))$ classical post processing time, putting these
groups in Class I. It is similar to the method of Ettinger and
Hoyer.

In \cite{EH99} Ettinger and H{\o}yer construct a quantum
observable for the graph isomorphism problem. Given two graphs of
$n$ vertices and an integer $m$, they define a quantum state on
$O(mn)$ qubits, that when observed, outputs ``yes" with certainty
if the graphs are isomorphic and ``no" with probability at least
$1-\frac{n!}{2^m}$ if they are not isomorphic. It is unknown if
this observable can be implemented efficiently.

Cleve and Watrous \cite{CleWat00} show how to reduce the
complexity and size of the QFT for $\mathbb{Z}_{2^n}$.
\begin{thm}
For any $m$ there is a quantum circuit that exactly computes the
QFT modulo $2^m$ that has size $O(m(\log m)^2\log \log m)$ and
depth $O(m)$.
\end{thm}
\begin{thm}
For any $m$ and $\epsilon$ there is a quantum circuit that
approximates the QFT modulo $2^m$ that has size $O(m\log
(m/\epsilon))$ and depth $O(\log m+\log\log(1/\epsilon))$.
\end{thm}
They give an upper bound.
\begin{thm}
Any quantum circuit consisting of one- and two- qubit gates that
approximates the QFT with precision $\frac{1}{10}$ or smaller must
have depth at least $\log n$.
\end{thm}

\subsubsection{Lattice Problems}
Regev \cite{Regev02} shows that an efficient algorithm solving the
HSP for dihedral groups would result in efficient algorithms for
solving the Unique Shortest Vector Problem (SVP) and the
subset-sum problem. First we sketch some definitions. A
\emph{lattice} is the set of all integral linear combinations of
$k$ linearly independent vectors in $\mathbb{R}^k$. This set of
$k$ vectors is called the \emph{basis} of the lattice. The SVP is
the problem of finding the shortest nonzero vector in this
lattice, given the basis. In the $f(k)$-unique-SVP we are given
the promise that the shortest vector is shorter by at least a
factor of $f(k)$ from all other non-parallel vectors. We also
define the Dihedral Coset problem (DCP). The input to the DCP for
the dihedral group $D_N$ of order $2N$ is a tensor product of
polynomially many (in $N$) registers, each with the state
$\ket{0,x}+\ket{1,(x+d\pmod N)}$ for some arbitrary
$x\in\left\{0,1,\dots,N-1\right\}$, and $d$ is the same for all
registers. The goal is to find $d$. We say the DCP has failure
parameter $\alpha$ if each of the registers with probability at
most $\frac{1}{(\log n)^\alpha}$ is in the state $\ket{b,x}$ for
arbitrary $b$. We take $N=k$, so the dihedral group size is
determined by the dimension of the lattice. The main theorem is
then
\begin{thm}
If there exists a solution to the DCP with failure parameter
$\alpha$ then there exists a quantum algorithm that solves the
$\Theta(k^{\frac{1}{2}+2\alpha})$-unique-SVP.
\end{thm}
Thus an efficient Dihedral HSP algorithm would give an efficient
$f(k)$-unique-SVP algorithm.

\subsubsection{Distinguishable Subgroups of $S_n$}
Kempe and Shalev \cite{KS04} analyze which subgroups of $S_n$ can
be distinguished efficiently using QFS. $H<S_n$ is
\emph{primitive} if it is transitive, and does not preserve a
non-trivial partition of the permutation domain. They show
\begin{thm}
Let $H\neq A_n,S_n$ be a subgroup of $S_n$, with $H$ a primitive
subgroup. Then $H$ is indistinguishable.
\end{thm}
\begin{thm}
A subgroup $H<S_n$ with property $\Upsilon$ (below) can be
efficiently distinguished from the identity subgroup using either
the weak or strong standard method with random basis only if it
contains an element of constant support (i.e., a permutation in
which all but a constant number of points are fixed). Property
$\Upsilon$ can be any of the following
\begin{itemize}
\item $H$ is of polynomial size,

\item $H$ is primitive.
\end{itemize}
\end{thm}
They also show other properties $\Upsilon$ for which the statement
is true, and conjecture it is true for all subgroups of $S_n$. If
their conjecture is true, which amounts to proving the following
conjecture, then QFS with random basis provides no advantage over
classical search. The \emph{minimal degree} of a subgroup $H<S_n$
is defined to be the minimal number of points moved by a
non-identity element of $H$. The \emph{support} of an element is
the number of points moved. Then the conjecture is
\begin{conj}
Every subgroup $H<S_n$ with non-constant minimal degree has at
most $n^{k/7}$ elements of support $k$.
\end{conj}

\subsection{Black-Box Group Algorithms}

\subsubsection{Black-box Group Algorithms}
Black-box groups were introduced by Babai and Szemer\'{e}di in
1984 \cite{BabSz84}. In the context of \emph{black-box groups},
each group element is encoded as a length $n=O(\log|G|)$ string,
and we assume group operations (multiplication, inverse, identity
testing) are preformed by a \emph{group oracle} (or
\emph{black-box}) in unit time. If each element is represented by
a unique string this is called the \emph{unique encoding} model,
otherwise it is \emph{not unique encoding}. A black-box group
without unique encoding augmented by an oracle that can recognize
any encoding of the identity element in unit time can compare
elements for equality in unit time. Any efficient algorithm in the
context of black-box groups remains efficient whenever the group
oracle can be replaced by an efficient process. It is provably
impossible to compute group orders in polynomial time in size log
of the group, even for abelian groups. This becomes possible using
quantum algorithms, as we will see. A black-box group $G$ is
defined by a set of $m$ generators, each of length $n$ bits, i.e.,
$G=\left<g_1,g_2,\dots,g_m\right>$. The quantity $mn$ is called
the \emph{input size} for the group. Throughout this section on
black-box group algorithms we reserve $n$ to denote the length of
the strings representing the finite group $G$, and all groups are
finite.

\subsubsection{Group Definitions}
To state results for black-box group algorithms we need more
definitions. Given a group $G$ and elements $g,h\in G$, we define
the \emph{commutator} of $g$ and $h$, denoted $[g,h]$, to be
$[g,h]=g^{-1}h^{-1}gh$, and for any two subgroups $H,K\leq G$ we
write $[H,K]$ to denote the subgroup of $G$ generated by all
commutators $[h,k]$ for $h\in H$ and $k\in K$. The \emph{derived
subgroup} (also known as the \emph{commutator subgroup}) of $G$ is
$G'=[G,G]$, and we write
\begin{eqnarray*}
G^{(0)}&=&G,\\
G^{(j)}&=&\left(G^{(j-1)}\right)', \text{ for }j\geq 1.
\end{eqnarray*}
A group $G$ is said to be \emph{solvable} if $G^{(m)}=\{1\}$ (the
trivial group) for some value of $m$.

A \emph{composition series} for $G$ is a sequence of subgroups of
$G=G_1\rhd G_2\dots\rhd G_t=1$ such that $G_{i+1}$ is normal in
$G_i$, and the \emph{factor groups} $G_i/G_{i+1}$ are simple. The
factor groups $G_i/G_{i+1}$ are unique up to isomorphism and
ordering. Beals and Babai \cite{BB93} define $v(G)$ as the
smallest natural number $v$ such that every nonabelian composition
factor of $G$ possesses a faithful permutation representation of
degree at most $v$. Thus for a solvable group $v(G)=1$ (solvable
implies factor groups are cyclic, hence abelian, hence have only
trivial irreducible representations). It is known that $v(G)$ is
polynomially bounded in the input size in many important cases,
such as permutation groups or matrix groups over algebraic number
fields.

A \emph{presentation} of $G$ is a sequence $g_1,\dots,g_s$ of
elements generating $G$, together with a set of group expressions
in variables $x_1,\dots,x_s$ called \emph{relations}, such that
$g_1,\dots,g_s$ generate $G$ and the kernel of the homomorphism
from the free group $F(x_1,\dots,x_s)\rightarrow G$ given by
$x_i\rightarrow g_i$ is the smallest normal subgroup of $F$
containing the relations. This gives a non-canonical yet very
concrete description of $G$ as the set of ``strings" of the $g_i$
and equivalence relations on those strings. Note the generators in
the presentation may differ from the original generators given for
$G$.

A \emph{nice representation} of a factor group $G_i/G_{i+1}$ means
a homomorphism from $G_i$ with kernel $G_{i+1}$ to either a
permutation group of degree polynomially bounded in the input size
+ $v(G)$ or to $\mathbb{Z}_p$, where $p$ is a prime dividing
$|G|$.

The \emph{exponent} of a group is the smallest integer $m$ such
that $g^m=e$ for all $g\in G$. Lagrange's theorem gives $m\leq
|G|$.

An abelian group (family) is \emph{smoothly abelian} if it can be
decomposed into the direct product of a subgroup of bounded
exponents and a subgroup of polylogarithmic size in the order of
the group. A solvable group (family) is \emph{smoothly solvable}
if its derived series is of bounded length and has smoothly
abelian factor groups.

A \emph{constructive membership test} is the following: given
pairwise commuting group elements $h_1,h_2,\dots,h_r,g$ of a group
$G$, either express $g$ as a product of powers of the $h_i$'s or
report that no such expression exists.

\subsubsection{Results}
Our first result \cite{CM02}, the basis for many later ones,
allows computing a canonical decomposition of a finite abelian
group from a generating set in polynomial time, i.e.,
\begin{thm}[Cheung, Mosca]\label{t:AbelDecomp}
Given a finite abelian black-box group $G$ with unique encoding,
the decomposition of $G$ into a direct sum of cyclic groups of
prime power order can be computed in time polynomial in the input
size by a quantum computer.
\end{thm}

Watrous \cite{Wal00} shows how to construct quantum certificates
proving group non-membership efficiently, and shows this is not
possible classically.

Watrous \cite{Wal01} gives a polynomial-time quantum algorithm for
computing the order of a solvable group, which gives
polynomial-time algorithms for membership testing of an element in
a subgroup, testing subgroup equality given two descriptions of
the subgroups, and testing subgroup normality, each for solvable
groups. The main result is
\begin{thm}[Group Order]
Given a finite, solvable black-box group $G$, there exists a
quantum algorithm that outputs the order of $G$ with probability
of error bounded by $\epsilon$ in time polynomial in the input
size $+\log(1/\epsilon)$. The algorithm produces a quantum state
$\phi$ that approximates the state $\ket{G}=|G|^{-1/2}\sum_{g\in
G}\ket{g}$ with accuracy $\epsilon$ in the trace norm metric.
\end{thm}

This result was also obtained using a different algorithm by
Ivanyos \emph{et. al.} \cite{IMS01} in a paper extending many of
the black-box group results from Beals-Babai \cite{BB93} to the
quantum setting. They obtain
\begin{thm}\label{t:groupTask}
Let $G$ be a finite black-box group with not necessarily unique
encoding. Assume the following are given:
\begin{enumerate}\item[]\begin{enumerate}
\item an oracle for computing the orders of elements of G,

\item an oracle for the constructive membership tests in
elementary abelian subgroups of $G$.
\end{enumerate}\end{enumerate}
Then the following tasks can be solved by quantum algorithms of
running time polynomial in the input size+$v(G)$:
\begin{enumerate}
\item constructive membership tests in subgroups of G,

\item computing the order of $G$ and a presentation for $G$,

\item finding generators for the center of $G$,

\item constructing a composition series $G=G_1\rhd
G_2\rhd\dots\rhd G_t=1$ for $G$, together with nice
representations of the composition factors $G_i/G_{i+1}$,

\item finding Sylow subgroups of $G$.
\end{enumerate}
\end{thm}

The hypotheses $(a)$ and $(b)$ can be met in many cases. For
example, using Shor's order finding method to compute element
orders, they give:
\begin{thm}
Assume $G$ is a black-box group with unique encoding. Then each
task in theorem \ref{t:groupTask} can be solved in time polynomial
in the input size + $v(G)$ by a quantum algorithm.
\end{thm}

\begin{thm}
Assume $G$ is a black-box group with not necessarily unique
encoding, and that $N$ is a normal subgroup given as a hidden
subgroup of $G$ (i.e., there is a $f$ hiding $N$). Then there are
quantum algorithms each with running time polynomial in the input
size + $v(G/N)$ that perform:
\begin{itemize}
\item all the tasks in theorem \ref{t:groupTask} for $G/N$,

\item finding generators for $N$. In particular, we can find
hidden normal subgroups of solvable black-box groups and
permutation groups in polynomial time in input size + $v(G/N)$
(note we do not need an efficient QFT as in Hallgren \emph{et.
al.} \cite{Hallgren00b}),
\end{itemize}
If instead of giving $N$ as a hidden subgroup, if $N$ is given by
generators, and $N$ is solvable or of polynomial size, then all
the tasks listed in theorem \ref{t:groupTask} can be solved for
$G/N$ in time polynomial in the input size + $v(G)$.
\end{thm}

\begin{thm}
Let $G$ be a black-box group with unique encoding. The HSP can be
solved by a quantum algorithm in time polynomial in the input size
+ $|G'|$, the size of the commutator subgroup of $G$.
\end{thm}
This includes the wreath products $\mathbb{Z}_2^k\wr \mathbb{Z}_2$
of R{\"{o}}tteler and Beth \cite{RottBeth98}.

A question remains: the above proofs only use the abelian QFT to
get the results. Does using the nonabelian QFTs give better
results?

~ 

Friedl \emph{et. al.} \cite{FIMSS03} introduced the \emph{Orbit
Coset} problem as a generalization of the hidden subgroup and
hidden shift\footnote{Hidden shift is called hidden translation in
their paper.} problems. Hidden shift was defined above in section
\ref{s:EquivalentDHSP}. As mentioned there, when $G$ is abelian,
hidden shift is equivalent to the HSP in the semidirect product
$G\rtimes \mathbb{Z}_2$.

\begin{defn}[\emph{Orbit Coset and Orbit Superposition}]Let $G$ be a finite group acting
on a finite set $\Gamma$ of mutually orthogonal quantum states.
\begin{itemize}
\item Given generators for $G$ and two quantum states
$\ket{\phi_0},\ket{\phi_1}\in\Gamma$, the problem
\textbf{\emph{Orbit Coset}} is to either reject the input if
$G(\ket{\phi_0})\cap G(\ket{\phi_1})=\emptyset$, or output a
generating set for $G_{\ket{\phi_1}}$ of size $O(\log|G|)$ and a
$u\in G$ such that $\ket{u\cdot\phi_1}=\ket{\phi_0}$.

\item Given generators for $G$ and a quantum state
$\ket{\phi}\in\Gamma$, the problem \emph{\textbf{Orbit
Superposition}} is to construct the uniform superposition
$\ket{G\cdot\phi}=\frac{1}{\sqrt{|G(\ket{\phi})|}}\sum_{\ket{\phi'}\in
G(\ket{\phi})}\ket{\phi'}$
\end{itemize}
\end{defn}

\begin{thm}
Let $p$ be a fixed prime. Then
\begin{itemize}

\item the problem of hidden shift over $\mathbb{Z}_p^m$ can be
solved in quantum polynomial time,

\item the problem of Hidden Subgroup over
$\mathbb{Z}_p^m\rtimes\mathbb{Z}_2$ can be solved in quantum
polynomial time.
\end{itemize}

\end{thm}
This gives that $Z_p^m\rtimes \mathbb{Z}_2$ is class I for any
prime $p$. 

\begin{thm}
Let $G$ be a smoothly solvable group and let $\alpha$ be a group
action of $G$. When $t=(\log^{\Omega(1)}|G|)\log(1/\epsilon)$,
Orbit Coset can be solved in $G$ for $\alpha^t$ in quantum time
$\text{\emph{poly}}(\log|G|)\log(1/\epsilon)$ with error
$\epsilon$.
\end{thm}
Using this they then show
\begin{thm}
Hidden shift can be solved over smoothly solvable groups in
quantum polynomial time. HSP can be solved in solvable groups
having smoothly solvable commutator subgroups quantum polynomial
time.
\end{thm}

Fenner and Zhang \cite{FZ04} also address black-box group
algorithms, obtaining efficient quantum algorithms for a few
classically hard problems, by reducing them to Orbit Coset
problems. The problems they study are Group Intersection (given
two subsets $S_1$ and $S_2$ of a group, determine if the groups
$\left<S_1\right>\cap\left<S_2\right>\neq\emptyset$), Coset
Intersection (given two subsets $S_1$ and $S_2$ of a group and a
group element $g$, determine if
$\left<S_1\right>g\cap\left<S_2\right>\neq\emptyset$), and
Double-Coset Membership (given two subsets $S_1$ and $S_2$ of a
group and group elements $g,h$, determine if
$g\in\left<S_1\right>h\left<S_2\right>$).

They obtain
\begin{thm}
Group Intersection over solvable groups can be solved efficiently
in quantum polynomial time if one of the underlying solvable
groups has a smoothly solvable commutator subgroup.
\end{thm}
\begin{thm}
Group Intersection over solvable groups is reducible to Orbit
Superposition in quantum polynomial time.
\end{thm}
\begin{thm}
Coset Intersection and Double-Coset Membership over solvable
groups can be solved in quantum polynomial time if one of the
underlying groups is smoothly solvable.
\end{thm}

van Dam, Hallgren, and Ip \cite{HvDI03} work on a hidden shift
problem They first obtain a superposition result (ignoring the
normalization constant):
\begin{thm}
Let $f:G\rightarrow\mathbb{C}$ be a complex valued function
defined on the set $G$ such that $f(x)$ has unit magnitude
whenever $f(x)$ is nonzero. Then there is an efficient algorithm
for creating the superposition $\sum_x f(x)\ket{x}$ with success
probability equal to the fraction of $x$ such that $f(x)$ is
nonzero and that uses only two queries to the function $f$.
\end{thm}
The proof idea computes the state $\sum_x\ket{x}\ket{f(x)}$, tests
if $f(x)$ is nonzero, moves the phase of $\ket{f(x)}$ into
$\ket{x}$ to high precision, and then applies the second $f$ to
undo the first.

Let $m$ be an integer, $m=p_1^{s_1}p_2^{s_2}\dots p_k^{s_k}$, then
by the Chinese Remainder Theorem, $(\mathbb{Z}/m\mathbb{Z})^*\cong
(\mathbb{Z}/p_1^{s_1}\mathbb{Z})^*\times(\mathbb{Z}/p_2^{s_2}\mathbb{Z})^*
\dots(\mathbb{Z}/p_k^{s_k}\mathbb{Z})^*$. A multiplicative
character $\chi$ on $\mathbb{Z}/m\mathbb{Z}$ can be written as
$\chi(x)=\chi_1(x_1)\chi_2(x_2)\dots\chi_k(x_k)$ using this
isomorphism, where $\chi_i(x_i)$ is a multiplicative character on
$(\mathbb{Z}/p_i^{s_i}\mathbb{Z})^*$. We say $\chi$ is
\emph{completely nontrivial} if each $\chi_i$ is nontrivial. With
this definition, they then solve some shifted character problems:
\begin{thm}
Given a nontrivial (resp. completely nontrivial) multiplicative
character $\chi$ of a finite field $\mathbb{F}_q$ (where $q=p^r$
for some prime $p$) (resp. over $\mathbb{Z}/m\mathbb{Z}$), and a
function $f$ for which there is a shift $s$ with $f(x)=\chi(x+s)$
for all $x\in \mathbb{F}_q$ (resp. $x\in \mathbb{Z}/m\mathbb{Z}$).
Then there is an efficient quantum algorithm finding $s$ with
probability $1-1/q^2$ (resp.
$(\frac{\phi(m)}{m})^3=\Omega((\frac{1}{\log\log m})^3)$).
\end{thm}

In the case where $m$ is unknown, this can still be done given a
bound on $m$.


\subsection{Hidden Subgroups are Distinguishable}\label{s:hidden}
In this section we show that at least information theoretically,
it is possible to find any hidden subgroup $H$ of a finite group
$G$ with only $\lceil 4\log |G|+2\rceil$ calls to the oracle
function $f$, following \cite{EHK99} and done differently in
\cite{EHK04}. Unfortunately, deducing $H$ from the resulting
quantum state requires exponential classical time, and it is still
open for which groups this can be reduced to a polynomial time
quantum algorithm. The idea is to create a quantum state that
contains enough information to deduce $H$ using few oracle calls,
and then use $|G|$ applications of various measurements to this
state to query each element of $G$. The technical work is to prove
the measurements do not perturb the state too much, which would
destroy information needed for later queries.

Precisely we prove:

\begin{thm}
Given a finite group $G$ and an oracle function $f:G\rightarrow X$
to a set $X$, such that $f$ separates cosets of a subgroup $H<G$
($f$ ``hides" $H$). Then there exists a quantum algorithm that
calls the oracle function $\lceil 4\log|G|+2\rceil$ times and
outputs a subset $S\subseteq G$, such that $S=H$ with probability
at least $1-1/|G|$.
\end{thm}
\begin{proof}
Fix a positive integer $m$. We work over the Hilbert space
$\mathcal{H}$ of dimension $|G|^m$, with orthonormal basis indexed
by $m$-tuples of elements of $G$. For any subset
$S=\{s_1,s_2,\dots,s_k\}\subseteq G$ let $\ket{S}$ be the
normalized superposition
$\ket{S}=\frac{1}{\sqrt{k}}\left(\ket{s_1}+\dots\ket{s_k}\right)$.
The first step is to prepare on $\mathcal{H}\otimes\mathcal{H}$
the state
\begin{equation}
\frac{1}{\sqrt{|G|^m}}\sum_{g_1,\dots,g_m\in
G}\ket{g_1,\dots,g_m}\ket{f(g_1),\dots,f(g_m)}
\end{equation}
where we define $\ket{f(g_i)}=\ket{g_iH}$. Note this required $m$
calls to the function $f$. Observing the second register leaves in
the first register the state \ket{\Psi} which is a tensor product
of random left cosets of H, uniformly distributed. We ignore the
second register for the rest of this proof. Let
$\ket{\Psi}=\ket{a_1H}\otimes\dots\otimes\ket{a_mH}$ denote the
first register, where the $a_i\in G$. For any (ordered) subset
$\{b_1,\dots,b_m\}\subseteq G$ and subgroup $K\leq G$ define
\begin{equation}
\ket{\Psi(K,\{b_i\})}=\ket{b_1K}\otimes\ket{b_2K}\otimes\dots\otimes\ket{b_mK}
\end{equation}
The key lemma, lemma \ref{l:ortho1}, shows for $K\nleq H$ that
$\braket{\Psi}{\Psi(K,\{g_i\})}$ is exponentially small for any
$m$ of the $g_i$.

Let $\mathcal{H}_K$ be the subspace of $\mathcal{H}$ spanned by
all vectors of the form \ket{\Psi(K,\{g_i\})} for all subsets
$\{g_1,\dots,g_m\}\subseteq G$. Let $P_K$ be the projection
operator\footnote{Thus $P_K=\sum_{(b_1,\dots,b_m)\in
G^m}\ket{\Psi(K,\{b_i\})}\bra{\Psi(K,\{b_i\})}$} onto
$\mathcal{H}_K$, and $P_K^\bot$ the projection onto the orthogonal
complement of $\mathcal{H}_K$ in $\mathcal{H}$. Define the
observable $A_K=P_K-P_K^\bot$, and fix an ordering
$g_1,g_2,\dots,g_{|G|}$ of $G$.

The algorithm then works as follows: First apply
$A_{\left<g_1\right>}$ to \ket{\Psi}, where $\left<g\right>\leq G$
denotes the cyclic subgroup generated by $g\in G$. If the outcome
is -1, then we know $g_1\not\in H$ with certainty, and if the
outcome is +1 we know $g_1\in H$ with high probability, by lemma
\ref{l:ortho1}. We then apply $A_{\left<g_2\right>}$ to the state
resulting from the first measurement. Continuing in this manner,
we test all elements of $G$ for membership in $H$ by sequentially
applying $A_{\left<g_2\right>}, A_{\left<g_3\right>}$, and so on
to the resulting states of the previous measurements. Of course if
we discover $g\in H$ then we can omit the tests for $g^j\in H$.
Note we may have to apply $O(|G|)$ operations to test each
element, making the algorithm complexity exponential in $\log
|G|$. All that remains to show is that each measurement alters the
state insignificantly with high probability, so that by the final
operator $A_{\left<g_{|G|}\right>}$ we have identified with high
probability exactly which elements are in $H$ and which are not.

We bound this probability of success. Let
$\ket{\Psi_0}=\ket{\Psi}$. For $1\leq i\leq |G|$, define the
unnormalized states
\begin{equation}
\ket{\Psi_i}=\left\{\begin{array}{ll}P_{\left<g_i\right>}\ket{\Psi_{i-1}}&\text{if
}g_i\in H\\P_{\left<g_i\right>}^\bot\ket{\Psi_{i-1}}&\text{if
}g_i\not\in H\end{array}\right.
\end{equation}
By induction and the definition of the probabilities,
\braket{\Psi_i}{\Psi_i} equals the probability that the algorithm
given above answers correctly whether $g_j\in H$ for all $1\leq
j\leq i$. Now for all $0\leq i\leq |G|$ let
$\ket{E_i}=\ket{\Psi}-\ket{\Psi_i}$ denote the error between the
original state and the desired state after testing
$\left<g_i\right>$.

Since $\ket{\Psi_{|G|}}=\ket{\Psi}-\ket{E_{|G|}}$, using
$\braket{E_{|G|}}{E_{|G|}}\leq\frac{|G|^2}{2^m}$ by lemma
\ref{l:ortho2} and the triangle inequality gives that the
probability for correctly determining all the elements of $H$ is
bounded below by $\braket{\Psi_{|G|}}{\Psi_{|G|}}\geq
1-\frac{2|G|}{2^{m/2}}$.

By choosing $m=\lceil 4\log|G|+2\rceil$ the main theorem follows
directly.
\end{proof}

\begin{lemma}\label{l:ortho1}
Use the notation above. Let $K\leq G$. If $K\nleq H$ then
$\brakett{\Psi}{P_K}{\Psi}\leq\frac{1}{2^m}$. If $K\leq H$ then
$\brakett{\Psi}{P_K}{\Psi}=1$.
\end{lemma}
\begin{proof}
Let $|H\cap K|=d$. Note that for all $g_1, g_2\in G$ we have
$|g_1H\cap g_2K|=d$ or $|g_1H\cap g_2K|=0$. This implies that if
$|g_1H\cap g_2K|=d$ then $\braket{g_1H}{g_2K}=d/\sqrt{|H||K|}$.
Therefore for any subset $\{b_1,\dots,b_m\}\subseteq G$
\begin{equation}
\braket{\Psi}{\Psi(K,\{b_i\})}=\left\{\begin{array}{ll}\left(\frac{d}{\sqrt{|H||K|}}\right)^m&\text{if
}|a_iH\cap b_iK|=d\text{ for }i=1,2,\dots,m
\\0&\text{otherwise}\end{array}\right.
\end{equation}
There exist exactly $(|H|/d)^m$ vectors of the form
\ket{\Psi(K,\{b_i\})} with \braket{\Psi}{\Psi(K,\{b_i\})} nonzero.
Hence
$\brakett{\Psi}{P_K}{\Psi}=\left(\frac{|H|}{d}\right)^m\left(\frac{d^2}{|H||K|}\right)^m=\left(\frac{d}{|K|}\right)^m$.
If $K\nleq H$ then $d/|K|\leq 1/2$ and if $K\leq H$ then $d=K$.
\end{proof}

\begin{lemma}\label{l:ortho2}
For all $0\leq i\leq |G|$ we have
$\braket{E_i}{E_i}\leq\frac{i^2}{2^m}$.
\end{lemma}
\begin{proof}
Proof by induction on $i$. Since $\ket{\Psi_0}=\ket{\Psi}$, by
definition $\ket{E_0}=0$. Now suppose
$\braket{E_i}{E_i}\leq\frac{i^2}{2^m}$. If $g_{i+1}\in H$, then
$\ket{\Psi_{i+1}}=P_{\left<g_{i+1}\right>}\left(\ket{\Psi}-\ket{E_i}\right)=\ket{\Psi}-P_{\left<g_{i+1}\right>}\ket{E_i}$.
Hence
$\braket{E_{i+1}}{E_{i+1}}\leq\braket{E_i}{E_i}\leq\frac{i^2}{2^m}$.
If $g_{i+1}\not\in H$, then
$\ket{\Psi_{i+1}}=P_{\left<g_{i+1}\right>}^\bot\left(\ket{\Psi}-\ket{E_i}\right)=\ket{\Psi}-P_{\left<g_{i+1}\right>}\ket{\Psi}-P_{\left<g_{i+1}\right>}^\bot\ket{E_i}$.
By lemma \ref{l:ortho1} we then have
$\braket{E_{i+1}}{E_{i+1}}=\brakett{\Psi}{P_{\left<g_i\right>}}{\Psi}+\braket{E_i}{E_i}
\leq\frac{1}{2^m}+\frac{i^2}{2^m}\leq\frac{(i+1)^2}{2^m}$.
\end{proof}



\section{Conclusion}
In conclusion, we have shown in great detail how to find hidden
subgroups in any finite abelian group. This was shown to be
efficient using a quantum computer, and is the basis for Shor's
factoring algorithm, as well as many other exponentially faster
quantum algorithms. The key ingredient was Fourier sampling - that
is, doing a quantum Fourier transform on a state encoding the
hidden subgroup, and then measuring (sampling) the resulting state
to gather information used to compute the hidden subgroup
generators.

Also, we described the nonabelian case of the HSP, using
representation theory to define the Fourier transform over
arbitrary finite groups, and then mimicking the abelian case in an
attempt to solve the HSP efficiently for any finite group. However
this case is much harder, and only partial results are known, many
of which we listed.

The main open problem in the field is finding an efficient quantum
algorithm for the symmetric group $S_n$, which would yield an
elusive (for over 30 years) efficient algorithm for determining
graph isomorphism. However it seems that quantum Fourier sampling
may not be up to the task since there are many negative results.
Yet there is hope that a clever basis choice for the irreducible
representations might turn this around. A second possibility, also
seemingly remote, is finding a new quantum algorithm which does
the trick, avoiding Fourier sampling completely.

\subsection{Other Quantum Algorithms}
There are many other areas where quantum algorithms are better
than classical ones. One of the earliest algorithms was Grover's
searching algorithm \cite{Grover96}, which reduces the classical
complexity of searching an unordered list of $N$ items from $O(N)$
to a provably best quantum $\Theta(\sqrt{N})$ oracle
queries\footnote{Many authors claim $O(\sqrt{N})$ is the algorithm
\emph{time} complexity. A careful look shows $O(\sqrt{N}\;\log N)$
is a more reasonable time complexity.}. See also \cite{BBHT96}.
This was exploited by \cite{RV01} to make a quantum string
matching algorithm much faster the best classical algorithms given
in \cite{Knuth77, BoyerMoore1977}.

Other quantum algorithms are found in \cite{Aaron02, BHMT00, DH96,
Hallgren02, HalDam00, HvDI, HvDI03, Joz03, Kempe02, vanDam00}.
More quantum algorithm overviews are in \cite{BARENCO96, BBBV97,
CEMM98, GalMD01, Lom02, EckMos98, Shor02}. Continuous variable
algorithms are considered in \cite{KaufLom03, AB02, ABL00}. A good
point to start learning quantum error correction is \cite{CRSS97}.

Another interesting direction is taken by Or\'{u}s,  Latorre, and
Mart\'{i}n-Delgado in \cite{LOM03a, LOM03b} where the authors
notice an invariant of efficient quantum algorithms labelled
``majorization," which they use to seek new algorithms.

A final direction is adiabatic quantum computation \cite{MvDV02},
another quantum computation computing model that may be physically
realizable. It has recently been shown to be equivalent to the
standard qubit model \cite{AvKLLR04}, but provides another
viewpoint for quantum computation.


\newpage
\appendix
These appendices contain results used above.
\section{The Cyclic Quantum Fourier Transform over $\mathbb{Z}_N$}\label{s:GenCyclicFT}
Here we give details on the cyclic QFT over $\mathbb{Z}_{2^n}$ and
over $\mathbb{Z}_N$ for $N$ odd.

\subsection{The Quantum Fourier Transform over $\mathbb{Z}_{2^n}$}\label{s:approxQFT}

This section follows Coppersmith \cite{Copper94}. Since we already
showed how to do the QFT over $\mathbb{Z}_{2^n}$ in section
\ref{s:powerNCase}, we only have to cover the approximate QFT. The
main result is
\begin{thm}\label{t:approxQFT}
Given an $\epsilon>0$ and a positive integer $n$, let $N=2^n$.
Then there is a quantum circuit approximating the Fourier
transform over $\mathbb{Z}_N$ using $O(\log N(\log\log N +
\log(1/\epsilon)))$ 2-qubit operations. The approximated quantum
state \ket{\phi} differs from the true Fourier transformed state
\ket{\psi} by $\left\|\;\ket{\phi}-\ket{\psi}\;\right\|<\epsilon$.
\end{thm}
\begin{proof}
Let $n$ be a positive integer. Let $a,c$ be $n$-bit integers. The
binary representations of $a$ and $c$ are
\begin{equation}
a=\sum_{i=0}^{n-1}a_i2^i,\;\;c=\sum_{i=0}^{n-1}c_i2^i.
\end{equation}
Let $X,Y$ be arrays of size $2^n$ indexed by $a$ or $c$. Let
$\omega=\omega_{2^n}=\exp(2\pi i/2^n)$ be the standard $2^n$ root
of unity.

The Fourier transform is defined as
\begin{equation}
Y_c=\frac{1}{\sqrt{2^n}}\sum_a X_a
\omega^{ac}=\frac{1}{\sqrt{2^n}}\sum_a
X_a\exp\left(\frac{2\pi}{2^n}ac\right),
\end{equation}
or, in binary notation,
\begin{equation}
Y_c=\frac{1}{\sqrt{2^n}}\sum_a
X_a\exp\left(\frac{2\pi}{2^n}\sum_{j,k=0}^{n-1}a_jc_k2^{j+k}\right).
\end{equation}
Whenever $j+k\geq n$, $\omega^{2^{j+k}}=1$, so we drop those
terms, giving the Fast Fourier Transform (FFT)
\begin{equation*}
(\text{FFT}) \;\;\;\;\; Y_c=\frac{1}{\sqrt{2^n}}\sum_a
X_a\exp\left(\frac{2\pi}{2^n}\sum_{\substack{0\leq j,k\leq n-1\\
j+k\leq n-1}} a_jc_k2^{j+k}\right).
\end{equation*}
Now we approximate. Instead of the summation range having a $0\leq
j+k\leq n-1$ bound, we parameterize on a positive integer $m<n$
and bound by $n-m\leq j+k\leq n-1$, giving the Approximate Fast
Fourier Transform ($\text{AFFT}_m$):
\begin{equation*}
(\text{AFFT}_m)\;\;\;\;\; Y_c=\frac{1}{\sqrt{2^n}}\sum_a
X_a\exp\left(\frac{2\pi}{2^n}\sum_{\substack{0\leq j,k\leq n-1\\
n-m\leq j+k\leq n-1}} a_jc_k2^{j+k}\right).
\end{equation*}

The argument of ``exp" in the AFFT differs from that in the FFT by
\begin{equation}
\frac{2\pi i}{2^n}\sum_{j+k<n-m} a_jc_k2^{j+k},
\end{equation}
and is bounded in magnitude by
\begin{eqnarray*}
\left|\frac{2\pi i}{2^n}\sum_{\substack{0\leq j,k\leq n-1\\
j+k<n-m}} a_jc_k2^{j+k}\right|&\leq&
\frac{2\pi}{2^n}\sum_{0\leq j<n-m}2^j\sum_{0\leq k<n-m-j} 2^k\\
&=&\frac{2\pi}{2^n}\sum_{0\leq j<n-m}2^j\left(2^{n-m-j}-1\right)\\
&=&\frac{2\pi}{2^n}\left((n-m)2^{n-m}-2^{n-m}+1\right)\\
&\leq&\frac{2\pi}{2^n}n2^{n-m}\\
&=&2\pi n2^{-m}.
\end{eqnarray*}
So the matrix entries of the AFFT differ from the FFT by a
multiplicative factor of $\exp(i \delta)$, where
$\left|\delta\right|\leq 2\pi n 2^{-m}$. Let this error be
$\exp(\delta_{j,k})$ in the $(j,k)$ entry. From arc length on a
circle, we have $|1-e^{i\delta}|\leq|\delta|$.

To compute the error between the quantum states resulting from the
FFT and AFFT, compute for any state $\ket{\psi}=\sum_ja_j\ket{j}$
\begin{eqnarray}
\left\|(\text{FFT}-\text{AFFT}_m)\ket{\psi}\right\|^2 &=&
\sum_{k=0}^{N-1}\left|\frac{1}{\sqrt{N}}\sum_{j=0}^{N-1}\omega_N^{jk}a_j (1-\exp(\delta_{j,k}))\right|^2\\
&\leq& (2\pi n2^{-m})^2\sum_{k=0}^{N-1}\left|\frac{1}{\sqrt{N}} \sum_{j=0}^{N-1}\omega_N^{jk}a_j\right|^2\\
&=& (2\pi n2^{-m})^2 \left\|\text{FFT}\ket{\psi}\right\|^2\\
&=& (2\pi n2^{-m})^2 \cdot 1
\end{eqnarray}
Thus for any $\epsilon>0$, taking $m\geq \log (2\pi) + \log n +
\log(1/\epsilon)$ gives that
\begin{equation}
\left\|(\text{FFT}-\text{AFFT}_m)\ket{\psi}\right\|<\epsilon
\end{equation}

Now we show how to compute the AFFT efficiently, similar to the
method in section \ref{s:powerNCase}. Let $Q^{(J,K)}$ be the
operation that multiplies the amplitude of those states with a 1
in positions $J$ and $K$ by a factor of $\omega^{2^{n-1-K-J}}$.
This is similar to the $R^{(a,b)}_k$ defined for the QFT earlier.
Let $H^{(J)}$ be the operation of applying the Hadamard matrix
$\frac{1}{\sqrt{2}}\left(\begin{matrix}1&1\\1&-1\end{matrix}\right)$
to qubit $J$. Then check that the operation
\begin{equation}
H^{(0)}Q^{(0,1)}Q^{(0,2)}\dots
Q^{(0,n-1)}H^{(1)}Q^{(1,2)}Q^{(1,3)}\dots Q^{n-2,n-1}H^{(n-1)}
\end{equation}
performs the QFT as earlier. To perform the AFFT we drop those
$Q^{(J,K)}$ with $K\geq J+m$, so it requires about $nm$ 2-qubit
operations. Taking $m=O(\log n + \log(1/\epsilon))$ to bound the
error as required, we obtain the complexity bound.
\end{proof}



\subsection{The Quantum Fourier Transform over $\mathbb{Z}_{N}$, $N$ Odd}\label{s:OddQFT}

\newcommand{\Hales}{\cite{Hal02}}       
\newcommand{\Hallgren}{\cite{HalHal00}} 

This section gives an algorithm to approximate the QFT over
$\mathbb{Z}_N$ efficiently. The algorithm is from the Hales thesis
\Hales~and the paper by Hallgren \emph{et .al} \Hallgren~, but
their proofs are incorrect. This section gives the proof from
Lomont \cite{Lomont04a}. The end result is a proof of the
correctness of their algorithm, with concrete bounds suitable for
quantum simulation instead of the asymptotic bounds listed in
their papers. The final result is theorem \ref{t:MainBound}. The
general idea of the algorithm is to make many copies of the
initial state vector and perform a $2^n$ style QFT for a large
value, and extract from this state period information for the
original odd $N$. The proof requires a lot tedious work; it is
more instructive to work through the algorithm until the general
idea is clear.

\subsubsection{Notation and Basic Facts} We fix three integers: an
odd integer $N\geq 3$, $L\geq 2$ a power of 2, and $M\geq LN$ a
power of 2. This gives $(M,N)=1$, which we need later.

Some notation and facts to clarify the presentation:
\begin{itemize}
\item $\sqrt{-1}$ will be written explicitly, as $i$ will always
denote an index.

\item For an integer $n>1$, let $\omega_n=e^{2\pi\sqrt{-1}/n}$
denote a primitive $n^\text{th}$ root of unity.

\item Fact:
$\left|1-e^{\theta\sqrt{-1}}\right|\leq\left|\theta\right|$ as can
be seen from arc length on the unit circle. If
$-\pi\leq\theta\leq\pi$ we also\footnote{This range can be
extended slightly.} have
$\left|\frac{\theta}{2}\right|\leq\left|1-e^{\theta\sqrt{-1}}\right|$.
Thus for real values $\alpha$ we have
$\left|1-\omega_M^\alpha\right|\leq\left|\frac{2\pi\alpha}{M}\right|$,
etc.

\item $\log n$ denotes $\log$ base 2, while $\ln n$ is the natural
log. Since $M$ and $L$ are powers of two, $\ceil{\log
M}=\floor{\log M}=\round{\log M}=\log M$, and similarly for $L$,
but we often leave the symbols to emphasize expressions are
integral.

\item For a real number $x$, \ceil{x} is the smallest integer
greater than or equal to $x$, \floor{x} is the largest integer
less than or equal to $x$, and \round{x} is the nearest integer,
with ties rounding up\footnote{We could break ties arbitrarily
with the same results.}. We often use the three relations:
\begin{align*}
x-\frac{1}{2}&\leq\round{x}\leq x+\frac{1}{2}\\
x-1&<\floor{x}\leq x\\
x&\leq\ceil{x}<x+1
\end{align*}

\item Indices: $i$ and $s$ will be indices from $0,1,\dots,N-1$.
$j$ will index from $0,1,\dots,L-1$. $k$ will index from
$0,1,\dots,M-1$. $a$ and $b$ will be arbitrary indices. $t$ will
index from a set $C_s$, defined in definition \ref{d:delta} below.

\item Given $i\in\{0,1,\dots,N-1\}$, let $i'=\round{\frac{M}{N}i}$
denote the nearest integer to $\frac{M}{N}i$ with ties broken as
above. Similarly for $s$ and $s'$. Note $0\leq i'\leq M-1$.

\item For a real number $x$ and positive real number $n$, let $x
\bmod n$ denote the real number $y$ such that $0\leq y<n$ and $y =
x+mn$ for an integer $m$. Note that we do not think of $x\bmod n$
as an equivalence class, but as a real number in $[0,n)$.

\item \ket{u} and \ket{v} are vectors in spaces defined later, and
given a vector \ket{u} denote its coefficients relative to the
standard (orthonormal) basis $\{\ket{0},\ket{1},\dots,\ket{n-1}\}$
by $u_0,u_1,\dots,u_{n-1}$, etc.

\item For a real number $x$, let
\begin{equation*}
\left|x\right|_M=\left\{\begin{matrix}x\bmod M&\text{ if }& 0\leq
(x\bmod M)\leq \frac{M}{2}\\-x\bmod M&\text{
otherwise}&\end{matrix}\right.
\end{equation*}
Thus $0\leq |x|_M\leq\frac{M}{2}$. Properties of this function are
easiest to see by noting it is a sawtooth function, with period
$M$, and height $M/2$.

\item For an integer $s$ set
$\delta_s=\round{\frac{M}{N}s}-\frac{M}{N}s$. Then
$\left|\delta_s\right|\leq\frac{1}{2}$.

\item The (unitary) Fourier transform over a cyclic group of order
$N$ is denoted $F_N$. Thus if
$\ket{u}=\sum_{i=0}^{N-1}u_i\ket{i}$, then
$F_N\ket{u}=\frac{1}{\sqrt{N}}\sum_{i,s=0}^{N-1}u_i\omega_N^{is}\ket{s}$.
We write $\ket{\hat{u}}=F_N\ket{u}$, with coefficients
$\hat{u}_i$.

\item $\sum_{i=0}^{N-1} |u_i|^2=1$ implies
$\sum_i|u_i|\leq\sqrt{N}$.
\end{itemize}
~\\

We define sets of integers which will play an important role:
\begin{defn}\label{d:intervals}
For $i=0,1,\dots,N-1$, let $(i)$ denote the set of integers in the
open interval
$\left(i'-\frac{M}{2N}+\frac{1}{2},i'+\frac{M}{2N}-\frac{1}{2}\right)$
taken $\bmod\;M$. Recall $i'=\round{\frac{M}{N}i}$.
\end{defn}

The second definition we make precise is a division and remainder
operation:
\begin{defn}\label{d:delta}
Given $M,N$ as above. Set
$\alpha=\floor{\frac{M}{2N}+\frac{1}{2}}$, and
$\beta=\ceil{\frac{M}{2N}-\frac{3}{2}}$. We define the map
$\Delta:\{0,1,\dots,M-1\}\rightarrow\{0,1,\dots,N-1\}\times\{-\alpha,-\alpha+1,\dots,\alpha\}$,
as follows: for any $k\in\{0,1,\dots,M-1\}$, let
$k\xrightarrow{\Delta}(s,t)$, via
\begin{eqnarray*}
k'&=&\round{k\frac{N}{M}}\\
t&=&k-\round{k'\frac{M}{N}}\\
s&=& k'\bmod N
\end{eqnarray*}
We extend this definition to a transform of basis elements
$\ket{k}$ via
\begin{eqnarray*}
\Delta\ket{k}=\ket{s}\ket{t+\alpha}
\end{eqnarray*}
and extend to all vectors by linearity.

Finally, from the image of $\Delta$, define $C_s=\{\;t\;\;
\big|\;\; (s,t)\in \text{\emph{Image }}\Delta\}$ to be those
values of $t$ appearing for a fixed $s$. Thus
$\sum_{k=0}^{M-1}\ket{k}\xrightarrow{\Delta}\sum_{s=0}^{N-1}\sum_{t\in
C_s}\ket{s}\ket{t+\alpha}$.
\end{defn}

We will show the integers $\{-\beta,\dots,\beta\}\subseteq
C_s\subseteq\{-\alpha,\dots,\alpha\}$ for all $s$, which is why we
defined $\beta$ with the $\Delta$ definition. $\alpha$ and $\beta$
remain fixed throughout the paper.

For the proofs to work, we need that the sets $(i)$ are disjoint
and have the same cardinality. Note also that the $\bmod\;M$
condition gives $M-1,0\in(0)$ when $M>3N$. We now show that the
sets defined here have the required properties:
\begin{lemma}\label{l:setProp}
For $i_1\neq i_2\in\{0,1,\dots,N-1\}$,
\begin{eqnarray}
\left|(i_1)\right| &=& \left|(i_2)\right|\\
(i_1)\bigcap(i_2)&=&\varnothing
\end{eqnarray}
\end{lemma}
\begin{proof}
Each set is defined using an interval of constant width, centered
at an integer, so the sets will have the same cardinality. To show
disjointness, for any integer $a$, take the rightmost bound
$R_a=\round{\frac{M}{N}a}+\frac{M}{2N}-\frac{1}{2}$ of an interval
and compare it to the leftmost bound
$L_{a+1}=\round{\frac{M}{N}(a+1)}-\frac{M}{2N}+\frac{1}{2}$ of the
next interval:
\begin{eqnarray}
L_{a+1}-R_a&=&\round{\frac{M}{N}(a+1)}-\round{\frac{M}{N}a}-\frac{M}{N}+1\\
&\geq&\left(\frac{M}{N}(a+1)-\frac{1}{2}\right)-\left(\frac{M}{N}a+\frac{1}{2}\right)-\frac{M}{N}+1\\
&=&0
\end{eqnarray}
giving that the open intervals are disjoint. Thus taking the
integers in the intervals $\bmod\;M$ remains disjoint (which
requires $i_1, i_2\leq N-1$).
\end{proof}

Note the image of $\Delta$ is not a cartesian product; the values
$t$ assumes depend on $s$, otherwise we would have that $M$ is a
multiple of $N$. In other words, the cardinality of $C_s$ depends
on $s$, with bounds given in the following lemma, where we show
that our definition works and list some properties:

\begin{lemma}\label{l:deltaProp}
Using the notation from definition \emph{\ref{d:delta}},

1) the map $\Delta$ is well defined, and a bijection with its
image,

2) $\alpha=\beta+1$,

3) the sets of integers satisfy $\{-\beta,\dots,\beta\}\subseteq
C_s\subseteq\{-\alpha,\dots,\alpha\}$ for all
$s\in\{0,1,\dots,N-1\}$.
\end{lemma}
\begin{proof}
Given a $k$ in $\{0,1,\dots,M-1\}$, let $\Delta(k)=(s,t)$. Clearly
$0\leq s\leq N-1$. Set $\alpha=\floor{\frac{M}{2N}+\frac{1}{2}}$.
To check that $-\alpha\leq t\leq \alpha$, note
\begin{equation}
\frac{N}{M}k-\frac{1}{2}\leq k'\leq\frac{N}{M}k+\frac{1}{2}
\end{equation}
giving
\begin{equation}
\frac{M}{2N}+\frac{1}{2}\geq t=k-\round{\frac{M}{N}k'}\geq
-\left(\frac{M}{2N}+\frac{1}{2}\right)
\end{equation}
and $t$ integral allows the rounding operation. Thus the
definition makes sense.

Next we check that both forms of $\Delta$ in the definition are
bijections. Suppose $k_1\neq k_2$ are both in $\{0,1,\dots,M-1\}$,
with images $\Delta(k_r)=(s_r,t_r),r=1,2$. Let
$k_r'=\round{\frac{N}{M}k_r},r=1,2$. Note $0\leq k_r'\leq N$.

Assume $(s_1,t_1)=(s_2,t_2)$. If $k_1'=k_2'$, then
\begin{align}
t_1&=k_1-\round{\frac{M}{N}k_1'}=k_1-\round{\frac{M}{N}k_2'}\\
&\neq k_2-\round{\frac{M}{N}k_2'}= t_2
\end{align}
a contradiction. So we are left with the case $k_1'\neq k_2'$. In
order for $s_1=s_2$ we have (without loss of generality) $k_1'=0,
k_2'=N$. But then $t_1=k_1\geq 0$ and $t_2=k_2-M\leq M-1-M=-1$, a
contradiction. Thus $\Delta$ in the first sense is a bijection.

The second interpretation follows easily, since $-\alpha\leq
t\leq\alpha$ gives $0\leq t+\alpha\leq2\alpha$. So the second
register needs to have a basis with at least $2\alpha+1$ elements,
which causes the number of qubits needed\footnote{This is proven
in theorem \ref{t:MainBound}.} to implement the algorithm to be
$\ceil{\log M} + 2$ instead of $\ceil{\log M}$.

To see $\alpha=\beta+1$, bound $\alpha-\beta$ using the methods
above, and\footnote{$(M,N)=1$ is used to get the strict
inequalities.} one obtains $2>\alpha-\beta>0$.

All integers between $\round{\frac{M}{N}(s+1)}$ and
$\round{\frac{M}{N}s}$ inclusive must be of the form
$t_1+\round{\frac{M}{N}s}$ for $t_1\in C_s$ or of the form
$t_2+\round{\frac{M}{N}(s+1)}$ for $t_2\in C_{s+1}$. This range
contains $\round{\frac{M}{N}(s+1)} - \round{\frac{M}{N}s}+1\geq
\frac{M}{N}$ integers, and at most $\alpha+1$ of these are of the
form $t_2+\round{\frac{M}{N}(s+1)}$ with $t_2\in C_{s+1}$. This
leaves at least
$\ceil{\frac{M}{N}}-\alpha\geq\frac{M}{2N}-\frac{3}{2}$ that have
to be of the form $t_1+\round{\frac{M}{N}s}$ with $t_1\in C_s$,
implying $\beta\in C_s$. Similar arguments give $\pm \beta\in
C_s$, thus $\{-\beta,\dots,\beta\}\subseteq
C_s\subseteq\{-\alpha,\dots,\alpha\}$ for all $s$.
\end{proof}

$\Delta$ is efficient to implement as a quantum operation, since
it is efficient classically \cite[Chapter 4]{ChuangNielsen}.
Finally we note that $\Delta$, being a bijection, can be extended
to a permutation of basis vectors $\ket{k}$, thus can be
considered an efficiently implementable unitary operation.

We define some vectors we will need. For
$i\in\left\{0,1,\dots,N-1\right\}$ define
\begin{eqnarray*}
\ket{A^i} &=& F_M F_{LN}^{-1}\ket{Li}\\
&=& \frac{1}{\sqrt{LMN}}\sum_{k=0}^{M-1}\sum_{a=0}^{LN-1}\omega_{N}^{-ai}\omega_M^{ak}\ket{k}\\
\ket{B^i} &=& \ket{A^i}\text{ restricted to integers in the set } (i)\\
&=&\sum_{b\in(i)}A^i_b\ket{b}\\
&=& \frac{1}{\sqrt{LMN}}\sum_{b\in(i)}\sum_{a=0}^{LN-1}
\omega_{N}^{-ai}\omega_M^{ab}\ket{b}\\
\ket{T^i} &=& \ket{A^i}\text{ restricted to integers outside the set } (i)\\
&=&\sum_{b\not\in(i)}A^i_b\ket{b}\\
&=&\ket{A^i}-\ket{B^i}\\
&=& \frac{1}{\sqrt{LMN}}\sum_{b\not\in(i)}\sum_{a=0}^{LN-1}
\omega_{N}^{-ai}\omega_M^{ab}\ket{b}\\
\end{eqnarray*}
Think $A^i$ for actual values, $B^i$ for bump functions, and $T^i$
for tail functions. Note that the coefficients $B^i_b$ and $T^i_b$
are just $A^i_b$ for $b$ in the proper ranges.

We also define three equivalent shifted versions of $\ket{B^0}$.
Note that to make these definitions equivalent we require the sets
$(i)$ to have the same cardinality. Let
$\ket{S^i}=\sum_{b\in(0)}B^0_b\ket{b+i'} =
\sum_{b\in(0)}A^0_b\ket{b+i'}=\sum_{b\in(i)}A^0_{b-i'}\ket{b}$,
where each $b\pm i'$ expression is taken $\bmod\;M$. The
$\ket{S^i}$ have \emph{disjoint support}, which follows from lemma
\ref{l:setProp}, and will be important for proving theorem
\ref{t:oddCyclicQFT}.

\subsubsection{The Algorithm} The algorithm takes a unit vector
(quantum state) $\ket{u}$ on $\ceil{\log N}$
qubits\footnote{Recall logs are base 2.}, does a Fourier transform
$F_L$, $L$ a power of two, on another register containing
$\ket{0}$ with $\ceil{\log M}-\ceil{\log N}+2$ qubits, to
create\footnote{Note it may be more efficient to apply the
Hadamard operator $H$ to each qubit in $\ket{0}$.} a
superposition, and then reindexes the basis to create $L$
(normalized) copies of the coefficients of $\ket{u}$, resulting in
$\ket{u_L}$. Then another power of two Fourier transform $F_M$ is
applied. The division $\Delta$ results in a vector very close to
the desired output $F_N\ket{u}$ in the first register, with
garbage in the second register (with some slight entanglement).
The point of this paper is to show how close the output is to this
tensor product. We use $\ceil{\log M}+2$ qubits, viewed in two
ways: as a single register \ket{k}, or as a $\ceil{\log N}$-qubit
first register, with the remaining qubits in the second register,
written \ket{s}\ket{t}. We note that merely $\ceil{\log M}$ qubits
may not be enough qubits to hold some of the intermediate results.
The algorithm is:

\subsubsection{The Odd Cyclic QFT Algorithm}\label{s:algorithm}
\begin{eqnarray}
\ket{u}\ket{0}&\xrightarrow{F_L}&\frac{1}{\sqrt{L}}
\sum_{i=0}^{N-1}\sum_{j=0}^{L-1}u_i\ket{i}\ket{j}\\
&\xrightarrow{\text{multiply}}&\frac{1}{\sqrt{L}}\sum_{i,j}u_i\ket{i+jN}\\
&=&\ket{u_L}\\
&\xrightarrow{F_M}&\frac{1}{\sqrt{LM}}\sum_{i,j}\sum_{k=0}^{M-1}u_i\omega_M^{\left(i+jN\right)k}\ket{k}\\
&\xrightarrow{\Delta}&\frac{1}{\sqrt{LM}}\sum_{i,j}u_i
\sum_{s=0}^{N-1}\sum_{t\in C_s}
\omega_M^{\left(i+jN\right)\left(t+\round{\frac{M}{N}s}\right)}\ket{s}\ket{t+\alpha}\\
&=&\frac{1}{\sqrt{N}}\sum_{i,s=0}^{N-1}u_i\omega_N^{is}\ket{s}\sqrt{\frac{N}{LM}}
\sum_{t\in C_s}\sum_{j=0}^{L-1}
\omega_M^{\left(i+jN\right)\left(t+\delta_s\right)}\ket{t+\alpha}\label{e:output}\\
&=&\ket{v}
\end{eqnarray}

$\ket{u_L}$ is the vector that is $L$ copies of the coefficients
from $\ket{u}$, normalized. $\ket{v}$ is the algorithm output.

Notice that $F_N\ket{u}$ appears in the output in line
\ref{e:output}, but the rest is unfortunately dependent on $s$ and
$i$. However the dependence is small: if $C_s$ were the same for
all $s$, if the $\delta_s$, which are bounded in magnitude by
$\frac{1}{2}$, were actually zero, and if the $i$ dependence were
dropped, then the output would leave $F_N\ket{u}$ in the first
register. The paper shows this is approximately true, and
quantifies the error.

\subsubsection{Initial Bounds}We need many bounds to reach the final
theorem, which we now begin proving.
\begin{lemma}\label{l:Mbounds}
For integers $N>2$, $M\geq 2N$, and any $i\in\{0,1,\dots,N-1\}$,
$k\in\{0,1,\dots,M-1\}$, with $k\not\in(i)$, we have
\begin{eqnarray}
\left|k-\frac{M}{N}i\right|_M&\geq&\frac{M}{2N}-1
\end{eqnarray}
\end{lemma}
\begin{proof}
The sets $(i)$ are disjoint, so we do two cases. If $i=0$, then
$k\not\in(0)$ implies
\begin{equation}
\frac{M}{2N}-\frac{1}{2}\leq k \leq M-\frac{M}{2N}+\frac{1}{2}
\end{equation}
from which it follows that
\begin{equation}
\left|k-\frac{M}{N}0\right|_M\geq\frac{M}{2N}-\frac{1}{2}>\frac{M}{2N}-1
\end{equation}

If $i\neq 0$, then either $k$ is less than the integers in $(i)$
or greater than the integers in $(i)$, giving two subcases.
Subcase 1:
\begin{equation}
0\leq k\leq\round{\frac{M}{N}i}-\frac{M}{2N}+\frac{1}{2}
\leq\frac{M}{N}i-\frac{M}{2N}+1
\end{equation}
implying
\begin{equation}
\frac{M}{2N}-1\leq\frac{M}{N}i-k\leq \frac{M}{N}i\leq
M-\frac{M}{N}
\end{equation}
which gives the bound. Subcase 2 is then
\begin{equation}
\frac{M}{N}i+\frac{M}{2N}-1\leq
\round{\frac{M}{N}i}+\frac{M}{2N}-\frac{1}{2}\leq k \leq M-1
\end{equation}
which implies
\begin{equation}
\frac{M}{2N}-1\leq k-\frac{M}{N}i \leq M-1-\frac{M}{N}i
\end{equation}
giving the bound and the proof.
\end{proof}

We now bound many of the $\ket{A^i}$ coefficients.

\begin{lemma}\label{l:approx} For
$k\in\left\{0,1,\dots,M-1\right\}$ and
$i\in\left\{0,1,\dots,N-1\right\}$, with $\frac{k}{M}-\frac{i}{N}$
not an integer, then
\begin{equation}
\left|A^i_k\right|\leq\sqrt{\frac{M}{LN}}\;\frac{2}{\pi\left|k
-\frac{M}{N}i\right|_M}
\end{equation}
\end{lemma}
\begin{proof}
We rewrite from the definition
\begin{eqnarray}
A^i_k &=& \frac{1}{\sqrt{LMN}}\sum_{a=0}^{LN-1}
\omega_M^{a\left(k-\frac{M}{N}i\right)}\\
\end{eqnarray}
which is a geometric series. By hypothesis,
$\omega_M^{\left(k-\frac{M}{N}i\right)}\neq 1$, so we can sum
as\footnote{Without this requirement, the sum would be $LN$, much
different than the claimed sum. The hypotheses avoid the resulting
divide by zero.}

\begin{eqnarray}
\left|A^i_k\right| &=&
\frac{1}{\sqrt{LMN}}\left|\frac{1-\omega_M^{LN\left(k-\frac{M}{N}i\right)}}
{1-\omega_M^{\left(k-\frac{M}{N}i\right)}}\right|
\end{eqnarray}
The numerator is bounded above by 2, and the denominator satisfies
\begin{eqnarray}
\left|1-\omega_M^{\left(k-\frac{M}{N}i\right)}\right| &=&
\left|1-\omega_M^{\left|k-\frac{M}{N}i\right|_M}\right|\\
&\geq& \frac{\pi \left|k-\frac{M}{N}i\right|_M}{M}
\end{eqnarray}
These together give
\begin{eqnarray}
\left|A^i_k\right| &\leq& \sqrt{\frac{M}{LN}}\;\frac{2}{\pi
\left|k-\frac{M}{N}i\right|_M}
\end{eqnarray}
\end{proof}

Note our initial requirement that $(M,N)=1$ is strong enough to
satisfy the non-integral hypothesis in lemma \ref{l:approx},
except for the case $i=k=0$, which we will avoid.

Next we bound a sum of these terms. We fix
$\gamma=\frac{1}{2}-\frac{N}{M}$ for the rest of this paper.

\begin{lemma}\label{l:sumGeneral}Given integers $N>2$ and $M>2N$, with $N$ odd. Let
$\gamma=\frac{1}{2}-\frac{N}{M}$. For a fixed integer
$k\in\left\{0,1,\dots,M-1\right\}$,
\begin{equation}
\sum_{\substack{i=0\\k\not\in(i)}}^{N-1}\frac{1}{\left|k-\frac{M}{N}i\right|_M}
\leq\frac{2N}{M}\left(\frac{1}{\gamma}+\ln\left|\frac{N-1}{2\gamma}+1\right|\;\right)
\end{equation}
\end{lemma}
\begin{proof}The minimum value of the
denominator is at least $\frac{M}{2N}-1$ by lemma \ref{l:Mbounds},
and the rest are spaced out by $\frac{M}{N}$, but can occur
twice\footnote{Both \Hales~and \Hallgren~appear to overlook this
fact.} since the denominator is a sawtooth function going over one
period, giving that
\begin{eqnarray}
\sum_{\substack{i=0\\k\not\in(i)}}^{N-1}\frac{1}{\left|k-\frac{M}{N}i\right|_M}
&\leq&2\sum_{a=0}^{\frac{N-1}{2}}\frac{1}{\frac{M}{2N}-1+\frac{M}{N}a}\\
&=&\frac{2N}{M}\left(\frac{1}{\gamma}+\sum_{a=1}^{\frac{N-1}{2}}\frac{1}{\gamma+a}\right)\\
&\leq&\frac{2N}{M}\left(\frac{1}{\gamma}+\int_0^{(N-1)/2}\frac{1}{x+\gamma}dx\right)\\
&=&\frac{2N}{M}\left(\frac{1}{\gamma}+\ln\left|\frac{N-1}{2\gamma}+1\right|\;\right)
\end{eqnarray}
\end{proof}

The generality of the above lemma would be useful where physically
adding more qubits than necessary would be costly, since the lemma
lets the bound tighten as $\frac{N}{M}$ decreases. However the
following corollary is what we will use in the final theorem.

\begin{cor}\label{l:sum} Given integers $N\geq 13$ and $M\geq16N$, with $N$
odd. For a fixed value $k\in\left\{0,1,\dots,M-1\right\}$,
\begin{equation}
\sum_{\substack{i=0\\k\not\in(i)}}^{N-1}\frac{1}{\left|k-\frac{M}{N}i\right|_M}
\leq\frac{4N\ln N}{M}
\end{equation}
\end{cor}
\begin{proof}
Using lemma \ref{l:sumGeneral}, $M\geq 16N$ gives
$\frac{1}{\gamma}\leq\frac{16}{7}$ and
\begin{eqnarray}
\frac{1}{\gamma}+\ln\left|\frac{N-1}{2\gamma}+1\right|&\leq&
\frac{16}{7}+\ln\left|\frac{8(N-1)}{7}+1\right|\\
&=&\ln\left(e^{\frac{16}{7}}\left(\frac{8(N-1)}{7}+1\right)\right)\\
&\leq& \ln\left(\frac{8}{7}\;e^{\frac{16}{7}}N\right)\\
&\leq&2\ln N
\end{eqnarray}
where the last step required $N\geq \left(
\frac{8}{7}\;e^{\frac{16}{7}}\right)>11.2$. The corollary follows.
\end{proof}

Next we prove a bound on a sum of the above terms, weighted with a
real unit vector. This will lead to a bound on the tails
$\left\|\sum_i\hat{u}_i\ket{T^i}\right\|$.

\newcommand{\matbound}{\frac{22N\ln^2 N}{M}+\frac{32N^3}{M^2}}

\begin{lemma}\label{l:matNorm}
Given integers $N\geq 13$ and $M\geq16N$, with $N$ odd. For any
unit vector $x\in \mathbb{R}^N$
\begin{equation}
\sum_{k=0}^{M-1}\left|\sum_{\substack{i=0\\k\not\in(i)}}^{N-1}\frac{x_i}{\left|k-\frac{M}{N}i\right|_M}\right|^2\leq
\matbound\label{e:matBound}
\end{equation}
\end{lemma}

\begin{proof} We split the expression into three parts, the first
of which we can bound using methods from \Hales~and \Hallgren, and
the other two terms we bound separately.

Using the $\Delta$ operator from definition \ref{d:delta}, along
with the values $\alpha$ and $\beta$ defined there, and using
lemma \ref{l:deltaProp}, we can rewrite each $k$ with
$k=t+\round{\frac{M}{N}k'}=t+\frac{M}{N}k'+\delta_s$. Since $s$
differs from $k'$ by a multiple of $N$, and the $|x|_M$ function
has period $M$, in $\left|\frac{M}{N}(k'-i)+t+\delta_s\right|_M$
we can replace $k'$ with $s$. Rewrite the left hand side of
inequality \ref{e:matBound} as
\begin{align}
\sum_{k=0}^{M-1}\left|\sum_{\substack{i=0\\k\not\in(i)}}^{N-1}\frac{x_i}{\left|k-\frac{M}{N}i\right|_M}\right|^2&=
\sum_{s=0}^{N-1}\sum_{t\in C_s}\left|\sum_{\substack{i=0\\s\neq
i}}^{N-1}\frac{x_i}{\left|\frac{M}{N}(s-i)+t+\delta_s\right|_M}\right|^2
\end{align}
Letting $\Delta k=(s,t)$, note that $k\not\in(i)$ if and only if
$s\neq i$, which can be shown from the definitions and the
rounding rules used earlier. To simplify notation, write
$q_{i,s}^t=\frac{M}{N}(s-i)+t+\delta_s$. We have not changed the
values of the denominators, so $|q_{i,s}^t|_M\geq\frac{M}{2N}-1$
by lemma \ref{l:Mbounds} for all $i,(s,t)$ in this proof.

We want to swap the $s$ and $t$ sums, but we need to remove the
$t$ dependence on $s$. Again using lemma \ref{l:deltaProp}, we can
split the expression into the three terms:
\begin{align}
\sum_{t=-\beta}^{\beta}\sum_{s=0}^{N-1}\left|\sum_{\substack{i=0\\s\neq
i}}^{N-1}\frac{x_i}{\left|q_{i,s}^t\right|_M}\right|^2\label{e:term1}\\
+\sum_{s\text{ with }\alpha\in
C_s}\left|\sum_{\substack{i=0\\s\neq
i}}^{N-1}\frac{x_i}{\left|q_{i,s}^\alpha\right|_M}\right|^2\label{e:term2}\\
+\sum_{s\text{ with }-\alpha\in
C_s}\left|\sum_{\substack{i=0\\s\neq
i}}^{N-1}\frac{x_i}{\left|q_{i,s}^{-\alpha}\right|_M}\right|^2\label{e:term3}
\end{align}
Next we bound the first term \ref{e:term1}. For a unit vector $x$
and fixed $t$ we rewrite the $s,i$ sum as the norm of a square
matrix $P_t$ acting on $x$, so that the sum over $s$ and $i$
becomes
\begin{align}
\left\|P_tx\right\|^2&=\sum_{s=0}^{N-1}\left|\sum_{\substack{i=0\\s\neq
i}}^{N-1}\frac{x_i}{\left|q_{i,s}^t\right|_M}\right|^2
\end{align}
We also define similarly to each $P_t$ a matrix $Q_t$ which is the
same except for minor modifications to the denominator:
\begin{align}
\left\|Q_tx\right\|^2&=\sum_{s=0}^{N-1}\left|\sum_{\substack{i=0\\s\neq
i}}^{N-1}\frac{x_i}{\left|q_{i,s}^t-\delta_s\right|_M}\right|^2
\end{align}
Note this matrix is circulant\footnote{That is, each row after the
first is the cyclic shift by one from the previous row.}, since
each entry in the matrix only depends on $s-i$. Also each entry is
nonnegative\footnote{$|q_{i,s}^t-\delta_s|_M\geq|q_{i,s}^t|_M-\frac{1}{2}\geq\frac{M}{2N}-\frac{3}{2}>0$
since $M>3N$}. Thus the expression is maximized by the vector
$y=\frac{1}{\sqrt{N}}\left(1,1,\dots,1\right)$ as shown in each of
\Hales, \Hallgren, and \cite{Hoy00}. Now we relate these matrix
expressions. Recall $|q_{i,s}^t|_M\geq\frac{M}{2N}-1$ and
$|\delta_s|\leq\frac{1}{2}$. Set $\lambda=\frac{N}{M-2N}$. Then we
find lower and upper bounds
\begin{align}
1-\lambda=1-\frac{1}{2(\frac{M}{2N}-1)}\leq
\frac{\left|q_{i,s}^t\right|_M-\frac{1}{2}}{\left|q_{i,s}^t\right|_M}
\leq\frac{\left|q_{i,s}^t-\delta_s\right|_M}{\left|q_{i,s}^t\right|_M}
\end{align}
and
\begin{align}
\frac{\left|q_{i,s}^t-\delta_s\right|_M}{\left|q_{i,s}^t\right|_M}
\leq\frac{\left|q_{i,s}^t\right|_M+\frac{1}{2}}{\left|q_{i,s}^t\right|_M}
\leq 1+\frac{1}{2(\frac{M}{2N}+1)} =1+\lambda
\end{align}
Rewriting
\begin{align}
\left\|P_tx\right\|^2&=\sum_{s=0}^{N-1}\left|\sum_{\substack{i=0\\s\neq
i}}^{N-1}\frac{x_i}{\left|q_{i,s}^t-\delta_s\right|_M}
\frac{\left|q_{i,s}^t-\delta_s\right|_M}
{\left|q_{i,s}^t\right|_M}\right|^2
\end{align}
and using the bounds gives
\begin{align}
(1-\lambda)^2\left\|Q_tx\right\|^2\leq\left\|P_tx\right\|^2\leq
(1+\lambda)^2\left\|Q_tx\right\|^2
\end{align}
Then since $y$ maximizes $\left\|Q_tx\right\|^2$,
\begin{align}
\left\|P_tx\right\|^2\leq
(1+\lambda)^2\left\|Q_tx\right\|^2\leq(1+\lambda)^2\left\|Q_ty\right\|^2
\leq\left(\frac{1+\lambda}{1-\lambda}\right)^2\left\|P_ty\right\|^2
\end{align}
giving that we can bound the leftmost term by
$\left(\frac{1+\lambda}{1-\lambda}\right)^2$ times the norm at
$y$. $\left(\frac{1+\lambda}{1-\lambda}\right)^2$ takes on values
between 1 and $\frac{225}{169}\approx 1.33$ for $M\geq 16N$,
better than the constant 4 in \Hales~ and \Hallgren.

Combined with corollary \ref{l:sum} this allows us to bound term
\ref{e:term1}:
\begin{align}
\sum_{t=-\beta}^{\beta}\sum_{s=0}^{N-1}\left|\sum_{\substack{i=0\\s\neq
i}}^{N-1}\frac{x_i}{\left|q_{i,s}^t\right|_M}\right|^2&\leq
\sum_t\frac{225}{169}\sum_{s=0}^{N-1}\left
|\sum_{\substack{i=0\\s\neq
i}}^{N-1}\frac{\frac{1}{\sqrt{N}}}{\left|q_{i,s}^t\right|_M}\right|^2\\
&\leq\left(2\beta+1\right)\frac{225}{169} \frac{N}{N} \left(
\frac{4N\ln N}{M}\right)^2\\
&\leq\frac{M}{N}\frac{225}{169} \left(
\frac{4N\ln N}{M}\right)^2\\
&\leq\frac{22N\ln^2 N}{M}
\end{align}

Now we bound the other two terms, \ref{e:term2} and \ref{e:term3}.
We need the following fact, which can be shown with calculus: the
expression $\left|\sum_{i=0}^{N-1}a_ix_i\right|$ subject to the
condition $\sum_{i=0}^{N-1}x_i^2=1$, has maximum value
$\sqrt{\sum_{i=0}^{N-1}a_i^2}$. Then term \ref{e:term2} can be
bounded using a similar technique as in the proof of lemma
\ref{l:sum}. Again we take $\gamma=\frac{1}{2}-\frac{N}{M}$.
\begin{align}
\sum_{s\text{ with }\alpha\in C_s}\left|\sum_{\substack{i=0\\s\neq
i}}^{N-1}\frac{x_i}{\left|q_{i,s}^\alpha\right|_M}\right|^2&\leq
\sum_{s}\left|\sqrt{\sum_{\substack{i=0\\s\neq
i}}^{N-1}\frac{1}{\left|q_{i,s}^\alpha\right|_M^2}}\right|^2\\
&\leq
N\frac{2N^2}{M^2}\left(\frac{1}{\gamma^2}+\sum_{a=1}^{\frac{N-1}{2}}\frac{1}
{\left(\frac{1}{2}-\frac{N}{M}+a\right)^2}\right)\\
&\leq \frac{2N^3}{M^2}\left(\frac{1}{\gamma^2}+\frac{1}{\gamma}
-\frac{1}{\frac{N-1}{2}+\gamma}\right)\\
&\leq\frac{16N^3}{M^2}
\end{align}
Term \ref{e:term3} is bound with the same method and result, and
adding these three bounds gives the desired inequality
\ref{e:matBound}.
\end{proof}

We now use these lemmata to bound the tails
$\left\|\sum_i\hat{u}_i\ket{T^i}\right\|$.

\begin{lemma}\label{l:tailBound}
Given three integers: an odd integer $N\geq 13$, $L\geq 2$ a power
of two, and $M\geq 16N$ a power of two, then
\begin{equation}
\left\|\sum_{i=0}^{N-1}\hat{u}_i\ket{T^i}\right\|\leq\tbound
\end{equation}
\end{lemma}
\begin{proof}
\begin{eqnarray}
\left\|\sum_{i=0}^{N-1}\hat{u}_i\ket{T^i}\right\|^2&=&
\sum_{k=0}^{M-1}\left|\sum_{\substack{i=0\\k\not\in(i)}}^{N-1}
\hat{u}_iT^i_k\right|^2\\
&\leq&\sum_k\frac{4M}{\pi^2LN}\left(\sum_{\substack{i=0\\k\not\in(i)}}^{N-1}
\frac{|\hat{u}_i|}{\left|k-\frac{M}{N}i\right|_M}\right)^2\label{e:bound1}\\
&\leq&\frac{4M}{\pi^2LN}\left(\matbound\right)
\end{eqnarray}
Taking square roots gives the result. Note that the requirements
of lemma \ref{l:approx} are satisfied when obtaining line
\ref{e:bound1}, since we avoid the $k=i=0$ case, and $(M,N)=1$.
\end{proof}

Next we show that the shifted $\ket{S^i}$ are close to the
$\ket{B^i}$, which will allow us to show the algorithm output is
close to a tensor product.

\begin{lemma}\label{l:bumpApprox}
\begin{equation}
\Big\|\ket{S^i}-\ket{B^i}\Big\|\leq\frac{\pi LN}{M\sqrt{3}}
\end{equation}

\end{lemma}
\begin{proof}

Recall $\ket{S^i}=\sum_{b\in(i)}A^0_{b-i'\bmod M}\ket{b}$ and
$\ket{B^i}=\sum_{b\in(i)}A^i_b\ket{b}$. It is important these are
supported on the same indices! Also recall that
$\ket{A^i}=F_MF_{LN}^{-1}\ket{Li}$ and that $F_M$ is unitary. Then
(dropping $\bmod\;M$ throughout for brevity)
\begin{eqnarray}
\Big\|\ket{S^i}-\ket{B^i}\Big\|^2&=& \Big\|\sum_{b\in(i)}A^0_{b-i'}\ket{b}-\sum_{b\in(i)}A^i_b\ket{b}\Big\|^2\\
&\leq&\Big\|\sum_{k=0}^{M-1}A^0_{k-i'}\ket{k}-\sum_{k=0}^{M-1}A^i_k\ket{k}\Big\|^2\label{e:sameIndices}\\
&=&\Big\|F_M^{-1}\left(\sum_{k=0}^{M-1}A^0_k\ket{k+i'}-\ket{A^i}\right)\Big\|^2\\
&=&\sum_{a=0}^{LN-1}\left|\frac{1}{\sqrt{LN}}\omega_M^{-ai'}-\frac{1}{\sqrt{LN}}\omega_N^{-ai}\right|^2\\
&=&\frac{1}{LN}\sum_{a=0}^{LN-1}\left|\omega_M^{-ai'}\left(1-\omega_M^{a\delta_i}\right)\right|^2
\end{eqnarray}
and this can be bounded by
\begin{eqnarray}
\frac{1}{LN}\sum_{a=0}^{LN-1}\left|\frac{2\pi a
\delta_i}{M}\right|^2 \leq \frac{\pi^2}{LNM^2}\sum_{a=0}^{LN-1}a^2
\leq\frac{\pi^2}{LNM^2}\frac{(LN)^3}{3}
\end{eqnarray}
Taking square roots gives the bound.
\end{proof}

In the above proof, to obtain line \ref{e:sameIndices} we needed
that \ket{S^i} and \ket{B^i} have the same support, but \ket{S^i}
is a shifted version of \ket{B^0}, so we implicitly needed all the
sets $(i)$ to have the same cardinality. This is not satisfied in
\Hallgren~(although it is needed) but is met in \Hales.

\newcommand{\bdd}{\floor{\frac{M}{2N}-\frac{1}{2}}}
For the rest of the section we need a set which is $(0)$ without
$\bmod\;M$ applied: let $\Lambda$ be those integers in the open
interval $(-\bdd,\bdd)$. Then

\begin{lemma}\label{l:deltaKet}
\begin{eqnarray}
\Delta \ket{S^i}&=&\ket{i}\sum_{t\in \Lambda} A^0_t\ket{t+\alpha}
\end{eqnarray}
\end{lemma}
\begin{proof}
By definition,
$\ket{S^i}=\sum_{b\in(0)}A^0_b\ket{b+\round{\frac{M}{N}i} \bmod
M}$. $\Delta\left(b+\round{\frac{M}{N}i}\right)=(i,b)$ (the proof
uses $(M,N)=1$), and $\Delta$ a bijection implies
$\Delta\ket{b+\round{\frac{M}{N}i}\bmod M}=\ket{i}\ket{b+\alpha}$.
The rest follows\footnote{It is tempting to use $C_0$ instead of
$\Lambda$, but this is not correct in all cases.}.
\end{proof}

\paragraph{Main results}\label{s:MainResults}
Now we are ready to use the above lemmata to prove the main
theorem.

\begin{theorem}\label{t:oddCyclicQFT}
Given three integers: an odd integer $N\geq 13$, $L\geq 16$ a
power of two, and $M\geq LN$ a power of two. Then the output
$\ket{v}$ of the algorithm in section \emph{\ref{s:algorithm}}
satisfies
\begin{equation}
\Big\|\ket{v}-F_N\ket{u}\otimes\sum_{t\in
\Lambda}A^0_t\ket{t+\alpha}\Big\|\leq \tbound+\frac{\pi
LN}{M\sqrt{3}}
\end{equation}
\end{theorem}

\begin{proof}
Note
\begin{align}
\ket{\hat{u}}&:=F_N\ket{u}=\sum_{i=0}^{N-1}\hat{u}_i\ket{i}&
F_M\ket{u_L}&=\sum_{i=0}^{N-1}\hat{u_i} \ket{A^i}
\end{align}

Using lemma \ref{l:deltaKet} and that $\Delta$ is unitary allows
us to rewrite the left hand side as
\begin{eqnarray}
\Big\|\ket{v}-\sum_{\substack{s=0\\t\in C_s}}^{N-1}\hat{u}_s
A^0_t\ket{s}\ket{t+\alpha}\Big\|
&=&\Big\|\Delta F_M\ket{u_L}-\sum_{s=0}^{N-1}\hat{u_s}\Delta\ket{S^s}\Big\|\\
&=&\Big\|\sum_{s=0}^{N-1}\hat{u_s}\ket{A^s}-\sum_{s=0}^{N-1}\hat{u_s}\ket{S^s}\Big\|\\
&=&\Big\|\sum_{s=0}^{N-1}\hat{u_s}(\ket{B^s}+\ket{T^s})-\sum_{s=0}^{N-1}\hat{u_s}\ket{S^s}\Big\|
\end{eqnarray}
By the triangle inequality this is bounded by
\begin{eqnarray}
\Big\|\sum_{s=0}^{N-1}\hat{u_s}\ket{T^s})\Big\|+
\Big\|\sum_{s=0}^{N-1}\hat{u_s}\ket{B^s})-\sum_{s=0}^{N-1}\hat{u_s}\ket{S^s}\Big\|
\end{eqnarray}
which in turn by lemmata \ref{l:tailBound} and \ref{l:bumpApprox}
is bounded by
\begin{eqnarray}
\tbound+\frac{\pi
LN}{M\sqrt{3}}\sqrt{\sum_s|\hat{u_s}|^2}\label{e:disjoint}
\end{eqnarray}
The last expression has $\left\|\ket{\hat{u}}\right\|=1$, which
gives the result. Note that to obtain line \ref{e:disjoint} we
needed the supports of the $\ket{B^s}$ disjoint, and that the
\ket{S^i} and \ket{B^i} have the same support\footnote{This is not
satisfied in \Hales, and the overlapping portions make that proof
invalid.}.
\end{proof}

This shows that the output of the algorithm in section
\ref{s:algorithm} is close to a tensor product of the desired
output $F_N\ket{u}$ and another vector (which is not in general a
unit vector). Since a quantum state is a unit vector, we compare
the output to a unit vector in the direction of our approximation
via:

\begin{lemma}\label{l:unitTriangle}
Let $\vec{a}$ be a unit vector in a finite dimensional vector
space, and $\vec{b}$ any vector in that space. For any
$0\leq\epsilon\leq 1$, if
$\left\|\vec{a}-\vec{b}\right\|\leq\epsilon$ then the unit vector
$\vec{b'}$ in the direction of $\vec{b}$ satisfies
$\left\|\vec{a}-\vec{b'}\;\right\|\leq\epsilon\sqrt{2}$.
\end{lemma}
\begin{proof}
Simple geometry shows the distance is bounded by
$\sqrt{2(1-\sqrt{1-\epsilon^2})}$, and this expression divided by
$\epsilon$ has maximum value $\sqrt{2}$ on $(0,1]$. The
$\epsilon=0$ case is direct.
\end{proof}

So we only need a $\sqrt{2}$ factor to compare the algorithm
output with a unit vector which is $F_N\ket{u}$ tensor another
unit vector. We let $\ket{\psi}$ denote the unit length vector in
the direction of $\sum_{t\in \Lambda}A^0_t\ket{t+\alpha}$ for the
rest of this paper.

For completeness, we repeat arguments from \cite{HalHal00,Hoy00}
to obtain the operation complexity and probability distribution,
and we show concrete choices for $M$ and $L$ achieving a desired
error bound.

To show that measuring the first register gives measurement
statistics which are very close to the desired distribution, we
need some notation. Given two probability distributions
$\mathcal{D}$ and $\mathcal{D}'$ over $\{0,1,\dots,M-1\}$, let
$\left|\mathcal{D}-\mathcal{D}'\right|=\sum_{k=0}^{M-1}\left|\mathcal{D}(k)-\mathcal{D}'(k)\right|$
denote the total variation distance. Then a result\footnote{Their
statement is a bound of $4\epsilon$, but their proof gives the
stronger result listed above. We choose the stronger form to help
minimize the number of qubits needed for simulations.} of
Bernstein and Vazirani \cite{BV97} states that if the distance
between any two states is small, then so are the
induced\footnote{The induced distribution from a state \ket{\phi}
is $\mathcal{D}(k)=|\braket{k}{\phi}|^2$.} probability
distributions:
\begin{lemma}[\cite{BV97}, Lemma 3.6]\label{l:prob}
Let \ket{\alpha} and \ket{\beta} be two normalized states,
inducing probability distributions $\mathcal{D}_\alpha$ and
$\mathcal{D}_\beta$. Then for any $\epsilon>0$
\begin{equation}
\left\|\ket{\alpha}-\ket{\beta}\right\|\leq\epsilon \Rightarrow
\left|\mathcal{D}_\alpha-\mathcal{D}_\beta\right|\leq
2\epsilon+\epsilon^2
\end{equation}
independent of what basis is used for measurement.
\end{lemma}

Combining this with theorem \ref{t:oddCyclicQFT} and lemmata
\ref{l:unitTriangle} and \ref{l:prob} gives the final result

\begin{thm}\label{t:MainBound}~\\

\emph{\textbf{1)}} Given an odd integer $N\geq 13$, and any
$\sqrt{2}\geq\epsilon>0$. Choose $L\geq 16$ and $M\geq L N$ both
integral powers of $2$ satisfying
\begin{equation}
\tbound+\frac{\pi LN}{M\sqrt{3}}\leq
\frac{\epsilon}{\sqrt{2}}\label{e:mainBound}
\end{equation}
Then there is a unit vector \ket{\psi} such that the output
\ket{v} of the algorithm in section \emph{\ref{s:algorithm}}
satisfies
\begin{equation}
\\|\ket{v}-F_N\ket{u}\otimes\ket{\psi}\|\leq\epsilon
\end{equation}

\emph{\textbf{2)}} We can always find such an $L$ and $M$ by
choosing
\begin{align}
L&=c_1\frac{\sqrt{N}}{\epsilon^2}\label{e:Lvalue}\\
M&=c_2\frac{N^\frac{3}{2}}{\epsilon^3}\label{e:Mvalue}
\end{align}
for some constants $c_1,c_2$ satisfying
\begin{align}
\minA \leq c_1 &\leq 2\times\minA\label{e:const1}\\
\minB \leq c_2 &\leq 2\times\minB\label{e:const2}
\end{align}

\emph{\textbf{3)}} The algorithm requires $\ceil{\log M}+2$
qubits. By claim $2$ a sufficient number of qubits is then
$\ceil{12.53+3\log\frac{\sqrt{N}}{\epsilon}}$. The algorithm has
operation complexity $O(\log M(\log \log M + \log1/\epsilon))$.
Again using claim $2$ yields an operation complexity of
\begin{equation}
O\left(\log \mr\left(\log \log \mr + \log1/\epsilon\right)\right)
\end{equation}

\emph{\textbf{4)}} The induced probability distributions
$\mathcal{D}_v$ from the output and $\mathcal{D}$ from
$F_N\ket{u}\otimes\ket{\psi}$ satisfy
\begin{equation}
\left|\mathcal{D}_v-\mathcal{D}\right|\leq 2\epsilon+\epsilon^2
\end{equation}
\end{thm}

\begin{proof}
Claim 1 follows directly from theorem \ref{t:oddCyclicQFT} and
lemma \ref{l:unitTriangle}. Claim 1 and lemma \ref{l:prob} give
claim 4.

To get claim 2, note that for the bound to be met, we must have
$\frac{\ln^2 N}{L}<\epsilon^2$, $\frac{N^2}{LM}<\epsilon^2$, and
$\frac{LN}{M}<\epsilon$. Trying to keep $M$ small as $N$ and
$\epsilon$ vary leads to the forms for $L$ and $M$ chosen. If we
substitute lines \ref{e:Lvalue} and \ref{e:Mvalue} into
\ref{e:mainBound} and simplify, we get
\begin{align}
\frac{4}{\pi}\sqrt{\frac{11\ln^2N}{c_1\sqrt{N}}+
\frac{16\epsilon^3}{c_1c_2}}
+\frac{\pi\sqrt{2}}{\sqrt{3}}\frac{c_1}{c_2}&\leq 1
\end{align}
The left hand side is largest when $\epsilon=\sqrt{2}$ and $N=55$,
so it is enough to find constants $c_1$ and $c_2$ such that
\begin{align}
\frac{4}{\pi}\sqrt{\frac{11\ln^2 55}{c_1\sqrt{55}}+
\frac{32\sqrt{2}}{c_1c_2}}
+\frac{\pi\sqrt{2}}{\sqrt{3}}\frac{c_1}{c_2}&\leq
1\label{e:c1c2bound}
\end{align}
Ultimately we want $L$ and $M$ to be powers of two, so we find a
range for each of $c_1$ and $c_2$ such that the upper bound is at
least twice the lower bound, and such that all pairs of values
$(c_1,c_2)$ in these ranges satisfy inequality \ref{e:c1c2bound}.
To check that the claimed ranges work, note that for a fixed
$c_1$, the expression increases as $c_2$ decreases, so it is
enough to check the bound for $c_2=\minB$. After replacing $c_2$
in the expression with $\minB$, the resulting expression has first
and second derivatives with respect to $c_1$ over the claimed
range, and the second derivative is positive, giving that the
maximum value is assumed at an endpoint. So we only need to check
inequality \ref{e:c1c2bound} at two points:
$(c_1,c_2)=(\minA,\minB)$ and $(2\times\minA,\minB)$, both of
which work. Thus the bound is met for all $(c_1,c_2)$ in the
ranges claimed. With these choices for $M$ and $L$, note that
$L\geq 16$ and $M\geq LN\Leftrightarrow c_2\geq\epsilon c_1$,
which is met over the claimed range, so all the hypothesis for
claim 1 are satisfied.

Finally, to prove claim 3, algorithm \ref{s:algorithm} and the
proof of lemma \ref{l:deltaProp} give that we need $\ceil{\log N}$
qubits in the first register and $\max\{\ceil{\log
L},\ceil{\log(2\alpha+1)}\}$ qubits in the second register. $L\leq
\frac{M}{N}< 2\alpha+1$ gives that it is enough to have
$\ceil{\log(2\alpha+1)}$ qubits in the second register. Then
$2\alpha+1\leq \frac{M}{2N}+2$ gives
\begin{align}
\ceil{\log(2\alpha+1)}\leq\ceil{1+\log M - \log N} =2+\ceil{\log
M}-\ceil{\log N}
\end{align}
Thus $\ceil{\log M}+2$ is enough qubits\footnote{An example
requiring $\ceil{\log M}+2$ qubits is $M=1024$, $N=65$, so the
bound is tight.} for the algorithm. By claim 2, we can take $M\leq
2\times \minB \frac{N^{3/2}}{\epsilon^3}$ giving $\ceil{\log
M}+2\leq\ceil{12.53+3\log\frac{\sqrt{N}}{\epsilon}}$.

As noted in \Hales~and \Hallgren, the most time consuming step in
algorithm \ref{s:algorithm} is the $F_M$ Fourier computation.
Coppersmith \cite{Copper94} (reproduced in section
\ref{s:approxQFT}) shows how to $\epsilon$ approximate the quantum
Fourier transform for order $M=2^m$ with operation complexity of
$O(\log M(\log \log M + \log1/\epsilon))$. Using this to
approximate our approximation within error $\epsilon$ gives the
time complexities in claim 3, finishing the proof.
\end{proof}


\section{Graph Reductions}\label{s:Graph}
\subsection{Basic Graph Algorithm Relations}
Note that $G$ in this section is no longer a group as in the rest
of the paper, but a \emph{graph}.

Following Mathon \cite{Math79}, we show several graph isomorphism
problems to be polynomially equivalent. If the ability to solve
problem $P_1$ allows solving problem $P_2$ with polynomially many
uses of $P_1$, we say $P_2$ is polynomially reducible to $P_1$,
and write $P_2\varpropto_p P_1$. If $P_2\varpropto_p P_1$ and
$P_1\varpropto_p P_2$ then we say $P_1$ and $P_2$ are
\emph{polynomially equivalent}.

Given two undirected graphs $G_1(V_1,E_1)$ and $G_2(V_2,E_2)$ with
vertex sets $V_i$ and edge sets $E_i$, $i=1,2$, we say $G_1$ is
isomorphic to $G_2$, written $G_1\cong G_2$, if there exists a
bijection $\rho:V_1\rightarrow V_2$ such that for all $x,y\in
V_1$, $(x,y)\in E_1$ if and only if $(\rho x, \rho y)\in E_2$ .

Denote the group of automorphisms of $G$ by $\text{aut}\;G$. The
automorphism partition $\mathcal{P}$ denotes the set of disjoint
orbits of each vertex under $\text{aut}\;G$.

We consider the following six problems:\\

\begin{tabular}{ll}
$\text{\textbf{ISO}}(G_1,G_2)$ & isomorphism recognition for $G_1$
and $G_2$, \\

$\text{\textbf{IMAP}}(G_1,G_2)$ & isomorphism map from $G_1$ onto
$G_2$ if
it exists, \\

$\text{\textbf{ICOUNT}}(G_1,G_2)$ & number of isomorphisms from
$G_1$
to $G_2$,\\

$\text{\textbf{ACOUNT}}(G)$ & number of automorphisms of $G$,\\

$\text{\textbf{AGEN}}(G)$ & generators of the automorphism group
of
$G$,\\

$\text{\textbf{APART}}(G)$ & automorphism partition of $G$.
\end{tabular}\\

Surprisingly,

\begin{thm}
The problems \emph{\textbf{ISO}}, \emph{\textbf{IMAP}},
\emph{\textbf{ICOUNT}}, \emph{\textbf{ACOUNT}},
\emph{\textbf{AGEN}}, and \emph{\textbf{APART}} are polynomially
equivalent.
\end{thm}

Before proving this we define some notation. Suppose $G(V,E)$ is a
graph with $n$ vertices. We define graph labels: Let
$G_{v_1,\dots,v_k}$ denote a copy of $G$ with unique distinct
labels attached to the vertices $v_1,\dots,v_k\in V$. This can be
accomplished in the following manner. To vertex $v_m$, $1\leq
m\leq k$, attach label ``$m$", which is a new graph using $2n+m+3$
vertices as follows:

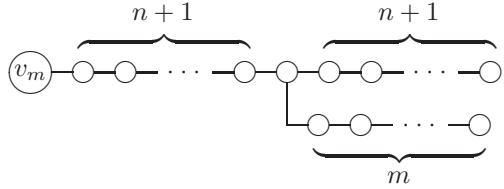
\begin{picture}(100,80)(-50,0)
\put(5,40){\makebox(10,10){$v_m$}} \put(10,45){\circle{16}}

\put(18,45){\line(1,0){8}}

\newsavebox{\subg}
\savebox{\subg}(10,10){

\put(0,0){\circle{8}}\put(4,0){\line(1,0){8}}

\put(16,0){\circle{8}}\put(20,0){\line(1,0){8}}

\put(33,0){\circle*{1}}\put(38,0){\circle*{1}}\put(43,0){\circle*{1}}

\put(49,0){\line(1,0){8}} \put(61,0){\circle{8}}

} 

\put(25,40){\usebox{\subg}} 

\put(95,45){\put(0,0){\line(1,0){8}}\put(12,0){\circle{8}}\put(16,0){\line(1,0){8}}

\put(12,-4){\line(0,-1){16}}\put(12,-20){\line(1,0){8}}

} 

\put(118,40){\usebox{\subg}} 

\put(114,20){\usebox{\subg}} 


\put(28,53){\put(0,0){$\overbrace{\hspace*{65pt}}$}\put(20.5,12){$n+1$}}

\put(121,53){\put(0,0){$\overbrace{\hspace*{65pt}}$}\put(20.5,12){$n+1$}}

\put(117,17){\put(0,0){$\underbrace{\hspace*{65pt}}$}\put(28,-14){$m$}}

\end{picture}

This modification has the property that vertices $v_1,\dots,v_k$
are fixed by any $\rho\in\text{aut}\;G_{v_1,\dots,v_k}$, and also
there is a natural inclusion
$\text{aut}\;G_{v_1,\dots,v_k}\subseteq\text{aut}\;G$, obtained by
ignoring the labels in $\text{aut}\;G$. Finally, labelling all
vertices adds $O(n^2)$ new vertices, retaining polynomial
algorithm equivalence between problems on $G$ and
$G_{v_1,\dots,v_k}$.

\begin{proof} (Following Mathon \cite{Math79})

\textbf{IMAP} $\varpropto_p$ \textbf{ISO}: Let $v_1,\dots,v_n$ be
the vertices of $G_1$. If $G_2$ does not have $n$ vertices then
there is no isomorphism. Otherwise use \textbf{ISO} at most $n$
times to find a $u_1\in V_2$ such that there is an isomorphism
$G_{1v_1}\cong G_{2u_1}$, otherwise there is no isomorphism. If
such a $u_1$ is found, there is an isomorphism $\rho$ mapping
$v_1\rightarrow u_1$. Continue fixing $v_1,\dots,v_j$,
$u_1,\dots,u_{j-1}$ and searching for $u_j\in V_2$. This
constructs an isomorphism if it exists, calling \textbf{ISO}
$O(n^2)$ times.

\textbf{ACOUNT} $\varpropto_p$ \textbf{ISO}: For a given labelling
$G_{v_1,\dots,v_k}$ of a graph $G$ let
$\text{aut}\;G_{v_1,\dots,v_k}$ be the corresponding automorphism
group, which is the subgroup of $\text{aut}\;G$ that fixes the
vertices $v_1,\dots,v_k$. We will show that
$|\text{aut}\;G_{v_1,\dots,v_{k-1}}|=d_k|\text{aut}\;G_{v_1,\dots,v_k}|$,
where $d_k$ is the size of the orbit $\pi_k$ of $v_k$ in
$\text{aut}\;G_{v_1,\dots,v_{k-1}}$. For $1\leq i \leq d_k$ let
$\phi_i\in \text{aut}\;G_{v_1,\dots,v_{k-1}}$ be an automorphism
which maps the $i^\text{th}$ vertex of $\pi_k$ onto $v_k$. Then
every $\tau\in\text{aut}\;G_{v_1,\dots,v_{k-1}}$ is a product of a
unique $\phi\in\{\phi_1,\dots,\phi_{d_k}\}$ and a unique
$\psi\in\text{aut}\;G_{v_1,\dots,v_k}$. Since
$|\text{aut}\;G_{v_1,\dots,v_n}|=1$, $|\text{aut}\;G|=d_1d_2\dots
d_n$, and each $d_k$ can be found by solving \textbf{ISO} at most
$n-k$ times. Thus we compute $|\text{aut}\;G|$ by calling
\textbf{ISO} at most $O(n^2)$ times.

\textbf{ICOUNT} $\varpropto_p$ \textbf{ISO}: Let $N_I$ be the
number of isomorphisms from $G_1$ onto $G_2$. If $G_1\ncong G_2$
then $N_I=0$ is determined with one call to \textbf{ISO}.
Otherwise we claim $N_I=|\text{aut}\;G_1|=|\text{aut}\;G_2|$, in
which case we use \textbf{ACOUNT} on $G_1$ and on $G_2$, calling
\textbf{ISO} $O(n^2)$ times as above. The claim is proved by the
fact that if $\sigma:V_1\rightarrow V_2$ is an isomorphism from
$G_1$ onto $G_2$ and $\rho$ is an automorphism of $G_2$ then
$\rho\circ\sigma$ is also a graph isomorphism. Moreover any
isomorphism $\sigma'$ can be uniquely expressed as
$\sigma'=\rho'\circ\sigma$ where $\rho'\in|\text{aut}\;G_2|$. This
$1-1$ correspondence between $|\text{aut}\;G_2|$ and the number of
isomorphisms $G_1\rightarrow G_2$ proves the claim.

\textbf{APART} $\varpropto_p$ \textbf{ISO}: Two vertices $u,v\in
V$ of a graph $G$ belong to the same cell of the automorphism
partition $\mathcal{P}$ of $G$ if $G_u\cong G_v$ for identical
labels of $u$ and $v$. Hence at most $O(n^2)$ calls to
\textbf{ISO} are needed to find $\mathcal{P}$, trying all
combinations of $u$ and $v$.

\textbf{AGEN} $\varpropto_p$ \textbf{ISO}: Applying \textbf{IMAP}
to the graphs $G_{v_1,\dots,v_k}$ and $G_{v_1,\dots,v_{k-1},v_l}$
with identical labels for $k+1\leq l\leq n$ we determine the sets
of automorphisms $\Phi_k=\{\phi_1,\dots,\phi_{d_k}\}$ at level $k$
(using notation from above). From the proof of \textbf{IMAP}
$\varpropto_p$ \textbf{ISO} it follows that the set
$\Phi_1\bigcup\dots\bigcup\Phi_n$ of maps generates
$\text{aut}\;G$. Since $d_k\leq n-k+1$ implies
$$\sum_{k=1}^nd_k\leq n^2$$ we see that at most $O(n^4)$ calls to
\textbf{ISO} solve \textbf{AGEN}. This order can be reduced to
$O(n^3)$ using \textbf{APART} to find the partition of
$G_{v_1,\dots,v_k}$ and by generating only one $\phi_i$ for every
feasible orbit in $V\setminus \{v_1,\dots,v_k\}$ at each level
$k$. It is easily shown at most $n$ generators are produced in
this case.

\textbf{ISO} $\varpropto_p$ \textbf{IMAP}, \textbf{ICOUNT}: A
single call to either \textbf{IMAP} or \textbf{ICOUNT} gives
\textbf{ISO}.

From now on assume $G_1$ and $G_2$ are each connected (otherwise
we may use their complements).

\textbf{ISO} $\varpropto_p$ \textbf{ACOUNT}: Apply \textbf{ACOUNT}
to $G_1$, $G_2$, and $G_3=G_1\bigcup G_2$. If
$|\text{aut}\;G_1|=|\text{aut}\;G_2|$ and
$|\text{aut}\;G_1|\cdot|\text{aut}\;G_2|\neq |\text{aut}\;G_3|$
then $G_1\cong G_2$, else $G_1\ncong G_2$.

\textbf{ISO} $\varpropto_p$ \textbf{AGEN}: Apply \textbf{AGEN} to
$G_3=G_1\bigcup G_2$. If $\sigma(v)=u$ for some $v\in V_1$, $u\in
V_2$, and $\sigma\in \text{aut}\;G_3$ then $G_1\cong G_2$, else
$G_1\ncong G_2$. From the proof of \textbf{AGEN} $\varpropto_p$
\textbf{ISO} we can assume we have at most $n^2$ generators of
$\text{aut}\;G$ to check, so this can be checked in at most $n^4 =
|V_1||V_2||n^2|$ operations, assuming constant time to check one.

\textbf{ISO} $\varpropto_p$ \textbf{APART}: Apply \textbf{APART}
to $G_3=G_1\bigcup G_2$. If $v,u$ belong to the same cell of the
partition $\mathcal{P}$ of $G_3$ for some $u\in V_1$, $u\in V_2$,
then $G_1\cong G_2$, otherwise $G_1\ncong G_2$. This can be
checked quickly by scanning the partition once.

This completes the proof of the theorem.
\end{proof}

Finally, following \cite[Theorem 1.31]{KST93}, we can reduce this
to efficient algorithms solving the following graph automorphism
questions:

\begin{itemize}
\item $\text{\textbf{GA}}(G)$ - Given a graph $G$, decide whether
its automorphism group has a nontrivial automorphism.

\item $\text{\textbf{GA1}}(G)$ - Given that
$|\text{aut}\;G|\in\{1,2\}$, determine $|\text{aut}\;G|$
\end{itemize}

We note that $\text{\textbf{GA}}(G)$ seems easier than
$\text{\textbf{ISO}}(G_1,G_2)$ \cite{KST93}.

As above, we are able to reduce the seemingly more complex
$\text{\textbf{GA}}$ to $\text{\textbf{GA1}}$:

\begin{thm}
$\textbf{\emph{GA}}\varpropto_p\textbf{\emph{GA1}}$
\end{thm}

For a proof, see \cite{KST93}.

So there are many ways to approach the graph isomorphism and graph
automorphism problems, some of which at first glance seem easier
than the original question. For the purposes of quantum
computation, and in particular reducing these questions to finding
hidden subgroups of $S_n$, see the next section
(\ref{s:isoReduct}).

As a final note, there are far reaching proofs that show
determining isomorphism between any finite algebraic structures
(such as rings, groups, fields, etc.) is polynomial-time
many-to-one reducible to \textbf{ISO}, making a fast \textbf{ISO}
algorithm extremely useful across many disciplines
\cite{Miller79}. These are a few of the reasons that an efficient
\textbf{ISO} algorithm has seen such strong research interest.

\subsection{Quantum HSP for Graph Isomorphism}\label{s:isoReduct}
We want to show how being able to find hidden subgroups $H$ of
$S_n$ allows solving \textbf{ISO}, which then gives efficient
algorithms for all the problems in the previous section. We define
our hidden function $f:S_n\rightarrow\left\{\text{permutations of
} G\right\}$ by $f(\pi)=\pi(G)$. So $f$ applies a permutation
$\pi$ to the vertices of $G$. We need to show $f$ separates cosets
of $H=\text{aut }G$, and that $f$ is efficiently computable. Then
an algorithm giving generators of $H$, i.e., giving an algorithm
for \textbf{AGEN}, gives the desired algorithm for \textbf{ISO}.

To make this precise, suppose $G$ is represented on a computer by
a list of pairs $(v_i,v_j)$ of vertices where there is an edge
from vertex $i$ to vertex $j$. Assume this list is sorted and each
pair is sorted. We define $f$ at the programming level as taking a
permutation (which can just be a list $\pi$ of $n$ pairs
$i\rightarrow\pi(i)$) and doing the following two steps: apply the
permutation to the integers $v_i$ in time $O(\text{\# edges})$,
then sort the result efficiently by usual methods (Quicksort,
etc.). Thus $f$ can be computed efficiently, and leaves $G$ in a
state where comparisons can be done quickly (that is, $G\cong
\pi(G)$ if and only if $G=f(\pi)$ using this encoding, which you
should check).

Let $S_n$ act on the $n$ vertices of $G$, and let
$H=\text{aut}(G)<S_n$. To show $f$ separates cosets of $H$, we
want $f(\pi_1)=f(\pi_2)$ if and only if $\pi_1 H=\pi_2 H$, which
follows from
\begin{eqnarray*}
f(\pi_1)=f(\pi_2)\Leftrightarrow \pi_1(G)=\pi_2(G) \Leftrightarrow
\pi_2^{-1}\pi_1 G=G \Leftrightarrow \\\pi_2^{-1}\pi_1\in \text{aut
}G \Leftrightarrow \pi_2^{-1}\pi_1 H=H\Leftrightarrow \pi_1
H=\pi_2 H.
\end{eqnarray*}
This shows $f$ can be used in the standard quantum Fourier
sampling algorithm to find generators for $H$. If this can be done
efficiently is an open question.

\section{Quantum Mechanics Details}\label{s:QuantumDetails}


\subsection{The Rules and Math of Quantum Mechanics}

\subsubsection{Enter the Qubit} First we start out with the basic
block of quantum computing. Analogous to the bit in classical
computing, there is a quantum bit in quantum computing. A
classical bit is a 2 state system, with the states denoted 0 and
1. A classical bit is always in one of those states or the other,
and measuring the state return a 0 or 1 with certainty. $n$ bits
can be in exactly one of $2^n$ different ordered states, usually
denoted $000\dots00$,
$000\dots01$,$\dots$,$111\dots11$.\footnote{``There are only 10
kinds of people in the world. Those who understand binary and
those who don't."}

Quantum bits (which we shall call qubits) similarly can exist in
two states, which we call \ket{0} and \ket{1}. However, they
behave as if existing in many ``in between" states. A quantum bit
can be physically represented by any two state (or more) system,
such as electron spin up and down, photon energy states, atomic
energy levels, molecular vibrational freedom, and many others. For
our purposes we assume physical representations are available
(they are).

To make the concept of a qubit precise, we define
\begin{defn}[Qubit]A \emph{\textbf{qubit}} (or quantum-bit) is a unit vector in
$\mathbb{C}^2$.
\end{defn}

\begin{defn}[State vector] The \emph{\textbf{state}} of a quantum
system is a (column) vector in some vector space, written
\ket{\psi}.
\end{defn}

With this definition, we fix an orthonormal basis of (column)
vectors, labelled $\ket{0} = \binom{1}{0}$ and
$\ket{1}=\binom{0}{1}$. It will turn out that physically, \emph{we
can only distinguish orthogonal quantum states}, thus the
orthogonal requirement. And considerations of probability will
make the normality convenient, thus we fix an orthonormal basis.
Any such basis of $\mathbb{C}^2$ will work, but we choose the
above representations since they are good to work with. Finally,
we make a qubit a unit vector because, again, it makes
calculations cleaner, and has some physical significance.

Now for the differences from classical bits. A qubit can be
\emph{any} unit vector, not just those corresponding to \ket{0}
and \ket{1}. A qubit can be in the state
\begin{equation}\label{e:qubit}
\alpha\ket{0}+\beta\ket{1}
\end{equation}
where $\alpha$ and $\beta$ are complex numbers, with
$|\alpha|^2+|\beta|^2=1$. While it only takes one ``bit" to fully
describe the state of a classical bit, it takes two complex
numbers to completely describe the state of one qubit, which
intuitively is infinitely more information! However we will see
there are practical limitations to the amount of ``information"
one can retrieve from a single qubit.

This gives us the first of four postulates of quantum mechanics:

\textbf{Quantum Mechanics Postulate 1: State Space} Associated to
an isolated physical system is a complex vector space with inner
product (a Hilbert space) known as the \emph{state space} of the
system. The system is completely described by its \emph{state
vector}, which is a unit vector in the system's state space. Thus
an $n$-qubit system is a unit vector in $\mathbb{C}^{2^n}$.

We will explain the inner product below (we can use the Euclidean
one).

\subsubsection{How to ``Measure" a Qubit} In principle you could
store the knowledge in the Library of Congress on one qubit, but
\emph{you could never retrieve it}. When you read out the value in
a qubit in the state in equation \ref{e:qubit}, it returns the
state \ket{0} with probability $|\alpha|^2$, or it returns the
state \ket{1} with probability $|\beta|^2$, and then the qubit
assumes the state just returned. Thus we can only get one state
back out from the qubit, which collapses (destroys) the rest of
the information in the qubit.

For example, suppose we have a qubit in the state
\begin{equation}
\ket{\psi}=\frac{1}{\sqrt{2}}\ket{0}+\frac{1}{\sqrt{2}}\ket{1}
\end{equation}
What are the odds that it returns a \ket{1} when measured? A
\ket{0}?

This generalizes to multiple qubits as we soon see.

One last point is worth mentioning - there is a useful way to
visualize operations on a single qubit, using the \textbf{Bloch
sphere}. It will turn out that under observation, states
\ket{\psi} and $e^{i\theta}\ket{\psi}$ have the same behavior, so
we can modify a state up to the phase $i\theta$, where
$i=\sqrt{-1}$. So given a single qubit state
$\alpha\ket{0}+\beta\ket{1}$, we can remove a phase to write
\begin{equation}
\alpha\ket{0}+\beta\ket{1} =
e^{i\gamma}\left(\cos\frac{\theta}{2}\ket{0}+e^{i\varphi}\sin\frac{\theta}{2}\ket{1}\right)
\end{equation}
Since the phase out front has no effect on measurements, we can
use $\theta$ and $\varphi$ for spherical coordinates
\begin{eqnarray}
x =& \cos\varphi\sin\theta\\
y =& \sin\varphi\sin\theta\\
z =& \cos\theta
\end{eqnarray}
This allows us to picture a qubit as a point on a three
dimensional sphere, and visualize operations upon a qubit.

Unfortunately, this has no known generalization to multiple qubits

\subsubsection{Qubits Galore} Similar to concatenating $n$ classical
bits to get ``bitstrings", we concatenate qubits to get larger
systems. Two qubits form a space spanned by four vectors
\begin{equation}
\ket{0}\otimes\ket{0}, \;\;\ket{0}\otimes\ket{1},
\;\;\ket{1}\otimes\ket{0}, \;\;\text{and }\ket{1}\otimes\ket{1}
\end{equation}
where we will define the ``tensor product" $\otimes$ in a moment.
Shorthand for the above expressions is
\begin{equation}
\ket{00}, \;\;\ket{01}, \;\;\ket{10}, \;\;\text{and }\ket{11}
\end{equation}

\begin{defn}
The \emph{\textbf{tensor product}} of two vectors
$x=(x_1,x_2,\dots,x_n)^T$ and $y=(y_1,y_2,\dots,y_m)^T$ as the
vector in $nm$ dimensional space given by
\begin{equation}
x\otimes y = \left(\begin{matrix}x_1 y \\ x_2 y\\\dots\\x_n y
\end{matrix}\right) = \left(\begin{matrix}x_1 y_1 \\ x_1 y_2\\\dots\\x_1
y_m\\x_2y_1\\\dots\\x_ny_m
\end{matrix}\right)
\end{equation}
\end{defn}

\begin{hwk}
Check this definition does not depend on a choice of basis.
\end{hwk}

Now we can check the second basis element (dictionary ordering)
\begin{eqnarray}
\ket{01} &=
\ket{0}\otimes\ket{1}\\
&=\binom{1}{0}\otimes\binom{0}{1}\\
&=\left(\begin{matrix}1\binom{0}{1}\\0\binom{0}{1}\end{matrix}\right)
&=\left(\begin{matrix}0\\1\\0\\0\end{matrix}\right)
\end{eqnarray}
and we get the second usual basis element of $\mathbb{C}^4$. This
works in general; that is, the vector corresponding to the state
$\ket{n}$ where $n$ is a binary number, is the $(n+1)^\text{th}$
standard basis element. We also use the decimal shorthand
sometimes: $\ket{32}$ is the 33rd standard basis vector in some
space which would be clear from context.

Back to the inner product from postulate 1: We write it using the
``braket" notation, where the symbol \ket{k} is called a ket, and
the dual \bra{j} is a bra. Given a state (ket)
$\ket{\psi}=\sum\alpha_j\ket{j}$, we define the dual (bra) as the
conjugate transpose, that is,
\begin{equation}
\bra{\psi}=\ket{\psi}^\dag=\sum\alpha_j^*\bra{j}
\end{equation}

Together we write \braket{j}{k}, which is the ``braket" of states
\ket{j} and \ket{k}. Since the states are orthonormal,
\braket{j}{k} is 1 if and only if $j=k$, otherwise it is zero. We
extend this inner product $\langle-,-\rangle$ to general states
via linearity. Thus states $\ket{\psi_1}=\sum \alpha_j\ket{j}$ and
$\ket{\psi_2}=\sum\beta_k\ket{k}$ give
\begin{eqnarray*}
\braket{\psi_1}{\psi_2}&=\sum_j\alpha_j^*\bra{j}\sum_k\beta_k\ket{k}\\
&=\sum_{j,k}\alpha_j^*\beta_k\braket{j}{k}
 &=\sum_m\alpha_m^*\beta_m
\end{eqnarray*}

So we have the equivalent notations for a 5-qubit state:
\begin{eqnarray*}
\ket{1}\otimes\ket{0}\otimes\ket{0}\otimes\ket{1}\otimes\ket{0} =&
\ket{10010}\\
 =& \ket{18}
\end{eqnarray*}

It is worth noting that not all composite states are simple tensor
products of single states. One of the simplest is one of the 2
qubit Bell states,
$\beta_{00}=\frac{\ket{00}+\ket{11}}{\sqrt{2}}$. This is an
example of an \emph{entangled state} which turns out to be a very
useful computational resource later.

\begin{hwk}
Prove $\beta_{00}$ is not of the form
$\ket{\psi}\otimes\ket{\varphi}$.
\end{hwk}

When appropriate, we may drop the normalization factor to clean up
calculations. Then we could write $\beta_{00}=\ket{00}+\ket{11}$,
with the understanding this needs to be normalized.

\subsubsection{Measuring Revisited}

 Now - how about measuring these states?
An arbitrary 2-qubit state is
$$\ket{\psi}=\alpha_{00}\ket{00}+\alpha_{01}\ket{01}+\alpha_{10}\ket{10}+\alpha_{11}\ket{11}$$
with complex valued $\alpha_{ij}$. Requiring
$\sum_{ij}|\alpha_{ij}|^2=1$ is called the ``normalization
requirement", and we assume all states are normalized. Sometimes
to avoid clutter we will drop the coefficients.

Suppose we only measure the first qubit of \ket{\psi}. We will
obtain \ket{0} with probability $|\alpha_{00}|^2+|\alpha_{01}|^2$,
that is, we obtain a state with probability equal to the sum of
the magnitudes of all states that contribute. After measuring, we
know the first qubit is \ket{0}, so only those type of states are
left, causing the new state to be
$$\ket{\psi^*}=\frac{\alpha_{00}\ket{00}+\alpha_{01}\ket{01}}{\sqrt{|\alpha_{00}|^2+|\alpha_{01}|^2}}$$
Notice the new normalization factor in the denominator. Again,
this idea generalizes to arbitrary (finite) dimension.

Thus we have a way to denote arbitrary quantum states on $n$
qubits:
\begin{equation}
\ket{\psi}=\sum_{j=0}^{2^n-1}\alpha_j\ket{j}
\end{equation}
where the $\alpha_i$ are complex numbers satisfying the
normalization requirement. Measuring \ket{\psi} returns state
$\ket{j}$ with probability $|\alpha_j|^2$, and then becomes state
$\ket{j}$

\subsubsection{Qubit Evolution} We would like our
quantum computers to work similar to classical computers.
Classically, a very basic operation at the bit level is the NOT
gate, which flips bits, that is $0$ becomes $1$ and $1$ becomes
$0$. So the quantum version would take the state
$\alpha\ket{0}+\beta\ket{1}\xrightarrow{NOT}\beta\ket{0}+\alpha\ket{1}$.
It is east to check the matrix
\begin{equation}
X=\left(\begin{array}{cc}0 & 1\\1 & 0\end{array}\right)
\end{equation}
performs the desired operation, by multiplying $X$ on the left of
the state. The name $X$ is historical, and we will see the
exponential of $X$ rotates qubits around the x-axis on the Bloch
sphere. Since $X$ acts like a NOT gate on a qubit, it is often
called the NOT operator.

For fun, we compute ``the square root of NOT." We want an operator
$\sqrt{\text{NOT}}$ that when applied twice to a qubit, has the
effect of NOT. This procedure will be useful when we need to
construct quantum circuits and when we explain exponentials.

In general, given a function $f(t)$ of one complex variable, we
extend this definition to diagonalizable matrices
$M=\text{diag}(m_1,m_2,\dots m_n)$ via:
\begin{equation}
f(M) = \text{diag}(f(m_1),f(m_2),\dots,f(m_n))
\end{equation}

Since we want $\sqrt{\text{X}}$, we need to diagonalize $X$. Note
the eigenvectors of $X$ are $\binom{1}{1}$ and $\binom{1}{-1}$.
Setting a matrix $P$ with these as column vectors, we have under
this basis change the diagonal matrix
\begin{eqnarray*}
PXP^{-1} =&\left(\begin{array}{rr}1&1\\1&-1\end{array}\right)\left(\begin{array}{rr}0&1\\1&0\end{array}\right)\left(\begin{array}{rr}\frac{1}{2}&\frac{1}{2}\\\frac{1}{2}&-\frac{1}{2}\end{array}\right)\\
=&\left(\begin{array}{rr}1&0\\0&-1\end{array}\right)
\end{eqnarray*}
Applying $f(t)=\sqrt{t}$, and changing the basis back gives
\begin{eqnarray*}
P^{-1}f\left(\begin{array}{rr}1&0\\0&-1\end{array}\right)P =&\left(\begin{array}{rr}\frac{1}{2}&\frac{1}{2}\\\frac{1}{2}&-\frac{1}{2}\end{array}\right)\left(\begin{array}{rr}1&0\\0&i\end{array}\right)\left(\begin{array}{rr}1&1\\1&-1\end{array}\right)\\
=&\frac{1}{2}\left(\begin{array}{rr}1+i&1-i\\1-i&1+i\end{array}\right)\\
=&\sqrt{\text{NOT}}
\end{eqnarray*}
It is an easy check to see that $\sqrt{\text{NOT}}^2=X$.

This process of diagonalizing an operator, applying a function,
and restoring the basis will be invaluable later.
\begin{hwk}
What is the effect of $e^{-i\theta X/2}$ on the Bloch sphere,
where $\theta$ is a real number?
\end{hwk}

\subsubsection{A Universal Quantum Gate?} It is a basic result in computer science that any circuit can be
built with NAND gates, which performs the following operation on
two bits $a$ and $b$:
$$
\begin{array}{c||c|c}
a\setminus b & 0 & 1 \\  \hline \hline 0 & 1 & 1\\ \hline 1 & 1 & 0 \\
\hline
\end{array}
$$
Any function on $n$ bits can be built up from NAND gates. However
the general function requires exponentially many gates, so in
practice we are restricted in the functions we utilize.

So is there a similar ``gate" for quantum computing? Yes, and no.
It will take a while to answer this precisely, but there are
finite (and small) sets of gates sufficient to \emph{approximate}
any desired quantum operation to any degree of accuracy in an
efficient manner.\footnote{The Solovay-Kitaev theorem says that
for any gate $U$ on a single qubit, and given any $\epsilon>0$, it
is possible to approximate $U$ to a precision $\epsilon$ using
$\Theta(\log^c(1/\epsilon))$ gates from a fixed, finite set, where
$1\leq c \leq 2$. Determining $c$ is an open problem.}

To understand what operations we can physically apply to a qubit
(or set of qubits), we are led to study rules from quantum
mechanics. It has become clear that abstract models of computation
and information theory should be derived from physical law, rather
than as standalone mathematical structures, since it is ultimately
physical law that determines computability and information.
Observation has led researchers to believe that at the quantum
level, the following two facts hold:
\begin{itemize}
\item All quantum evolution is reversible. That is very unlike the
classical case, where for example NAND is not
reversible.\footnote{Charles Bennett of IBM research showed in the
1970's that energy is used in computations to \emph{destroy}
information. Lossless computation can theoretically be done with
no energy usage whatsoever!} This is illustrated by the fact that
an electron in orbit does not emit radiation and spiral into the
nucleus.

\item Quantum evolution is linear. That is, if an experiment is
done on the state \ket{0} and on the state \ket{1}, then when
performed on mixed states the resulting state is the same state as
if the initial two answers were added.
\end{itemize}

So we are left with ``reversible" linear operators on the states,
that is, matrices! Since the resulting state should satisfy the
normalization requirement also, it turns out that any
\textbf{unitary} operation is allowed. Recall $U$ unitary means
$UU^\dag=I$. We now have :

\textbf{Quantum Mechanics Postulate 2: State Evolution} The
evolution of a \emph{closed} quantum system is described by a
\emph{unitary transformation}. That is, the state of a system
\ket{\psi} at time $t_1$ is related to the state \ket{\psi'} at
time $t_2$ by a unitary operator $U$ which depends only on the
times $t_1$ and $t_2$,
\begin{equation}
\ket{\psi}=U\ket{\psi'}
\end{equation}

Now we know how to specify quantum states and what is legal for
manipulating the state.

\subsubsection{Intermission - Linear Algebra Review} We will need several
facts, terms, and theorems from linear algebra. It will be easiest
to just fire them off: (we also combine some previous facts here
for the heck of it)

\begin{defn}{~} Let $H,A,B,U$ be linear operators on a vector space $V$.
\begin{enumerate}
\item $H^\dag$ is the conjugate transpose of H.

\item $H$ is \textbf{Hermitian} or self-adjoint if $H=H^\dag$.

\item \ket{\psi} is a column vector.

\item \bra{\psi} is the dual to \ket{\psi}, defined
$\bra{v}\equiv\ket{v}^\dag$.

\item
$\ket{\psi\phi}=\ket{\psi}\ket{\phi}=\ket{\psi}\otimes\ket{\phi}$.

\item $[A,B]=AB-BA$.

\item $\{A,B\}=AB+BA$.

\item $A$ is \textbf{normal} if $A^\dag A = A A^\dag$.

\item $U$ is \textbf{unitary} if $U^\dag U=I$.

\item $A$ is \textbf{positive} if $\brakett{\psi}{A}{\psi}\geq 0$
for all $\psi$.

\item $\brakett{\psi}{A}{\phi}$ is the inner product of $\psi$ and
$A\ket{\phi}$.

\item We define specific matrices (the first 4 are the Pauli matrices)\\
$\sigma_0=I$,

$\sigma_1=\sigma_x=X=\left(\begin{array}{rr}0&1\\1&0\end{array}\right)$,

$\sigma_2=\sigma_y=Y=\left(\begin{array}{rr}0&-i\\i&0\end{array}\right)$,

$\sigma_3=\sigma_z=Z=\left(\begin{array}{rr}1&0\\0&-1\end{array}\right)$,

$H=\frac{1}{\sqrt{2}}\left(\begin{array}{rr}1&1\\1&-1\end{array}\right)$,
$S=\left(\begin{array}{rr}1&0\\0&i\end{array}\right)$,
$T=\left(\begin{array}{rr}1&0\\0&e^\frac{i\pi}{4}\end{array}\right)$

\item For a unit vector $\vec{n}=(n_x,n_y,n_z)\in \mathbb{R}^3$,
define $\vec{n}.\vec{\sigma}\equiv
n_x\sigma_x+n_y\sigma_y+n_z\sigma_z$.

\item \textbf{Bloch Sphere} Given a state $a\ket{0}+b\ket{1}$ we
may assume $a$ is real by phase rotation. Then define for
$\phi\in[0,2\pi]$ and $\theta\in[0,\pi]$
\begin{eqnarray}
\cos\left(\frac{\theta}{2}\right)&=&a\\
e^{i\phi}\sin\left(\frac{\theta}{2}\right)&=&b
\end{eqnarray}
Then the point on the Bloch Sphere is
$(\cos\phi\sin\theta,\sin\phi\sin\theta,\cos\theta)$.

\item Define the three rotation matrices: $R_x(\theta)=e^{-\theta
X
i/2}=\cos\frac{\theta}{2}I-i\sin\frac{\theta}{2}X=\left(\begin{array}{rr}\cos(\theta/2)&-i\sin(\theta/2)\\-i\sin(\theta/2)&\cos(\theta/2)\end{array}\right)$

$R_y(\theta)=e^{-\theta Y
i/2}=\cos\frac{\theta}{2}I-i\sin\frac{\theta}{2}Y=\left(\begin{array}{rr}\cos(\theta/2)&-\sin(\theta/2)\\\sin(\theta/2)&\cos(\theta/2)\end{array}\right)$

$R_z(\theta)=e^{-\theta Z
i/2}=\cos\frac{\theta}{2}I-i\sin\frac{\theta}{2}Z=\left(\begin{array}{cc}e^{-i\theta/2}&0\\0&e^{i\theta/2}\end{array}\right)$

\item For a composite quantum system $AB$, the \textbf{partial
trace} is an operator from density operators on $AB$ to density
operators on $A$ defined for
$\text{tr}_B(\ket{a_1}\bra{a_2}\otimes\ket{b_1}\bra{b_2})=\braket{b_2}{b_1}\ket{a_1}\bra{a_2}$,
and extended by linearity. On matrices: let $\dim A=n$, $\dim
B=m$, then it takes a $mn$ by $mn$ matrix, and replaces each $m$
by $m$ sub-block with its trace to give a $n$ by $n$ matrix.

\item The \textbf{Bell States} are the 2-qubit basis states
\begin{align}
\ket{\beta_{00}}=&\;\frac{\ket{00}+\ket{11}}{\sqrt{2}}\\
\ket{\beta_{01}}=&\;\frac{\ket{01}+\ket{10}}{\sqrt{2}}\\
\ket{\beta_{10}}=&\;\frac{\ket{00}-\ket{11}}{\sqrt{2}}\\
\ket{\beta_{11}}=&\;\frac{\ket{01}-\ket{10}}{\sqrt{2}}
\end{align}

\end{enumerate}
\end{defn}

\textbf{Note:} The four Pauli matrices ($I$, $X$, $Y$, and $Z$)
have significance, since they form a basis of all linear operators
on one qubit, and correspond to similarly named actions on the
Bloch sphere.

We can write operators like $X$ in an equivalent operator
notation, which is often convenient to use in calculations. Noting
that $\bra{0}$ is a row vector, then $\ket{0}\bra{0}$ is a
$2\times 2$ matrix. We can write $X$ as:
\begin{eqnarray}
X =& \ket{0}\bra{1} + \ket{1}\bra{0}\\
 =& \binom{1}{0} (0\text{~}1) + \binom{0}{1} (1\text{~}0)\\
 =&\left(\begin{array}{cc}0&1\\0&0\end{array}\right)+\left(\begin{array}{cc}0&0\\1&0\end{array}\right)
 =& \left(\begin{array}{cc}0&1\\1&0\end{array}\right)
\end{eqnarray}
This is interpreted quickly: $X$ sends state 0 to 1, and vice
versa.

\textbf{Example:} As an example calculation, we compute
$\brakett{\beta_{00}}{I_2\otimes X}{\beta_{10}}$ two different
ways. The first way is matrix multiplication: Noting that
$\ket{00}=(1,0,0,0)^T$ and $\ket{11}=(0,0,0,1)^T$, we have
\begin{eqnarray}
\brakett{\beta_{00}}{I\otimes X}{\beta_{10}} =&
\left(\frac{\ket{00}+\ket{11}}{\sqrt{2}}\right)^\dag
\left(\begin{array}{cc}1 & 0\\ 0 & 1\end{array}\right) \otimes \left(\begin{array}{cc} 0 & 1\\ 1 & 0\end{array}\right) \left(\frac{\ket{00}-\ket{11}}{\sqrt{2}}\right)\\
=& \left(\frac{1}{\sqrt{2}}\right)^2
\left(\begin{array}{cccc}1&0&0&1\end{array}\right)\left(\begin{array}{cccc}0&1&0&0\\1&0&0&0\\0&0&0&1\\0&0&1&0
\end{array}\right)\left(\begin{array}{r}1\\0\\0\\-1\end{array}\right)\\
=& 0
\end{eqnarray}
For the other method, note as operators we can write
$I=\ket{0}\bra{0}+\ket{1}\bra{1}$, and $X$ swaps basis vectors,
giving $X=\ket{0}\bra{1}+\ket{1}\bra{0}$. Then we have
\begin{eqnarray}
I\otimes X =&
\left(\ket{0}\bra{0}+\ket{1}\bra{1}\right)\otimes\left(\ket{0}\bra{1}+\ket{1}\bra{0}\right)\\
=&\ket{00}\bra{01}+\ket{01}\bra{00}+\ket{10}\bra{11}+\ket{11}\bra{10}
\end{eqnarray}
where we used the fact
$\ket{a}\bra{b}\otimes\ket{c}\bra{d}=\ket{ac}\bra{bd}$. Apply this
and use orthonormality,
\begin{eqnarray}
\brakett{\beta_{00}}{I\otimes X}{\beta_{10}} =&
\left(\frac{\bra{00}+\bra{11}}{\sqrt{2}}\right)
\left(\ket{00}\bra{01}+\ket{01}\bra{00}+\ket{10}\bra{11}+\ket{11}\bra{10}\right)\left(\frac{\ket{00}-\ket{11}}{\sqrt{2}}\right)\\
=& \left(\frac{1}{\sqrt{2}}\right)^2(0+0+0+\dots+0)\\
=& 0
\end{eqnarray}
where we get terms like $\braket{00}{00}\braket{01}{00}=1\cdot
0=0$.

\begin{hwk}Write the matrices above in operator form for practice.\end{hwk}
\begin{hwk}Compute the eigen-values and eigen-vectors for the matrices defined above. They will be useful.\end{hwk}
\begin{hwk}Understand the behavior of each matrix above on the Bloch sphere  representation of a qubit.\end{hwk}

\subsubsection{Useful Linear Algebra Theorems}
\begin{theorem}[Cauchy Schwartz Inequality]
$|\braket{v}{w}|^2\leq\braket{v}{v}\braket{w}{w}$
\end{theorem}
\begin{theorem}[Spectral Decomposition]
Any normal operator $M$ on a vector space $V$ is diagonal with
respect to some orthonormal basis for $V$. Conversely, any
diagonalizable operator is normal.
\end{theorem}
\begin{proof}
Sketch: Induct on $d=\dim V$. $d=1$ is trivial. Let $\lambda$ be
an eigenvalue of $M$, $P$ the projector onto the $\lambda$
eigenspace, and $Q$ the projector onto the orthogonal complement.
$M=PMP+QMQ$ is diagonal with respect to some basis (strip off an
eigenvalue one at a time...)
\end{proof}

Check: There is a matrix $P$, with unit eigenvectors as columns,
so that $PMP^\dag$ is diagonal, with entries the eigenvalues.


\begin{theorem}[Simultaneous diagonalization]
Suppose $A$ and $B$ are Hermitian operators on a vector space V.
Then $[A,B]=0 \Leftrightarrow $ there exists an orthonormal basis
such that both $A$ and $B$ are diagonal with respect to that
basis.
\end{theorem}

\begin{theorem}[Polar decomposition]
Let $A$ be a linear operator on a vector space $V$. Then  there
exists a unitary $U$ and positive operators $J$ and $K$ such that
$$A=UJ=KU$$
where the unique $J$ and $K$ are given by $J\equiv\sqrt{A^\dag A}$
and $K\equiv\sqrt{AA^\dag}$. Moreover, $A$ invertible implies $U$
is unique.
\end{theorem}
\begin{proof}
$J\equiv \sqrt{A^\dag A}$ is positive, so spectral gives
$J=\sum_i\lambda_i\ket{i}\bra{i}$, $(\lambda_i\geq 0)$. Let
$\ket{\phi_i}=A\ket{i}$. For $\lambda_i\neq 0$, let
$\ket{e_i}=\ket{\phi_i}/\lambda_i$. Extend to orthogonal basis
$\ket{e_i}$, and define unitary $U\equiv\sum_i \ket{e_i}\bra{i}$.
This satisfies $A=UJ$. Multiply on left by adjoint
$A^\dag=JU^\dag$ giving $J^2=A^\dag A$, so $J=\sqrt{A^\dag A}$.

Then $A=UJ=UJU^\dag U=KU$ with $K=UJU^\dag$. This
$K=\sqrt{AA^\dag}$.
\end{proof}

\begin{theorem}[Singular value decomposition]
Let $A$ be a square matrix. Then there exists unitary $U$ and $V$,
and diagonal $D$, such that $$A=UDV$$ The diagonal elements of $D$
are called singular values of $A$.
\end{theorem}
\begin{proof}
By polar decomposition, $A=SJ$ for $S$ unitary and $J$ positive.
By spectral $J=TDT^\dag$, $T$ unitary, $D$ diagonal with
nonnegative entries. $U\equiv ST$ and $V\equiv T^\dag$ completes
the proof.
\end{proof}

\begin{theorem}
Every unitary $2\times2$ matrix can be expressed as
\begin{equation}
\left(\begin{matrix} e^{i\alpha} & 0 \\ 0 & e^{i\alpha}
\end{matrix}\right) \cdot
\left(\begin{matrix} e^{\frac{i\beta}{2}} & 0 \\ 0 &
e^{-\frac{i\beta}{2}}
\end{matrix}\right) \cdot
\left(\begin{matrix} \cos\frac{\gamma}{2} & -\sin\frac{\gamma}{2} \\
\sin\frac{\gamma}{2} & \;\;\;\cos\frac{\gamma}{2}
\end{matrix}\right) \cdot
\left(\begin{matrix} e^{\frac{i\delta}{2}} & 0 \\ 0 &
e^{-\frac{i\delta}{2}}
\end{matrix}\right)
\end{equation}
\end{theorem}

\textbf{Note}: Notice the third matrix is a usual rotation in the
plane. The 2nd and 4th matrices are Z-axis rotation on the Bloch
sphere, and the first matrix is merely a phase shift of the entire
state. This decomposition gives some intuition of how a single
qubit operator acts.

\begin{theorem}[Z-Y decomposition for a single
qubit]\label{t:ZYdecomp} $U$ is a unitary operation on a single
qubit. Then there are real numbers $\alpha,\beta,\delta,\gamma$
such that
$$U=e^{i\alpha}R_z(\beta)R_y(\gamma)R_z(\delta)
$$
\end{theorem}

\textbf{Note}: Similarly there are \textbf{X-Y}, \textbf{Z-X},
etc. decomposition theorems.

\begin{theorem}[ABC corollary]
Suppose $U$ is a unitary gate on a single qubit. Then there are
unitary operators $A$, $B$, and $C$, such that $ABC=I$, and
$U=e^{i\alpha}AXBXC$, where $\alpha$ is some overall phase factor.
\end{theorem}
\begin{proof} Apply theorem \ref{t:ZYdecomp} with $A\equiv
R_z(\beta)R_y(\gamma/2)$, $B\equiv
R_y(-\gamma/2)R_z(-(\delta+\beta)/2)$, and $C\equiv
R_z((\delta-\beta)/2)$.
\end{proof}

This weird looking theorem becomes very useful when trying to
construct quantum circuits. It allows one to use a Controlled NOT
gate (a circuit that flips a qubit based on the state of another
qubit) to contract arbitrary controlled $U$ gates.

\subsubsection{Useful Linear Algebra Facts!} Here are some facts that help in
computations and proofs when dealing with quantum computing.
\begin{enumerate}
\item Any complex $n\times n$ matrix $A$ can be written as a sum
of 4 positive Hermitian matrices: $A=B+iC$ with $B,C$ Hermitian
$B=\frac{1}{2}\left(A^*+A\right)$, and $C$ accordingly. Then any
Hermitian $B$ can be written as the sum of 2 positive Hermitian
matrices $B=\left(B+\lambda I\right)-\lambda I$ where $-\lambda$
is the most negative eigenvalue of $B$.

\item Every positive $A$ is of the form $BB^*$.

\item
$\ket{a_1}\bra{a_2}\otimes\ket{b_1}\bra{b_2}=\ket{a_1b_1}\bra{a_2b_2}$
(useful in partial trace operations).

\item Trace of kets: $\ket{\psi}=\sum_{i,j}a_{ij}\ket{ij}$, when
converted to a density matrix $\rho=\ket{\psi}\bra{\psi}$, and
then the trace is taken over the $j$, gives
$$tr_B(p)=\sum_i\left(\sum_j|a_{i,j}|^2\right)\ket{i}\bra{i},$$ so
it seems $tr_B(\ket{\psi})$ should be something like
$\sum_i\sqrt{\sum_j|a_{i,j}|^2}\;\ket{i}$. In particular, tracing
out some columns in $\ket{011010}$ removes those columns, but the
new kets are not a simple sum of the previous ones... It may be ok
to sum probabilities, then sqrt when collapsing, but I am not
clear.

\item Unitary also satisfies $U U^\dag=I$, so $U$ is normal and
has spectral decomposition (all QC ops unitary!).

\item Unitary preserves inner products.

\item \textbf{Positive $\Rightarrow$ Hermitian $\Rightarrow$
normal}.

\item $A^\dag A$ is positive for any linear operator $A$.

\item Tensor of unitary (resp Hermitian, positive, projector) is
unitary (resp,...).

\item If $P=\left(\begin{array}{rr}a &b\\c& d\end{array}\right)$
is invertible, then
$P^{-1}=\frac{1}{ad-bc}\left(\begin{array}{rr}d &-b\\-c&
a\end{array}\right)$.

\item Given eigenvectors $v_1$ and $v_2$ of $B$, with eigenvalues
$\lambda_1$ and $\lambda_2$, create the change of basis matrix
\mbox{$P=\left(\begin{array}{cc} v_1 & v_2\end{array}\right)$}.
Then the diagonal matrix $D$ is
$$D=\left(\begin{array}{rr}\lambda_1&0\\0&\lambda_2\end{array}\right)=P^{-1}BP$$

\item $W$ is a subspace of $V$ with basis \ket{i}. Projection to
$W$ is $P=\sum_i \ket{i}\bra{i}$. $Q=I-P$ is the orthogonal
complement.

\item Eigenvectors with distinct eigenvalues of a Hermitian
operator are orthogonal.

\item $\vec{n}.\vec{\sigma}$ has eigenvalues $\pm 1$ with
corresponding
eigenvectors $\left(\begin{array}{c}n_z\pm 1\\
n_x+in_y\end{array}\right)$.

\item $U$ unitary $\Rightarrow$ $U$ has a spectral decomposition
$\Rightarrow$ $U$ is diagonal in some orthonormal basis
$\Rightarrow$
$U=\text{diag}(e^{i\alpha_1},e^{i\alpha_2},\dots,e^{i\alpha_n})\Rightarrow
U$ has a \emph{unitary} $n^{th}$ root $V$, $V^n=U$.

\item tr $(\ket{\psi}\bra{\phi})=\braket{\phi}{\psi}$.

\item For unit vectors $\vec{r}$ and $\vec{s}$,
$(\vec{r}.\vec{\sigma})\cdot(\vec{s}.\vec{\sigma})=\vec{r}\cdot\vec{s}I+(\vec{r}\times\vec{s}).\vec{\sigma}$.

\end{enumerate}

\subsubsection{Some Basic Identities} There are lots of identities between
the operators we have above which will be useful in reducing
circuits later on. This is a good place to list some.
$$
[X,Y]=2iZ \quad [Y,Z]=2iX \quad [Z,X]= 2iY $$
$$
\{\sigma_i,\sigma_j\}=2\delta{ij} \;\;\text{if}\;\; i,j\neq0 \quad
\sigma_i^2=I
$$
$$R_z(\frac{\pi}{2})R_x(\frac{\pi}{2})R_z(\frac{\pi}{2})=e^{-i\pi/2}H$$
$$XYX=-Y\Rightarrow XR_y(\theta)X=R_y(-\theta)$$
$$HXH=Z \quad HYH=-Y \quad HZH=X$$
$$HTH=phase*R_x(\frac{\pi}{4})$$
$C$ is CNOT, $X_j$ is $X$ acting on qubit $j$, etc.
$$
\begin{array}{rclcrcl}
CX_1X&=&X_1X_2 && CY_1C&=&Y_1X_2 \\
CZ_1C&=&Z_1 &&  CX_2C&=&X_2 \\
CY_2C&=&Z_1Y_2 && CZ_2C&=&Z_1Z_2\\
R_{z,1}(\theta)C&=&CR_{z,1}(\theta) &&
R_{x,2}(\theta)C&=&CR_{x,2}(\theta)
\end{array}
$$

For $i,j=1,2,3$,
$\sigma_j\sigma_k=\delta_{jk}I+i\sum_{l=1}^3\epsilon_{jkl}\sigma_l$
where $\epsilon_{jkl}$ is the antisymmetric tensor on 3
indices.\footnote{Exercise 2.43 in Neilsen and Chuang. All of
these identities appear in the book, as exercises or in the text.}

\begin{hwk}
Check these identities using the matrix form and the operator form
to gain mastery of these calculations.
\end{hwk}

\subsubsection{Measuring the Qubits}
The final operation we need to understand about qubits is, how can
we get information back out of them? The process is called
measurement, and there are several equivalent ways to think about
it. We will cover the easiest to understand, intuitively and
mathematically. However, to gain the precise control over
measurements, we will have to resort later to an equivalent, yet
more complicated, measurement framework.

\textbf{Quantum Mechanics Postulate 3: State Measurement} Quantum
measurements are described by a collection $\{M_m\}$ of
\emph{measurement operators}. These are operators acting on the
state space of a system being measured. The index $m$ refers to
the measurement outcomes that may occur in the experiment. If the
state of the system is \ket{\psi} immediately before the
measurement, then the probability that result $m$ occurs is given
by
\begin{equation}
p(m) = \brakett{\psi}{M_m^\dag M_m}{\psi}
\end{equation}
and the state of the system after the measurement is
\begin{equation}
\frac{M_m\ket{\psi}}{\sqrt{p(m)}}
\end{equation}
The measurement operators satisfy the \emph{completeness equation}
\begin{equation}
\sum_m M_m^\dag M_m = I
\end{equation}

Finally, note cascaded measurements are single measurements. Thus
if your algorithm calls for a succession of measurements, this is
equivalent to a single measurement.

\subsubsection{Combining States and Partial States}

\textbf{Quantum Mechanics Postulate 4: State Combining} The state
space of a composite physical system is the tensor product of the
state spaces of the component systems. Moreover, if we have
systems numbered 1 through $n$, and system number $j$ is prepared
in the state \ket{\psi_j}, then the joint state of the total
system is
$\ket{\psi_1}\otimes\ket{\psi_2}\otimes\dots\ket{\psi_n}$.

And that is all there is to quantum mechanics (as far as we are
concerned). These four postulates form the basis of all that is
known about quantum mechanics, a physical theory that has stood
for over seven decades, and is used to explain phenomena at many
scales.

However, quantum mechanics does not mesh well with the other main
intellectual achievement in theoretical physics in the 20th
century, relativity. Combining these two theories into a unified
framework has occupied the best minds for over 50 years, and
currently superstring theory is the best candidate for this
unification.

Using the above postulates gives us an important theorem from
Wootters and Zurek \cite{WZ82}:

\subsubsection{The No Cloning Theorem}
\begin{thm}{\textbf{The No Cloning Theorem.}}It is impossible to
build a machine that can clone any given quantum state.
\end{thm}
This is in stark contrast to the classical case, where we copy
information all the time.
\begin{proof}
Suppose we have a machine with two slots: $A$ for the quantum
state \ket{\psi} to be cloned, and $B$ in some fixed initial state
\ket{s}, and the machine makes a copy of the quantum state $A$. By
the rules of quantum mechanics, the evolution $U$ is unitary, so
we have
\begin{equation}
\ket{\psi}\otimes\ket{s}\xrightarrow{U}\ket{\psi}\otimes\ket{\psi}
\end{equation}
Now suppose we have two states we wish to clone, \ket{\psi} and
\ket{\varphi}, giving
\begin{eqnarray*}
U\left(\ket{\psi}\otimes\ket{s}\right) =& \ket{\psi}\otimes\ket{\psi}\\
U\left(\ket{\varphi}\otimes\ket{s}\right) =&
\ket{\varphi}\otimes\ket{\varphi}
\end{eqnarray*}
Taking the inner product of these two equations, and using $U^\dag
U+=$:
\begin{eqnarray*}
\left(\bra{\varphi}\otimes\bra{s}\right)U^\dag U\left(\ket{\psi}\otimes\ket{s}\right) =& \left(\bra{\varphi}\otimes\bra{\varphi}\right)\left(\ket{\psi}\otimes\ket{\psi}\right)\\
\braket{\varphi}{\psi}\braket{s}{s} =& \braket{\varphi}{\psi}\braket{\varphi}{\psi}\\
\braket{\varphi}{\psi} =& \left(\braket{\varphi}{\psi}\right)^2
\end{eqnarray*}
This has solutions if and only if \braket{\varphi}{\psi} is 0 or
1, so cloning cannot be done for general states.\footnote{There is
a lot of research on what can be cloned, how much information can
be cloned, etc.}

\end{proof}

This ends the quantum mechanics for quantum computing primer.


\section{Random Group Generation}\label{s:RandomGroup}
This section is derived from Igor Pak's online lecture notes
\cite{Pak01}. The point of this section is to prove
\begin{theorem}\label{t:randomGroupGen}
Let $G$ be a finite group. For an integer $t\geq 0$, the
probability that $t+\lceil\log |G|\rceil$ elements chosen
uniformly at random from $G$ will generate $G$ is bounded by
\begin{equation}
\text{\emph{prob}}\{\left<g_1,g_2,\dots,g_{t+\lceil\log|G|\rceil}\right>=G\}\geq
1-\frac{1}{2^t}\text{ for }t\geq 0
\end{equation}
\end{theorem}

We will need some preliminaries to prove this. The idea will be to
bound the number of elements that should generate $G$ by the
number needed by the ``hardest'' to generate group, which can be
shown to be $\mathbb{Z}_2^r$, and then estimate how many elements
are needed to generate the latter group. First some notation:

\begin{defn}
\item Given a finite group $G$, and elements $g_1,g_2,\dots,g_t$
chosen uniformly at random from $G$, denote the probability that
the $g_i$ generate $G$ by
\begin{equation*}
\psi_t(G)=\text{\emph{prob}}\{\left<g_1,g_2,\dots,g_t\right>=G\}.
\end{equation*}
\end{defn}

First a reduction to a simpler group:

\begin{lemma}\label{l:gReduction}
Let $|G|\leq 2^r$, $r\geq 1$. Then for all $t\geq 1$,
$\psi_t(G)\geq \psi_t(\mathbb{Z}_2^r)$, where $\mathbb{Z}_2^r$ is
the additive group of binary $r$-tuples.
\end{lemma}
\begin{proof}
Fix $t$ and a subgroup $H\subsetneq G$. For a given sequence
$g_1,g_2,\dots,g_t$ of $G$, define subgroups $H_j$ of $G$ as
$H_1=\left<g_1\right>$, $H_2=\left<g_1,g_2\right>$,
$H_3=\left<g_1,g_2,g_3\right>$, etc. Let $H_j'$ be the similarly
defined subgroups of $\mathbb{Z}_2^r$. Let
$\tau_1,\tau_2,\dots,\tau_L$ be the indices $j$ where $H_j\neq
H_{j-1}$, and define similarly $\tau_1',\tau_2',\dots,\tau_R'$ for
the $H_j'$. We will induct on $|G|$. When $|G|=1$, the theorem is
true. Let $s=\tau_{L-1}$. We compute
\begin{eqnarray*}
\text{prob}\left(\tau_L-\tau_{L-1}\leq t\;\;|\;\; H_s=H\right)&=&1-\left(\frac{|H|}{|G|}\right)^t\\
&\geq&1-\frac{1}{2^t}\\
&=&1-\text{prob}\left(\tau_R'-\tau_{R-1}'>t\right)\\
&=&\text{prob}\left(\tau_R'-\tau_{R-1}'\leq t\right)
\end{eqnarray*}
This, combined with the induction assumption
$\text{prob}\left(\tau_{L-1}\leq
t\right)\geq\text{prob}\left(\tau_{R-1}'\leq t\right)$, gives
\begin{equation}
\text{prob}\left(\tau_L\leq t | H_s = H\right)\geq
\text{prob}\left(\tau_R'\leq t\right)=
\psi_t\left(\mathbb{Z}^r_2\right)
\end{equation}
This holds for any fixed $t$ and $H$, so the theorem follows.
\end{proof}

\begin{lemma}\label{l:z2rbound}\footnote{The article \cite{Pak01} proved a stronger form, but this is sufficient for our purposes.}
$$\psi_{r+t}(\mathbb{Z}_2^r)\geq 1-\frac{1}{2^t}\text{ for }t\geq 0$$
\end{lemma}
\begin{proof}
View $\mathbb{Z}^r_2$ as the $r$ dimensional vector space over the
2 element field $\mathbb{Z}_2$. Then $\psi_{r+t}(\mathbb{Z}^r_2)$
is the probability that $r+t$ randomly chosen vectors spans the
entire $r$ dimensional space $\mathbb{Z}_2^r$. If we write the
$r+t$ vectors as rows of a $(r+t)\times r$ matrix, then this is
the probability that the matrix has column rank $r$. This happens
if and only if all $r$ columns are linearly independent.

The first column (which has $r+t$ entries) is nonzero with
probability $\left(1-\frac{1}{2^{r+t}}\right)$. The probability
that the second column is linearly independent of the first is
$\left(1-\frac{1}{2^{r+t-1}}\right)$, and so on. Thus for $t\geq
0$ we get that
\begin{eqnarray*}
\psi_{r+t}(\mathbb{Z}_2^r)&=&\left(1-\frac{1}{2^{t+r}}\right)\left(1-\frac{1}{2^{t+r-1}}\right)\dots\left(1-\frac{1}{2^{t+1}}\right)\\
&=&1-\frac{1}{2^t}\sum_{a=1}^r\frac{1}{2^a}+\frac{1}{4^t}\sum_{\substack{a,b=1\\a\neq
b}}^r \frac{1}{2^a}\frac{1}{2^b}-
\frac{1}{8^t}\sum_{\substack{a,b,c=1\\a\neq b\neq c}}^r\frac{1}{2^{a+b+c}}+\dots\\
&=&1-\frac{1}{2^t}\left(1-\frac{1}{2^r}\right) +
\frac{1}{4^{t}}\sum_{\substack{a,b=1\\a\neq
b}}^r\left(\frac{1}{2^{a+b}}-\sum_{\substack{c=1\\c\neq a\neq
b}}^r\frac{1}{2^{a+b+c}}\right)+\dots\\
&\geq&1-\frac{1}{2^t}+\frac{1}{4^t}\sum_{a\neq
b}\left(\frac{1}{2^{a+b}}-\sum_{c=1}^r\frac{1}{2^{a+b+c}}\right)+\dots\\
&=&1-\frac{1}{2^t}+\frac{1}{4^t}\sum_{a\neq
b}\left(\frac{1}{2^{a+b}}\left(1-\sum_{c=1}^r\frac{1}{2^{c}}\right)\right)+\dots\\
&\geq&1-\frac{1}{2^t}
\end{eqnarray*}
Note that in the lines above that the ellipses denotes a finite
number of terms, which can be paired up similarly to the two terms
shown, with at most one final positive term which can then be
dropped in the inequality.
\end{proof}
Now we prove theorem \ref{t:randomGroupGen}.
\begin{proof}
Set $r=\lceil\log |G|\rceil$, giving $|G|\leq 2^r$. Then for
$t\geq 0$ we have $\psi_{t+r}(G)\geq\psi_{t+r}(\mathbb{Z}_2^r)$ by
lemma \ref{l:gReduction}, and then this is $\geq 1-\frac{1}{2^t}$
by lemma \ref{l:z2rbound}, which proves theorem
\ref{t:randomGroupGen}.
\end{proof}

Finally, note there are much better bounds, but this one gives the
exponential performance we need for our purposes.


\section{GCD Probabilities}\label{s:GCDProbabilities}
This appendix shows the proof that the probability of the GCD of
integers uniformly sampled from a fixed range becomes
exponentially close to 1 in terms of the number of samples. The
formal result is lemma \ref{l:trials}.

Unfortunately we need the next result without proof to start off
the result.

\begin{lemma}[\cite{Tamb36}]\label{l:phibound}
Let $\varphi(n)$ be the Euler totient function\footnote{For a
positive integer $n$, $\varphi(n)$ returns the number of positive
integers less than $n$ and relatively prime to $n$.}. Then for any
positive integer $n$,
\begin{equation}
\left|\sum_{c=1}^n\varphi(c)-\frac{3 n^2}{\pi^2}\right| < n\ln n
\end{equation}
where $\ln n$ is log base $e$.
\end{lemma}

\begin{lemma}\label{l:gcd2}
Fix an integer $n>0$. Choose two nonnegative integers $a,b\leq n$
uniformly at random. Then the probability that $\gcd(a,b)=1$ is
$\geq\frac{1}{2}$.
\end{lemma}
\begin{proof}
Given the uniformly randomly chosen integers $a,b$, the
probability that $\max\{a,b\}=c$ is $\frac{2c+1}{(n+1)^2}$. This
can be seen by looking at a matrix with $a_{ij}$ entry $(i,j)$,
and counting elements, for $i,j\in\{0,1,\dots,n\}$. Assuming
$c>0$, which happens with probability
$p_0=\frac{(n+1)^2-1}{(n+1)^2}$, the probability that the second
integer is relatively prime to the largest one $c$ is precisely
$\frac{\varphi(c)}{c}$. So the probability $p_n$ that
$\gcd(a,b)=1$ is exactly
\begin{eqnarray}
p_n&=&p_0\sum_{c=1}^n \frac{2c+1}{(n+1)^2}\frac{\varphi(c)}{c}\\
&=&\frac{n^2+2n}{(n+1)^4}\sum_{c=1}^n\left(2+\frac{1}{c}\right)\varphi(c)\label{e:gcdeqn}\\
&\geq&\frac{2n^2+4n}{(n+1)^4}\sum_{c=1}^n\varphi(c)
\end{eqnarray}
By lemma \ref{l:phibound} $\sum\varphi(c)>\frac{3n^2}{\pi^2}-n\log
n$, giving
\begin{equation}
p_n\geq\left(\frac{2n^2+4n}{(n+1)^4}\right)\left(\frac{3n^2-\pi^2
n\log n}{\pi^2}\right)
\end{equation}
Denoting the right hand side by $f(n)$, it is easy to check $f$ is
increasing\footnote{$\lim_{n\rightarrow\infty}f(n)=6/\pi^2$,
agreeing with Dirichlet's 1849 theorem to that effect.} for $n\geq
4$ and that $f(94)>0.5$, proving the proposition for integers
$n\geq 94$. The remaining cases $n=1,2,\dots,93$ can be easily
(yet tediously) checked using equation \ref{e:gcdeqn}. I recommend
Mathematica or Maple.
\end{proof}

\begin{lemma}\label{l:trials}
Suppose we have $k\geq 2$ uniformly random samples
$t_1,t_2,\dots,t_k$ from the integers $\{0,1,\dots,d-1\}$ for an
integer $d\geq 2$. Then
$$\text{\emph{prob}}\left(\gcd(t_1,t_2,\dots,t_k)=1\right)\geq
1-\left(\frac{1}{2}\right)^{k/2}$$
\end{lemma}
\begin{proof}
Consider the samples taken as pairs. Certainly if any pair
$t_{2j-1}$ and $t_{2j}$ are relatively prime, then
$\gcd(t_1,t_2,\dots,t_k)=1$. By lemma \ref{l:gcd2} the probability
that $\gcd(t_{2j-1},t_{2j})>1$ is $\leq\frac{1}{2}$, so the
probability that every such pair, $j=1,2,\dots,\lfloor
k/2\rfloor$, has $\gcd>1$ is
$\leq\left(\frac{1}{2}\right)^{\lfloor
k/2\rfloor}\leq\left(\frac{1}{2}\right)^{k/2}$. Thus the
probability that $\gcd(t_1,t_2,\dots,t_k)=1$ is $\geq
1-\left(\frac{1}{2}\right)^{k/2}$.
\end{proof}

Finally we note that the above estimates and probabilities are
very conservative, yet yield the essential fact that the
probability of success increases exponentially with the number of
trials.


\nocite{AbramsLloyd98, Bennett73, EH99b} 

\bibliographystyle{amsplain}
\bibliography{qbib,cbib,mbib}


\end{document}